\hsize=31pc
\vsize=49pc
\lineskip=0pt
\parskip=0pt plus 1pt
\hfuzz=1pt
\vfuzz=2pt
\pretolerance=2500
\tolerance=5000
\vbadness=5000
\hbadness=5000
\widowpenalty=500
\clubpenalty=200
\brokenpenalty=500
\predisplaypenalty=200
\voffset=-1pc
\nopagenumbers
\catcode`@=11
\newif\ifams
\amsfalse
%\amstrue
%
%%%%%%%%%%%%%%%%%%%%%%%%%%%%%%%%%%%%%%%%%%%%%%%%%%%%%%%%%%%%%
%                                                           %
%  The following section may be commented out and           %
%  \ifams set to either \amstrue to use the AMS fonts       %
%  or \amsfalse if they are not available                   %
%                                                           %
%%%%%%%%%%%%%%%%%%%%%%%%%%%%%%%%%%%%%%%%%%%%%%%%%%%%%%%%%%%%%
%
%\def\Yesreply{Y }
%\def\Noreply{N }
%\def\yesreply{y }
%\def\noreply{n }
%\newif\ifnotyorn
%\message{Do you want to use AMSfonts, msam and msbm? Y or N: }%
%\loop
%\read-1 to \reply
%\ifx\reply\yesreply\global\amstrue\notyornfalse
%\else\ifx\reply\Yesreply\global\amstrue\notyornfalse
%\else\ifx\reply\noreply\global\amsfalse\notyornfalse
%\else\ifx\reply\Noreply\global\amsfalse\notyornfalse
%\else\notyorntrue
%\message{Please type y or Y  (Yes) or n or N (No)}\fi\fi\fi\fi
%\ifnotyorn\repeat
%
%%%%%%%%%%%%%%%%%%%%%%%%%%%%%%%%%%%%%%%%%%%%%%%%%%%%%%%%%%%%
%
\newfam\bdifam
\newfam\bsyfam
\newfam\bssfam
%\newfam\msafam
%\newfam\msbfam
%
\newif\ifxxpt
\newif\ifxviipt
\newif\ifxivpt
\newif\ifxiipt
\newif\ifxipt
\newif\ifxpt
\newif\ifixpt
\newif\ifviiipt
\newif\ifviipt
\newif\ifvipt
\newif\ifvpt
%
% Headings in 20pt, 17pt or 14pt
%
\def\headsize#1#2{\def\headb@seline{#2}%
                \ifnum#1=20\def\HEAD{twenty}%
                           \def\smHEAD{twelve}%
                           \def\vsHEAD{nine}%
                           \ifxxpt\else\xdef\f@ntsize{\HEAD}%
                           \def\m@g{4}\def\s@ze{20.74}%
                           \loadheadfonts\xxpttrue\fi
                           \ifxiipt\else\xdef\f@ntsize{\smHEAD}%
                           \def\m@g{1}\def\s@ze{12}%
                           \loadxiiptfonts\xiipttrue\fi
                           \ifixpt\else\xdef\f@ntsize{\vsHEAD}%
                           \def\s@ze{9}%
                           \loadsmallfonts\ixpttrue\fi
                      \else
                \ifnum#1=17\def\HEAD{seventeen}%
                           \def\smHEAD{eleven}%
                           \def\vsHEAD{eight}%
                           \ifxviipt\else\xdef\f@ntsize{\HEAD}%
                           \def\m@g{3}\def\s@ze{17.28}%
                           \loadheadfonts\xviipttrue\fi
                           \ifxipt\else\xdef\f@ntsize{\smHEAD}%
                           \loadxiptfonts\xipttrue\fi
                           \ifviiipt\else\xdef\f@ntsize{\vsHEAD}%
                           \def\s@ze{8}%
                           \loadsmallfonts\viiipttrue\fi
                      \else\def\HEAD{fourteen}%
                           \def\smHEAD{ten}%
                           \def\vsHEAD{seven}%
                           \ifxivpt\else\xdef\f@ntsize{\HEAD}%
                           \def\m@g{2}\def\s@ze{14.4}%
                           \loadheadfonts\xivpttrue\fi
                           \ifxpt\else\xdef\f@ntsize{\smHEAD}%
                           \def\s@ze{10}%
                           \loadxptfonts\xpttrue\fi
                           \ifviipt\else\xdef\f@ntsize{\vsHEAD}%
                           \def\s@ze{7}%
                           \loadviiptfonts\viipttrue\fi
                \ifnum#1=14\else
                \message{Header size should be 20, 17 or 14 point
                              will now default to 14pt}\fi
                \fi\fi\headfonts}
%
% Text in 12pt, 11pt or 10pt
%
\def\textsize#1#2{\def\textb@seline{#2}%
                 \ifnum#1=12\def\TEXT{twelve}%
                           \def\smTEXT{eight}%
                           \def\vsTEXT{six}%
                           \ifxiipt\else\xdef\f@ntsize{\TEXT}%
                           \def\m@g{1}\def\s@ze{12}%
                           \loadxiiptfonts\xiipttrue\fi
                           \ifviiipt\else\xdef\f@ntsize{\smTEXT}%
                           \def\s@ze{8}%
                           \loadsmallfonts\viiipttrue\fi
                           \ifvipt\else\xdef\f@ntsize{\vsTEXT}%
                           \def\s@ze{6}%
                           \loadviptfonts\vipttrue\fi
                      \else
                \ifnum#1=11\def\TEXT{eleven}%
                           \def\smTEXT{seven}%
                           \def\vsTEXT{five}%
                           \ifxipt\else\xdef\f@ntsize{\TEXT}%
                           \def\s@ze{11}%
                           \loadxiptfonts\xipttrue\fi
                           \ifviipt\else\xdef\f@ntsize{\smTEXT}%
                           \loadviiptfonts\viipttrue\fi
                           \ifvpt\else\xdef\f@ntsize{\vsTEXT}%
                           \def\s@ze{5}%
                           \loadvptfonts\vpttrue\fi
                      \else\def\TEXT{ten}%
                           \def\smTEXT{seven}%
                           \def\vsTEXT{five}%
                           \ifxpt\else\xdef\f@ntsize{\TEXT}%
                           \loadxptfonts\xpttrue\fi
                           \ifviipt\else\xdef\f@ntsize{\smTEXT}%
                           \def\s@ze{7}%
                           \loadviiptfonts\viipttrue\fi
                           \ifvpt\else\xdef\f@ntsize{\vsTEXT}%
                           \def\s@ze{5}%
                           \loadvptfonts\vpttrue\fi
                \ifnum#1=10\else
                \message{Text size should be 12, 11 or 10 point
                              will now default to 10pt}\fi
                \fi\fi\textfonts}
%
% Small sized material in 10pt, 9pt or 8pt
%
\def\smallsize#1#2{\def\smallb@seline{#2}%
                 \ifnum#1=10\def\SMALL{ten}%
                           \def\smSMALL{seven}%
                           \def\vsSMALL{five}%
                           \ifxpt\else\xdef\f@ntsize{\SMALL}%
                           \loadxptfonts\xpttrue\fi
                           \ifviipt\else\xdef\f@ntsize{\smSMALL}%
                           \def\s@ze{7}%
                           \loadviiptfonts\viipttrue\fi
                           \ifvpt\else\xdef\f@ntsize{\vsSMALL}%
                           \def\s@ze{5}%
                           \loadvptfonts\vpttrue\fi
                       \else
                 \ifnum#1=9\def\SMALL{nine}%
                           \def\smSMALL{six}%
                           \def\vsSMALL{five}%
                           \ifixpt\else\xdef\f@ntsize{\SMALL}%
                           \def\s@ze{9}%
                           \loadsmallfonts\ixpttrue\fi
                           \ifvipt\else\xdef\f@ntsize{\smSMALL}%
                           \def\s@ze{6}%
                           \loadviptfonts\vipttrue\fi
                           \ifvpt\else\xdef\f@ntsize{\vsSMALL}%
                           \def\s@ze{5}%
                           \loadvptfonts\vpttrue\fi
                       \else
                           \def\SMALL{eight}%
                           \def\smSMALL{six}%
                           \def\vsSMALL{five}%
                           \ifviiipt\else\xdef\f@ntsize{\SMALL}%
                           \def\s@ze{8}%
                           \loadsmallfonts\viiipttrue\fi
                           \ifvipt\else\xdef\f@ntsize{\smSMALL}%
                           \def\s@ze{6}%
                           \loadviptfonts\vipttrue\fi
                           \ifvpt\else\xdef\f@ntsize{\vsSMALL}%
                           \def\s@ze{5}%
                           \loadvptfonts\vpttrue\fi
                 \ifnum#1=8\else\message{Small size should be 10, 9 or
                            8 point will now default to 8pt}\fi
                \fi\fi\smallfonts}
\def\F@nt{\expandafter\font\csname}
\def\Sk@w{\expandafter\skewchar\csname}
\def\@nd{\endcsname}
\def\@step#1{ scaled \magstep#1}
\def\@half{ scaled \magstephalf}
\def\@t#1{ at #1pt}
%
% For 14, 17 and 20 point fonts use \loadheadfonts
%
\def\loadheadfonts{\bigf@nts
\F@nt \f@ntsize bdi\@nd=cmmib10 \@t{\s@ze}%
\Sk@w \f@ntsize bdi\@nd='177
\F@nt \f@ntsize bsy\@nd=cmbsy10 \@t{\s@ze}%
\Sk@w \f@ntsize bsy\@nd='60
\F@nt \f@ntsize bss\@nd=cmssbx10 \@t{\s@ze}}
%
% For 12 point fonts use \loadxiiptfonts
%
\def\loadxiiptfonts{\bigf@nts
\F@nt \f@ntsize bdi\@nd=cmmib10 \@step{\m@g}%
\Sk@w \f@ntsize bdi\@nd='177
\F@nt \f@ntsize bsy\@nd=cmbsy10 \@step{\m@g}%
\Sk@w \f@ntsize bsy\@nd='60
\F@nt \f@ntsize bss\@nd=cmssbx10 \@step{\m@g}}
%
% For 11 point fonts use \loadxiptfonts
%
\def\loadxiptfonts{%
\font\elevenrm=cmr10 \@half
\font\eleveni=cmmi10 \@half
\skewchar\eleveni='177
\font\elevensy=cmsy10 \@half
\skewchar\elevensy='60
\font\elevenex=cmex10 \@half
\font\elevenit=cmti10 \@half
\font\elevensl=cmsl10 \@half
\font\elevenbf=cmbx10 \@half
\font\eleventt=cmtt10 \@half
\ifams\font\elevenmsa=msam10 \@half
\font\elevenmsb=msbm10 \@half\else\fi
\font\elevenbdi=cmmib10 \@half
\skewchar\elevenbdi='177
\font\elevenbsy=cmbsy10 \@half
\skewchar\elevenbsy='60
\font\elevenbss=cmssbx10 \@half}
%
% For 10 point fonts use \loadxptfonts
%
\def\loadxptfonts{%
\font\tenbdi=cmmib10
\skewchar\tenbdi='177
\font\tenbsy=cmbsy10
\skewchar\tenbsy='60
\ifams\font\tenmsa=msam10
\font\tenmsb=msbm10\else\fi
\font\tenbss=cmssbx10}%
%
% For 8 and 9 point fonts use \loadsmallfonts
%
\def\loadsmallfonts{\smallf@nts
\ifams
\F@nt \f@ntsize ex\@nd=cmex\s@ze
\else
\F@nt \f@ntsize ex\@nd=cmex10\fi
\F@nt \f@ntsize it\@nd=cmti\s@ze
\F@nt \f@ntsize sl\@nd=cmsl\s@ze
\F@nt \f@ntsize tt\@nd=cmtt\s@ze}
%
% For 7 point fonts use \loadviiptfonts
%
\def\loadviiptfonts{%
\font\sevenit=cmti7
\font\sevensl=cmsl8 at 7pt
\ifams\font\sevenmsa=msam7
\font\sevenmsb=msbm7
\font\sevenex=cmex7
\font\sevenbsy=cmbsy7
\font\sevenbdi=cmmib7\else
\font\sevenex=cmex10
\font\sevenbsy=cmbsy10 at 7pt
\font\sevenbdi=cmmib10 at 7pt\fi
\skewchar\sevenbsy='60
\skewchar\sevenbdi='177
\font\sevenbss=cmssbx10 at 7pt}%
%
%  For 6 point fonts use \loadviptfonts
%
\def\loadviptfonts{\smallf@nts
\ifams\font\sixex=cmex7 at 6pt\else
\font\sixex=cmex10\fi
\font\sixit=cmti7 at 6pt}
%
% For 5 point fonts use \loadvptfonts
%
\def\loadvptfonts{%
\font\fiveit=cmti7 at 5pt
\ifams\font\fiveex=cmex7 at 5pt
\font\fivebdi=cmmib5
\font\fivebsy=cmbsy5
\font\fivemsa=msam5
\font\fivemsb=msbm5\else
\font\fiveex=cmex10
\font\fivebdi=cmmib10 at 5pt
\font\fivebsy=cmbsy10 at 5pt\fi
\skewchar\fivebdi='177
\skewchar\fivebsy='60
\font\fivebss=cmssbx10 at 5pt}
\def\bigf@nts{%
\F@nt \f@ntsize rm\@nd=cmr10 \@step{\m@g}%
\F@nt \f@ntsize i\@nd=cmmi10 \@step{\m@g}%
\Sk@w \f@ntsize i\@nd='177
\F@nt \f@ntsize sy\@nd=cmsy10 \@step{\m@g}%
\Sk@w \f@ntsize sy\@nd='60
\F@nt \f@ntsize ex\@nd=cmex10 \@step{\m@g}%
\F@nt \f@ntsize it\@nd=cmti10 \@step{\m@g}%
\F@nt \f@ntsize sl\@nd=cmsl10 \@step{\m@g}%
\F@nt \f@ntsize bf\@nd=cmbx10 \@step{\m@g}%
\F@nt \f@ntsize tt\@nd=cmtt10 \@step{\m@g}%
\ifams
\F@nt \f@ntsize msa\@nd=msam10 \@step{\m@g}%
\F@nt \f@ntsize msb\@nd=msbm10 \@step{\m@g}\else\fi}
\def\smallf@nts{%
\F@nt \f@ntsize rm\@nd=cmr\s@ze
\F@nt \f@ntsize i\@nd=cmmi\s@ze
\Sk@w \f@ntsize i\@nd='177
\F@nt \f@ntsize sy\@nd=cmsy\s@ze
\Sk@w \f@ntsize sy\@nd='60
\F@nt \f@ntsize bf\@nd=cmbx\s@ze
\ifams
\F@nt \f@ntsize bdi\@nd=cmmib\s@ze
\F@nt \f@ntsize bsy\@nd=cmbsy\s@ze
\F@nt \f@ntsize msa\@nd=msam\s@ze
\F@nt \f@ntsize msb\@nd=msbm\s@ze
\else
\F@nt \f@ntsize bdi\@nd=cmmib10 \@t{\s@ze}%
\F@nt \f@ntsize bsy\@nd=cmbsy10 \@t{\s@ze}\fi
\Sk@w \f@ntsize bdi\@nd='177
\Sk@w \f@ntsize bsy\@nd='60
\F@nt \f@ntsize bss\@nd=cmssbx10 \@t{\s@ze}}%
%
% Fonts for headings
%
\def\headfonts{%
\textfont0=\csname\HEAD rm\@nd
\scriptfont0=\csname\smHEAD rm\@nd
\scriptscriptfont0=\csname\vsHEAD rm\@nd
\def\rm{\fam0\csname\HEAD rm\@nd
\def\sc{\csname\smHEAD rm\@nd}}%
\textfont1=\csname\HEAD i\@nd
\scriptfont1=\csname\smHEAD i\@nd
\scriptscriptfont1=\csname\vsHEAD i\@nd
\textfont2=\csname\HEAD sy\@nd
\scriptfont2=\csname\smHEAD sy\@nd
\scriptscriptfont2=\csname\vsHEAD sy\@nd
\textfont3=\csname\HEAD ex\@nd
\scriptfont3=\csname\smHEAD ex\@nd
\scriptscriptfont3=\csname\smHEAD ex\@nd
\textfont\itfam=\csname\HEAD it\@nd
\scriptfont\itfam=\csname\smHEAD it\@nd
\scriptscriptfont\itfam=\csname\vsHEAD it\@nd
\def\it{\fam\itfam\csname\HEAD it\@nd
\def\sc{\csname\smHEAD it\@nd}}%
\textfont\slfam=\csname\HEAD sl\@nd
\def\sl{\fam\slfam\csname\HEAD sl\@nd
\def\sc{\csname\smHEAD sl\@nd}}%
\textfont\bffam=\csname\HEAD bf\@nd
\scriptfont\bffam=\csname\smHEAD bf\@nd
\scriptscriptfont\bffam=\csname\vsHEAD bf\@nd
\def\bf{\fam\bffam\csname\HEAD bf\@nd
\def\sc{\csname\smHEAD bf\@nd}}%
\textfont\ttfam=\csname\HEAD tt\@nd
\def\tt{\fam\ttfam\csname\HEAD tt\@nd}%
\textfont\bdifam=\csname\HEAD bdi\@nd
\scriptfont\bdifam=\csname\smHEAD bdi\@nd
\scriptscriptfont\bdifam=\csname\vsHEAD bdi\@nd
\def\bdi{\fam\bdifam\csname\HEAD bdi\@nd}%
\textfont\bsyfam=\csname\HEAD bsy\@nd
\scriptfont\bsyfam=\csname\smHEAD bsy\@nd
\def\bsy{\fam\bsyfam\csname\HEAD bsy\@nd}%
\textfont\bssfam=\csname\HEAD bss\@nd
\scriptfont\bssfam=\csname\smHEAD bss\@nd
\scriptscriptfont\bssfam=\csname\vsHEAD bss\@nd
\def\bss{\fam\bssfam\csname\HEAD bss\@nd}%
\ifams
\textfont\msafam=\csname\HEAD msa\@nd
\scriptfont\msafam=\csname\smHEAD msa\@nd
\scriptscriptfont\msafam=\csname\vsHEAD msa\@nd
\textfont\msbfam=\csname\HEAD msb\@nd
\scriptfont\msbfam=\csname\smHEAD msb\@nd
\scriptscriptfont\msbfam=\csname\vsHEAD msb\@nd
\else\fi
\normalbaselineskip=\headb@seline pt%
\setbox\strutbox=\hbox{\vrule height.7\normalbaselineskip
depth.3\baselineskip width0pt}%
\def\sc{\csname\smHEAD rm\@nd}\normalbaselines\bf}
%
% Fonts for text
%
\def\textfonts{%
\textfont0=\csname\TEXT rm\@nd
\scriptfont0=\csname\smTEXT rm\@nd
\scriptscriptfont0=\csname\vsTEXT rm\@nd
\def\rm{\fam0\csname\TEXT rm\@nd
\def\sc{\csname\smTEXT rm\@nd}}%
\textfont1=\csname\TEXT i\@nd
\scriptfont1=\csname\smTEXT i\@nd
\scriptscriptfont1=\csname\vsTEXT i\@nd
\textfont2=\csname\TEXT sy\@nd
\scriptfont2=\csname\smTEXT sy\@nd
\scriptscriptfont2=\csname\vsTEXT sy\@nd
\textfont3=\csname\TEXT ex\@nd
\scriptfont3=\csname\smTEXT ex\@nd
\scriptscriptfont3=\csname\smTEXT ex\@nd
\textfont\itfam=\csname\TEXT it\@nd
\scriptfont\itfam=\csname\smTEXT it\@nd
\scriptscriptfont\itfam=\csname\vsTEXT it\@nd
\def\it{\fam\itfam\csname\TEXT it\@nd
\def\sc{\csname\smTEXT it\@nd}}%
\textfont\slfam=\csname\TEXT sl\@nd
\def\sl{\fam\slfam\csname\TEXT sl\@nd
\def\sc{\csname\smTEXT sl\@nd}}%
\textfont\bffam=\csname\TEXT bf\@nd
\scriptfont\bffam=\csname\smTEXT bf\@nd
\scriptscriptfont\bffam=\csname\vsTEXT bf\@nd
\def\bf{\fam\bffam\csname\TEXT bf\@nd
\def\sc{\csname\smTEXT bf\@nd}}%
\textfont\ttfam=\csname\TEXT tt\@nd
\def\tt{\fam\ttfam\csname\TEXT tt\@nd}%
\textfont\bdifam=\csname\TEXT bdi\@nd
\scriptfont\bdifam=\csname\smTEXT bdi\@nd
\scriptscriptfont\bdifam=\csname\vsTEXT bdi\@nd
\def\bdi{\fam\bdifam\csname\TEXT bdi\@nd}%
\textfont\bsyfam=\csname\TEXT bsy\@nd
\scriptfont\bsyfam=\csname\smTEXT bsy\@nd
\def\bsy{\fam\bsyfam\csname\TEXT bsy\@nd}%
\textfont\bssfam=\csname\TEXT bss\@nd
\scriptfont\bssfam=\csname\smTEXT bss\@nd
\scriptscriptfont\bssfam=\csname\vsTEXT bss\@nd
\def\bss{\fam\bssfam\csname\TEXT bss\@nd}%
\ifams
\textfont\msafam=\csname\TEXT msa\@nd
\scriptfont\msafam=\csname\smTEXT msa\@nd
\scriptscriptfont\msafam=\csname\vsTEXT msa\@nd
\textfont\msbfam=\csname\TEXT msb\@nd
\scriptfont\msbfam=\csname\smTEXT msb\@nd
\scriptscriptfont\msbfam=\csname\vsTEXT msb\@nd
\else\fi
\normalbaselineskip=\textb@seline pt
\setbox\strutbox=\hbox{\vrule height.7\normalbaselineskip
depth.3\baselineskip width0pt}%
\everymath{}%
\def\sc{\csname\smTEXT rm\@nd}\normalbaselines\rm}
%
% Fonts for small material (captions, footnotes etc)
%
\def\smallfonts{%
\textfont0=\csname\SMALL rm\@nd
\scriptfont0=\csname\smSMALL rm\@nd
\scriptscriptfont0=\csname\vsSMALL rm\@nd
\def\rm{\fam0\csname\SMALL rm\@nd
\def\sc{\csname\smSMALL rm\@nd}}%
\textfont1=\csname\SMALL i\@nd
\scriptfont1=\csname\smSMALL i\@nd
\scriptscriptfont1=\csname\vsSMALL i\@nd
\textfont2=\csname\SMALL sy\@nd
\scriptfont2=\csname\smSMALL sy\@nd
\scriptscriptfont2=\csname\vsSMALL sy\@nd
\textfont3=\csname\SMALL ex\@nd
\scriptfont3=\csname\smSMALL ex\@nd
\scriptscriptfont3=\csname\smSMALL ex\@nd
\textfont\itfam=\csname\SMALL it\@nd
\scriptfont\itfam=\csname\smSMALL it\@nd
\scriptscriptfont\itfam=\csname\vsSMALL it\@nd
\def\it{\fam\itfam\csname\SMALL it\@nd
\def\sc{\csname\smSMALL it\@nd}}%
\textfont\slfam=\csname\SMALL sl\@nd
\def\sl{\fam\slfam\csname\SMALL sl\@nd
\def\sc{\csname\smSMALL sl\@nd}}%
\textfont\bffam=\csname\SMALL bf\@nd
\scriptfont\bffam=\csname\smSMALL bf\@nd
\scriptscriptfont\bffam=\csname\vsSMALL bf\@nd
\def\bf{\fam\bffam\csname\SMALL bf\@nd
\def\sc{\csname\smSMALL bf\@nd}}%
\textfont\ttfam=\csname\SMALL tt\@nd
\def\tt{\fam\ttfam\csname\SMALL tt\@nd}%
\textfont\bdifam=\csname\SMALL bdi\@nd
\scriptfont\bdifam=\csname\smSMALL bdi\@nd
\scriptscriptfont\bdifam=\csname\vsSMALL bdi\@nd
\def\bdi{\fam\bdifam\csname\SMALL bdi\@nd}%
\textfont\bsyfam=\csname\SMALL bsy\@nd
\scriptfont\bsyfam=\csname\smSMALL bsy\@nd
\def\bsy{\fam\bsyfam\csname\SMALL bsy\@nd}%
\textfont\bssfam=\csname\SMALL bss\@nd
\scriptfont\bssfam=\csname\smSMALL bss\@nd
\scriptscriptfont\bssfam=\csname\vsSMALL bss\@nd
\def\bss{\fam\bssfam\csname\SMALL bss\@nd}%
\ifams
\textfont\msafam=\csname\SMALL msa\@nd
\scriptfont\msafam=\csname\smSMALL msa\@nd
\scriptscriptfont\msafam=\csname\vsSMALL msa\@nd
\textfont\msbfam=\csname\SMALL msb\@nd
\scriptfont\msbfam=\csname\smSMALL msb\@nd
\scriptscriptfont\msbfam=\csname\vsSMALL msb\@nd
\else\fi
\normalbaselineskip=\smallb@seline pt%
\setbox\strutbox=\hbox{\vrule height.7\normalbaselineskip
depth.3\baselineskip width0pt}%
\everymath{}%
\def\sc{\csname\smSMALL rm\@nd}\normalbaselines\rm}%
\everydisplay{\indenteddisplay
   \gdef\labeltype{\eqlabel}}%
%
%%%%%%%%%%%%%%%%%%%%%%%%%%%%%%%%%%%%%%%%%%%%%%%%%%%%%%%%%%%
%                                                         %
%  Macros to define extra maths symbols                   %
%                                                         %
%%%%%%%%%%%%%%%%%%%%%%%%%%%%%%%%%%%%%%%%%%%%%%%%%%%%%%%%%%%
%
\def\hexnumber@#1{\ifcase#1 0\or 1\or 2\or 3\or 4\or 5\or 6\or 7\or 8\or
 9\or A\or B\or C\or D\or E\or F\fi}
\edef\bffam@{\hexnumber@\bffam}
\edef\bdifam@{\hexnumber@\bdifam}
\edef\bsyfam@{\hexnumber@\bsyfam}
\def\undefine#1{\let#1\undefined}
\def\newsymbol#1#2#3#4#5{\let\next@\relax
 \ifnum#2=\thr@@\let\next@\bdifam@\else
 \ifams
 \ifnum#2=\@ne\let\next@\msafam@\else
 \ifnum#2=\tw@\let\next@\msbfam@\fi\fi
 \fi\fi
 \mathchardef#1="#3\next@#4#5}
\def\mathhexbox@#1#2#3{\relax
 \ifmmode\mathpalette{}{\m@th\mathchar"#1#2#3}%
 \else\leavevmode\hbox{$\m@th\mathchar"#1#2#3$}\fi}

\def\bi#1{{\fam\bdifam\relax#1}}
%
% If file amsmacro is not in current directory
% or somewhere with set path add path before
% file name in following line
%
\ifams\input amsmacro\fi
%
% Bold italic Greek characters
%
\newsymbol\bitGamma 3000
\newsymbol\bitDelta 3001
\newsymbol\bitTheta 3002
\newsymbol\bitLambda 3003
\newsymbol\bitXi 3004
\newsymbol\bitPi 3005
\newsymbol\bitSigma 3006
\newsymbol\bitUpsilon 3007
\newsymbol\bitPhi 3008
\newsymbol\bitPsi 3009
\newsymbol\bitOmega 300A
\newsymbol\balpha 300B
\newsymbol\bbeta 300C
\newsymbol\bgamma 300D
\newsymbol\bdelta 300E
\newsymbol\bepsilon 300F
\newsymbol\bzeta 3010
\newsymbol\bfeta 3011
\newsymbol\btheta 3012
\newsymbol\biota 3013
\newsymbol\bkappa 3014
\newsymbol\blambda 3015
\newsymbol\bmu 3016
\newsymbol\bnu 3017
\newsymbol\bxi 3018
\newsymbol\bpi 3019
\newsymbol\brho 301A
\newsymbol\bsigma 301B
\newsymbol\btau 301C
\newsymbol\bupsilon 301D
\newsymbol\bphi 301E
\newsymbol\bchi 301F
\newsymbol\bpsi 3020
\newsymbol\bomega 3021
\newsymbol\bvarepsilon 3022
\newsymbol\bvartheta 3023
\newsymbol\bvaromega 3024
\newsymbol\bvarrho 3025
\newsymbol\bvarzeta 3026
\newsymbol\bvarphi 3027
\newsymbol\bpartial 3040
\newsymbol\bell 3060
\newsymbol\bimath 307B
\newsymbol\bjmath 307C
\mathchardef\binfty "0\bsyfam@31
\mathchardef\bnabla "0\bsyfam@72
\mathchardef\bdot "2\bsyfam@01
\mathchardef\bGamma "0\bffam@00
\mathchardef\bDelta "0\bffam@01
\mathchardef\bTheta "0\bffam@02
\mathchardef\bLambda "0\bffam@03
\mathchardef\bXi "0\bffam@04
\mathchardef\bPi "0\bffam@05
\mathchardef\bSigma "0\bffam@06
\mathchardef\bUpsilon "0\bffam@07
\mathchardef\bPhi "0\bffam@08
\mathchardef\bPsi "0\bffam@09
\mathchardef\bOmega "0\bffam@0A
\mathchardef\itGamma "0100
\mathchardef\itDelta "0101
\mathchardef\itTheta "0102
\mathchardef\itLambda "0103
\mathchardef\itXi "0104
\mathchardef\itPi "0105
\mathchardef\itSigma "0106
\mathchardef\itUpsilon "0107
\mathchardef\itPhi "0108
\mathchardef\itPsi "0109
\mathchardef\itOmega "010A
\mathchardef\Gamma "0000
\mathchardef\Delta "0001
\mathchardef\Theta "0002
\mathchardef\Lambda "0003
\mathchardef\Xi "0004
\mathchardef\Pi "0005
\mathchardef\Sigma "0006
\mathchardef\Upsilon "0007
\mathchardef\Phi "0008
\mathchardef\Psi "0009
\mathchardef\Omega "000A
%
% Counter definitions
%
\newcount\firstpage  \firstpage=1  % start page no
\newcount\jnl                      % journal no
\newcount\secno                    % section number
\newcount\subno                    % number of subsection
\newcount\subsubno                 % number of subsubsection
\newcount\appno                    % appendix number
\newcount\tabno                    % table number
\newcount\figno                    % figure number
\newcount\countno                  % equation numbers
\newcount\refno                    % reference number
\newcount\eqlett     \eqlett=97    % equation letter
\newif\ifletter
\newif\ifwide
\newif\ifnotfull
\newif\ifaligned
\newif\ifnumbysec
\newif\ifappendix
\newif\ifnumapp
\newif\ifssf
\newif\ifppt
\newdimen\t@bwidth
\newdimen\c@pwidth
\newdimen\digitwidth                    %character width
\newdimen\argwidth                      %argument width
\newdimen\secindent    \secindent=5pc   %indentation of maths
\newdimen\textind    \textind=16pt      %indentation of text
\newdimen\tempval                       %temporary value
\newskip\beforesecskip
\def\beforesecspace{\vskip\beforesecskip\relax}
\newskip\beforesubskip
\def\beforesubspace{\vskip\beforesubskip\relax}
\newskip\beforesubsubskip
\def\beforesubsubspace{\vskip\beforesubsubskip\relax}
\newskip\secskip
\def\secspace{\vskip\secskip\relax}
\newskip\subskip
\def\subspace{\vskip\subskip\relax}
\newskip\insertskip
\def\insertspace{\vskip\insertskip\relax}
\def\sp@ce{\ifx\next*\let\next=\@ssf
               \else\let\next=\@nossf\fi\next}
\def\@ssf#1{\nobreak\secspace\global\ssftrue\nobreak}
\def\@nossf{\nobreak\secspace\nobreak\noindent\ignorespaces}
\def\subsp@ce{\ifx\next*\let\next=\@sssf
               \else\let\next=\@nosssf\fi\next}
\def\@sssf#1{\nobreak\subspace\global\ssftrue\nobreak}
\def\@nosssf{\nobreak\subspace\nobreak\noindent\ignorespaces}
\beforesecskip=24pt plus12pt minus8pt
\beforesubskip=12pt plus6pt minus4pt
\beforesubsubskip=12pt plus6pt minus4pt
\secskip=12pt plus 2pt minus 2pt
\subskip=6pt plus3pt minus2pt
\insertskip=18pt plus6pt minus6pt%
\fontdimen16\tensy=2.7pt
\fontdimen17\tensy=2.7pt
%
% Labels etc for cross referencing macros
%
\def\eqlabel{(\ifappendix\applett
               \ifnumbysec\ifnum\secno>0 \the\secno\fi.\fi
               \else\ifnumbysec\the\secno.\fi\fi\the\countno)}
\def\seclabel{\ifappendix\ifnumapp\else\applett\fi
    \ifnum\secno>0 \the\secno
    \ifnumbysec\ifnum\subno>0.\the\subno\fi\fi\fi
    \else\the\secno\fi\ifnum\subno>0.\the\subno
         \ifnum\subsubno>0.\the\subsubno\fi\fi}
\def\tablabel{\ifappendix\applett\fi\the\tabno}
\def\figlabel{\ifappendix\applett\fi\the\figno}
\def\gac{\global\advance\countno by 1}
%
% Redefinition of footnote macros to lose rule and remove indentation
%

\def\vfootnote#1{\insert\footins\bgroup
\interlinepenalty=\interfootnotelinepenalty
\splittopskip=\ht\strutbox % top baseline for broken footnotes
\splitmaxdepth=\dp\strutbox \floatingpenalty=20000
\leftskip=0pt \rightskip=0pt \spaceskip=0pt \xspaceskip=0pt%
\noindent\smallfonts\rm #1\ \ignorespaces\footstrut\futurelet\next\fo@t}
%
% Redefinition of endinsert to give more controllable
% space around  tables and figures
%
\def\endinsert{\egroup
    \if@mid \dimen@=\ht0 \advance\dimen@ by\dp0
       \advance\dimen@ by12\p@ \advance\dimen@ by\pagetotal
       \ifdim\dimen@>\pagegoal \@midfalse\p@gefalse\fi\fi
    \if@mid \insertspace \box0 \par \ifdim\lastskip<\insertskip
    \removelastskip \penalty-200 \insertspace \fi
    \else\insert\topins{\penalty100
       \splittopskip=0pt \splitmaxdepth=\maxdimen
       \floatingpenalty=0
       \ifp@ge \dimen@=\dp0
       \vbox to\vsize{\unvbox0 \kern-\dimen@}%
       \else\box0\nobreak\insertspace\fi}\fi\endgroup}
%
% special macros for display equations
%
% for indentation of turned over lines in mathematics
%
\def\ind{\hbox to \secindent{\hfill}}
%
% for turned over equals sign to left of maths indent
%

%
% for other signs to left of maths indent
%

%
% displayed equation indented
%
\def\indeqn#1{\alignedfalse\displ@y\halign{\hbox to \displaywidth
    {$\ind\@lign\displaystyle##\hfil$}\crcr #1\crcr}}
%
% displayed equation indented with alignments
%
\def\indalign#1{\alignedtrue\displ@y \tabskip=0pt
  \halign to\displaywidth{\ind$\@lign\displaystyle{##}$\tabskip=0pt
    &$\@lign\displaystyle{{}##}$\hfill\tabskip=\centering
    &\llap{$\@lign\hbox{\rm##}$}\tabskip=0pt\crcr
    #1\crcr}}
\def\fl{{\hskip-\secindent}}
\def\indenteddisplay#1$${\indispl@y{#1 }}
\def\indispl@y#1{\disptest#1\eqalignno\eqalignno\disptest}
\def\disptest#1\eqalignno#2\eqalignno#3\disptest{%
    \ifx#3\eqalignno
    \indalign#2%
    \else\indeqn{#1}\fi$$}
%
% Roman small caps (if in Roman \sc gives small caps)
%

%
% Italic small caps (if in italic \sc gives italic small caps)
%

%
% Bold small caps (if in bold \sc gives bold small caps)
%

%
% Small caps in maths
%

%
% Miscellaneous definitions
%

\def\ns{\noalign{\vskip-3pt}}

%

%
% Bold h bar
%
\def\bhbar{\rlap{\kern1pt\raise.4ex\hbox{\bf\char'40}}\bi{h}}

\def\dash{---{}--- }

\def\e{{\rm e}}
\def\etal{{\it et al\/}\ }
\def\frac#1#2{{#1\over#2}}
\ifams
\def\lap{\lesssim}
\def\gap{\gtrsim}
\let\le=\leqslant

\else

\def\gap{\;\lower3pt\hbox{$\buildrel > \over \sim$}\;}%
\def\lap{\;\lower3pt\hbox{$\buildrel < \over \sim$}\;}\fi

\chardef\ii="10
\def\tqs{\hbox to 25pt{\hfil}}

\def\Or{\mathop{\rm O}\nolimits}

\def\Bbbone{1\kern-.22em {\rm l}}
%
% Primes to display summations and products
% which also have sub or superscripts
%
\def\rp{\raise8pt\hbox{$\scriptstyle\prime$}}
%
% then use \sum^{...}_{...}\rp or \prod^{...}_{...}\rp.
%
% Shadow brackets
%
% Single brackets for normal size only
%

%
% Variable size for display style
%
\def\[#1\]{\setbox0=\hbox{$\dsty#1$}\argwidth=\wd0
    \setbox0=\hbox{$\left[\box0\right]$}\advance\argwidth by -\wd0
    \left[\kern.3\argwidth\box0\kern.3\argwidth\right]}
%
% Variable size for text style
%
\def\lsb#1\rsb{\setbox0=\hbox{$#1$}\argwidth=\wd0
    \setbox0=\hbox{$\left[\box0\right]$}\advance\argwidth by -\wd0
    \left[\kern.3\argwidth\box0\kern.3\argwidth\right]}
\def\œ{{\tenu \$}}
%
% Square for end of theorems
%

%
\def\pt(#1){({\it #1\/})}
\def\ts{{\thinspace}}%

\let\dsty=\displaystyle

%
% Definition for Nuclear Physics Keyword abstract
%
\def\reactions#1{\vskip 12pt plus2pt minus2pt%
\vbox{\hbox{\kern\secindent\vrule\kern12pt%
\vbox{\kern0.5pt\vbox{\hsize=24pc\parindent=0pt\smallfonts\rm NUCLEAR
REACTIONS\strut\quad #1\strut}\kern0.5pt}\kern12pt\vrule}}}
%
% Definition for slashed characters
%
\def\slashchar#1{\setbox0=\hbox{$#1$}\dimen0=\wd0%
\setbox1=\hbox{/}\dimen1=\wd1%
\ifdim\dimen0>\dimen1%
\rlap{\hbox to \dimen0{\hfil/\hfil}}#1\else
\rlap{\hbox to \dimen1{\hfil$#1$\hfil}}/\fi}
%
% Redefine \textindent for use in \item
%
\def\textindent#1{\noindent\hbox to \parindent{#1\hss}\ignorespaces}
%
% Symbols and curves for use in figure captions
%
\def\opencirc{\raise1pt\hbox{$\scriptstyle{\bigcirc}$}}

\ifams
\def\opensqr{\hbox{$\square$}}

\def\opentridown{\hbox{$\triangledown$}}

\else
\def\opensqr{\vbox{\hrule height.4pt\hbox{\vrule width.4pt height3.5pt
    \kern3.5pt\vrule width.4pt}\hrule height.4pt}}

\def\opentridown{\raise1pt\hbox{$\scriptstyle\bigtriangledown$}}

           %  These produce the
                   %  equivalent open character
           %  to be filled in.
\fi
\def\dotted{\hbox{${\mathinner{\cdotp\cdotp\cdotp\cdotp\cdotp\cdotp}}$}}
\def\dashed{\hbox{-\ts -\ts -\ts -}}

\def\full{\hbox{------}}
%
% Redefinition of \cases
%
\def\m@th{\mathsurround=0pt}
%
% Displaystyle now used for first term
%
\def\cases#1{%
\left\{\,\vcenter{\normalbaselines\openup1\jot\m@th%
     \ialign{$\displaystyle##\hfil$&\rm\tqs##\hfil\crcr#1\crcr}}\right.}%
%
% Original version of cases now called \oldcases
%
\def\oldcases#1{\left\{\,\vcenter{\normalbaselines\m@th
    \ialign{$##\hfil$&\rm\quad##\hfil\crcr#1\crcr}}\right.}
%
% Cases with number at end each line (using automatic numbering)
%
\def\numcases#1{\left\{\,\vcenter{\baselineskip=15pt\m@th%
     \ialign{$\displaystyle##\hfil$&\rm\tqs##\hfil
     \crcr#1\crcr}}\right.\hfill
     \vcenter{\baselineskip=15pt\m@th%
     \ialign{\rlap{$\phantom{\displaystyle##\hfil}$}\tabskip=0pt&\en
     \rlap{\phantom{##\hfil}}\crcr#1\crcr}}}
\def\ptnumcases#1{\left\{\,\vcenter{\baselineskip=15pt\m@th%
     \ialign{$\displaystyle##\hfil$&\rm\tqs##\hfil
     \crcr#1\crcr}}\right.\hfill
     \vcenter{\baselineskip=15pt\m@th%
     \ialign{\rlap{$\phantom{\displaystyle##\hfil}$}\tabskip=0pt&\enpt
     \rlap{\phantom{##\hfil}}\crcr#1\crcr}}\global\eqlett=97
     \global\advance\countno by 1}
%
% for equation numbers instead of \eqno
%
\def\eq(#1){\ifaligned\@mp(#1)\else\hfill\llap{{\rm (#1)}}\fi}
\def\ceq(#1){\ns\ns\ifaligned\@mp\fi\eq(#1)\cr\ns\ns}
\def\eqpt(#1#2){\ifaligned\@mp(#1{\it #2\/})
                    \else\hfill\llap{{\rm (#1{\it #2\/})}}\fi}

%
% Automatic numbering of equations
%
\countno=1
\def\eqnobysec{\numbysectrue}
\def\aleq{&\rm(\ifappendix\applett
               \ifnumbysec\ifnum\secno>0 \the\secno\fi.\fi
               \else\ifnumbysec\the\secno.\fi\fi\the\countno}
\def\noaleq{\hfill\llap\bgroup\rm(\ifappendix\applett
               \ifnumbysec\ifnum\secno>0 \the\secno\fi.\fi
               \else\ifnumbysec\the\secno.\fi\fi\the\countno}
\def\@mp{&}
\def\en{\ifaligned\aleq)\else\noaleq)\egroup\fi\gac}
\def\cen{\ns\ns\ifaligned\@mp\fi\en\cr\ns\ns}
\def\enpt{\ifaligned\aleq{\it\char\the\eqlett})\else
    \noaleq{\it\char\the\eqlett})\egroup\fi
    \global\advance\eqlett by 1}
\def\endpt{\ifaligned\aleq{\it\char\the\eqlett})\else
    \noaleq{\it\char\the\eqlett})\egroup\fi
    \global\eqlett=97\gac}
%
% abbreviations for Institute of Physics Publishing journals
%

        %1968-87
   %1988 and onwards
     %1968--1988
        %1989 and onwards

           %1975--1988
     %1989 and onwards

                 %1990 and onwards

\def\RPP{{\it Rep. Prog. Phys.}}

%
% Other commonly quoted journals
%

\def\AP{{\it Ann. Phys., Lpz.}}
\def\APNY{{\it Ann. Phys., NY\/}}

\def\NC{{\it Nuovo Cimento\/}}

\def\NP{{\it Nucl. Phys.}}
\def\PL{{\it Phys. Lett.}}
\def\PR{{\it Phys. Rev.}}
\def\PRL{{\it Phys. Rev. Lett.}}

\def\RMP{{\it Rev. Mod. Phys.}}

\def\ZP{{\it Z. Phys.}}
\headline={\ifodd\pageno{\ifnum\pageno=\firstpage\hfill
   \else\rrhead\fi}\else\lrhead\fi}
\def\rrhead{\textfonts\hskip\secindent\it
    \shorttitle\hfill\rm\folio}
\def\lrhead{\textfonts\hbox to\secindent{\rm\folio\hss}%
    \it\aunames\hss}
\footline={\ifnum\pageno=\firstpage \hfill\textfonts\rm\folio\fi}
\def\@rticle#1#2{\vglue.5pc
    %{\parindent=\secindent \bf #1\par}
     \line{\textfonts\rm \hfil \vbox{\hbox{FTUV/95-4}    %my modification
           \hbox{IFIC/95-4}\hbox{hep-ph/9502366}}        %my modification
                        \par} \vskip 5pc                 %my modification
     \vskip2.5pc
    {\exhyphenpenalty=10000\hyphenpenalty=10000
     \baselineskip=18pt\raggedright\noindent
     \headfonts\bf#2\par}\futurelet\next\sh@rttitle}%
\def\title#1{\gdef\shorttitle{#1}
    \vglue4pc{\exhyphenpenalty=10000\hyphenpenalty=10000
    \baselineskip=18pt
    \raggedright\parindent=0pt
    \headfonts\bf#1\par}\futurelet\next\sh@rttitle}

\def\article#1#2{\gdef\shorttitle{#2}\@rticle{#1}{#2}}
\def\review#1{\gdef\shorttitle{#1}%
     \@rticle{REVIEW}{#1}}  % my modification
    %\@rticle{REVIEW \ifpbm\else ARTICLE\fi}{#1}}
\def\topical#1{\gdef\shorttitle{#1}%
    \@rticle{TOPICAL REVIEW}{#1}}
\def\comment#1{\gdef\shorttitle{#1}%
    \@rticle{COMMENT}{#1}}
\def\note#1{\gdef\shorttitle{#1}%
    \@rticle{NOTE}{#1}}
\def\prelim#1{\gdef\shorttitle{#1}%
    \@rticle{PRELIMINARY COMMUNICATION}{#1}}
\def\letter#1{\gdef\shorttitle{Letter to the Editor}%
     \gdef\aunames{Letter to the Editor}
     \global\lettertrue\ifnum\jnl=7\global\letterfalse\fi
     \@rticle{LETTER TO THE EDITOR}{#1}}
\def\sh@rttitle{\ifx\next[\let\next=\sh@rt
                \else\let\next=\f@ll\fi\next}
\def\sh@rt[#1]{\gdef\shorttitle{#1}}
\def\f@ll{}
\def\author#1{\ifletter\else\gdef\aunames{#1}\fi\vskip1.5pc
    {\parindent=\secindent
     \hang\textfonts
     \ifppt\bf\else\rm\fi#1\par}
     \ifppt\bigskip\else\smallskip\fi
     \futurelet\next\@unames}
\def\@unames{\ifx\next[\let\next=\short@uthor
                 \else\let\next=\@uthor\fi\next}
\def\short@uthor[#1]{\gdef\aunames{#1}}
\def\@uthor{}
\def\address#1{{\parindent=\secindent
    \exhyphenpenalty=10000\hyphenpenalty=10000
\ifppt\textfonts\else\smallfonts\fi\hang\raggedright\rm#1\par}%
    \ifppt\bigskip\fi}
\def\jl#1{\global\jnl=#1}
\jl{0}%
\def\journal{\ifnum\jnl=1 J. Phys.\ A: Math.\ Gen.\
        \else\ifnum\jnl=2 J. Phys.\ B: At.\ Mol.\ Opt.\ Phys.\
        \else\ifnum\jnl=3 J. Phys.:\ Condens. Matter\
        \else\ifnum\jnl=4 J. Phys.\ G: Nucl.\ Part.\ Phys.\
        \else\ifnum\jnl=5 Inverse Problems\
        \else\ifnum\jnl=6 Class. Quantum Grav.\
        \else\ifnum\jnl=7 Network\
        \else\ifnum\jnl=8 Nonlinearity\
        \else\ifnum\jnl=9 Quantum Opt.\
        \else\ifnum\jnl=10 Waves in Random Media\
        \else\ifnum\jnl=11 Pure Appl. Opt.\
        \else\ifnum\jnl=12 Phys. Med. Biol.\
        \else\ifnum\jnl=13 Modelling Simulation Mater.\ Sci.\ Eng.\
        \else\ifnum\jnl=14 Plasma Phys. Control. Fusion\
        \else\ifnum\jnl=15 Physiol. Meas.\
        \else\ifnum\jnl=16 Sov.\ Lightwave Commun.\
        \else\ifnum\jnl=17 J. Phys.\ D: Appl.\ Phys.\
        \else\ifnum\jnl=18 Supercond.\ Sci.\ Technol.\
        \else\ifnum\jnl=19 Semicond.\ Sci.\ Technol.\
        \else\ifnum\jnl=20 Nanotechnology\
        \else\ifnum\jnl=21 Meas.\ Sci.\ Technol.\
        \else\ifnum\jnl=22 Plasma Sources Sci.\ Technol.\
        \else\ifnum\jnl=23 Smart Mater.\ Struct.\
        \else\ifnum\jnl=24 J.\ Micromech.\ Microeng.\
   \else Institute of Physics Publishing\
   \fi\fi\fi\fi\fi\fi\fi\fi\fi\fi\fi\fi\fi\fi\fi
   \fi\fi\fi\fi\fi\fi\fi\fi\fi}
\let\abs=\beginabstract

\let\endabs=\endabstract
\def\submitted{\ifppt\noindent\textfonts\rm
   To be published in Reports on Progress in Physics   %my modification
   %Submitted to \journal\par
     \bigskip\fi}
\def\today{\number\day\ \ifcase\month\or
     January\or February\or March\or April\or May\or June\or
     July\or August\or September\or October\or November\or
     December\fi\space \number\year}
\def\date{\ifppt\noindent\textfonts\rm
     %Date: \today\par\goodbreak\bigskip\fi}
     February 1995 \par\goodbreak\bigskip\fi}  % my modification
%
% Physics Abstracts classification numbers
%

%

%
%%%%%%%%%%%%%%%%%%%%%%%%%%%%%%%%%%%%%%%%%%%%%%%%%%%%%%%%%%%%
%                                                          %
%  Sections, subsections, etc                              %
%                                                          %
%%%%%%%%%%%%%%%%%%%%%%%%%%%%%%%%%%%%%%%%%%%%%%%%%%%%%%%%%%%%
%
\def\section#1{\ifppt\ifnum\secno=0\eject\fi\fi
    \subno=0\subsubno=0\global\advance\secno by 1
    \gdef\labeltype{\seclabel}\ifnumbysec\countno=1\fi
    \goodbreak\beforesecspace\nobreak
    \noindent{\bf \the\secno. #1}\par\futurelet\next\sp@ce}
\def\subsection#1{\subsubno=0\global\advance\subno by 1
     \gdef\labeltype{\seclabel}%
     \ifssf\else\goodbreak\beforesubspace\fi
     \global\ssffalse\nobreak
     \noindent{\it \the\secno.\the\subno. #1\par}%
     \futurelet\next\subsp@ce}
\def\subsubsection#1{\global\advance\subsubno by 1
     \gdef\labeltype{\seclabel}%
     \ifssf\else\goodbreak\beforesubsubspace\fi
     \global\ssffalse\nobreak
     \noindent{\it \the\secno.\the\subno.\the\subsubno. #1}\null.
     \ignorespaces}
%

%
%%%%%%%%%%%%%%%%%%%%%%%%%%%%%%%%%%%%%%%%%%%%%%%%%%%%%%%%%%%%
%                                                          %
%  Appendices                                              %
%                                                          %
%%%%%%%%%%%%%%%%%%%%%%%%%%%%%%%%%%%%%%%%%%%%%%%%%%%%%%%%%%%%
%
\def\numappendix#1{\ifappendix\ifnumbysec\countno=1\fi\else
    \countno=1\figno=0\tabno=0\fi
    \subno=0\global\advance\appno by 1
    \secno=\appno\gdef\applett{A}\gdef\labeltype{\seclabel}%
    \global\appendixtrue\global\numapptrue
    \goodbreak\beforesecspace\nobreak
    \noindent{\bf Appendix \the\appno. #1\par}%
    \futurelet\next\sp@ce}
\def\numsubappendix#1{\global\advance\subno by 1\subsubno=0
    \gdef\labeltype{\seclabel}%
    \ifssf\else\goodbreak\beforesubspace\fi
    \global\ssffalse\nobreak
    \noindent{\it A\the\appno.\the\subno. #1\par}%
    \futurelet\next\subsp@ce}
\def\@ppendix#1#2#3{\countno=1\subno=0\subsubno=0\secno=0\figno=0\tabno=0
    \gdef\applett{#1}\gdef\labeltype{\seclabel}\global\appendixtrue
    \goodbreak\beforesecspace\nobreak
    \noindent{\bf Appendix#2#3\par}\futurelet\next\sp@ce}
\def\Appendix#1{\@ppendix{A}{. }{#1}}
\def\appendix#1#2{\@ppendix{#1}{ #1. }{#2}}
\def\App#1{\@ppendix{A}{ }{#1}}
\def\app{\@ppendix{A}{}{}}
\def\subappendix#1#2{\global\advance\subno by 1\subsubno=0
    \gdef\labeltype{\seclabel}%
    \ifssf\else\goodbreak\beforesubspace\fi
    \global\ssffalse\nobreak
    \noindent{\it #1\the\subno. #2\par}%
    \nobreak\subspace\noindent\ignorespaces}
%
%%%%%%%%%%%%%%%%%%%%%%%%%%%%%%%%%%%%%%%%%%%%%%%%%%%%%%%%%%%%
%                                                          %
%  Acknowledgments, notes added and foreign abstracts      %
%                                                          %
%%%%%%%%%%%%%%%%%%%%%%%%%%%%%%%%%%%%%%%%%%%%%%%%%%%%%%%%%%%%
%
\def\@ck#1{\ifletter\bigskip\noindent\ignorespaces\else
    \goodbreak\beforesecspace\nobreak
    \noindent{\bf Acknowledgment#1\par}%
    \nobreak\secspace\noindent\ignorespaces\fi}
\def\ack{\@ck{s}}
\def\ackn{\@ck{}}
\def\n@ip#1{\goodbreak\beforesecspace\nobreak
    \noindent\smallfonts{\it #1}. \rm\ignorespaces}
\def\naip{\n@ip{Note added in proof}}
\def\na{\n@ip{Note added}}

%
%  \resume and \zus in Physics in Medicine and Biology only
%

%

%
%%%%%%%%%%%%%%%%%%%%%%%%%%%%%%%%%%%%%%%%%%%%%%%%%%%%%%%%%%%%
%                                                          %
%  Tables                                                  %
%                                                          %
%%%%%%%%%%%%%%%%%%%%%%%%%%%%%%%%%%%%%%%%%%%%%%%%%%%%%%%%%%%
%

%

%
\def\table#1{\tablecaption{#1}}
\def\tablecont{\topinsert\global\advance\tabno by -1
    \tablecaption{(continued)}}
\def\tablecaption#1{\gdef\labeltype{\tablabel}\global\widefalse
    \leftskip=\secindent\parindent=0pt
    \global\advance\tabno by 1
    \smallfonts{\bf Table \ifappendix\applett\fi\the\tabno.} \rm #1\par
    \smallskip\futurelet\next\t@b}
\def\endtable{\vfill\goodbreak}
\def\t@b{\ifx\next*\let\next=\widet@b
             \else\ifx\next[\let\next=\fullwidet@b
                      \else\let\next=\narrowt@b\fi\fi
             \next}
\def\widet@b#1{\global\widetrue\global\notfulltrue
    \t@bwidth=\hsize\advance\t@bwidth by -\secindent}
\def\fullwidet@b[#1]{\global\widetrue\global\notfullfalse
    \leftskip=0pt\t@bwidth=\hsize}
\def\narrowt@b{\global\notfulltrue}
\def\align{\catcode`?=13\ifnotfull\moveright\secindent\fi
    \vbox\bgroup\halign\ifwide to \t@bwidth\fi
    \bgroup\strut\tabskip=1.2pc plus1pc minus.5pc}
\def\endalign{\egroup\egroup\catcode`?=12}

%
% Use \L{#}, \R{#} and \C{#} to specify left, right or centred
% columns immediately after \table. For example
% \align\L{#}&&\L{#}\cr gives the preamble for a table with
% all columns aligned left, \align\L{#}&\C{#}&\R{#}\cr
% gives a table with 3 columns, the first aligned left, the second
% centred and the third aligned right.
%

\def\C#1{\hfill#1\hfill}
%
%  Rules for tables \br at top and bottom
%  \mr to separate headings from entries
%
\def\br{\noalign{\vskip2pt\hrule height1pt\vskip2pt}}
\def\mr{\noalign{\vskip2pt\hrule\vskip2pt}}
\def\tabnote#1{\vskip-\lastskip\noindent #1\par}
%
% Definitions for centring headings over several columns
% \centre{4}{Results for helium} will centre
% Results for helium over four columns
% \crule{4} will produce a rule centred over four columns
% to go below a centred heading
%

%

\catcode`?=13
\def\lineup{\setbox0=\hbox{\smallfonts\rm 0}%
    \digitwidth=\wd0%
    \def?{\kern\digitwidth}%
    \def\\{\hbox{$\phantom{-}$}}%
    \def\-{\llap{$-$}}}
\catcode`?=12
%
% Macros for two parts of a table of equal width side by side
% \table{caption}[w]
% \sidetable{first part}{second part}
% \endtable
% Use \table preamble for tables of 31picas width
%
\def\sidetable#1#2{\hbox{\ifppt\hsize=18pc\t@bwidth=18pc
                          \else\hsize=15pc\t@bwidth=15pc\fi
    \parindent=0pt\vtop{\null #1\par}%
    \ifppt\hskip1.2pc\else\hskip1pc\fi
    \vtop{\null #2\par}}}
\def\lstable#1#2{\everypar{}\tempval=\hsize\hsize=\vsize
    \vsize=\tempval\hoffset=-3pc
    \global\tabno=#1\gdef\labeltype{\tablabel}%
    \noindent\smallfonts{\bf Table \ifappendix\applett\fi
    \the\tabno.} \rm #2\par
    \smallskip\futurelet\next\t@b}
\def\inctabno{\global\advance\tabno by 1}
%
%%%%%%%%%%%%%%%%%%%%%%%%%%%%%%%%%%%%%%%%%%%%%%%%%%%%%%%%%%%%
%                                                          %
%  Figures                                                 %
%                                                          %
%%%%%%%%%%%%%%%%%%%%%%%%%%%%%%%%%%%%%%%%%%%%%%%%%%%%%%%%%%%%
%
\def\Figures{\vfill\eject\global\appendixfalse\textfonts\rm
    \everypar{}\noindent{\bf Figure captions}\par
    \bigskip}
\def\figure#1{\figc@ption{#1}\bigskip}
\def\figc@ption#1{\global\advance\figno by 1\gdef\labeltype{\figlabel}%
   {\parindent=\secindent\smallfonts\hang
    {\bf Figure \ifappendix\applett\fi\the\figno.} \rm #1\par}}
%
%%%%%%%%%%%%%%%%%%%%%%%%%%%%%%%%%%%%%%%%%%%%%%%%%%%%%%%%%%%%
%                                                          %
%  Reference lists                                         %
%                                                          %
%%%%%%%%%%%%%%%%%%%%%%%%%%%%%%%%%%%%%%%%%%%%%%%%%%%%%%%%%%%%
%
\def\refHEAD{\goodbreak\beforesecspace
     \noindent\textfonts{\bf References}\par
     \let\ref=\rf
     \nobreak\smallfonts\rm}
\def\references{\refHEAD\parindent=0pt
     \everypar{\hangindent=18pt\hangafter=1
     \frenchspacing\rm}%
     \secspace}
\def\rf#1{\par\noindent\hbox to 21pt{\hss #1\quad}\ignorespaces}
\def\refjl#1#2#3#4{\noindent #1 {\it #2 \bf #3} #4\par}
\def\refbk#1#2#3{\noindent #1 {\it #2} #3\par}
%
% reference to a journal article in numerical system
%

%
% reference to a book or report in numerical system
%

%
%%%%%%%%%%%%%%%%%%%%%%%%%%%%%%%%%%%%%%%%%%%%%%%%%%%%%%%%%%%%
%                                                          %
%  Theorems, lemmas, etc                                   %
%                                                          %
%%%%%%%%%%%%%%%%%%%%%%%%%%%%%%%%%%%%%%%%%%%%%%%%%%%%%%%%%%%%
%

%
% NB \note#1 is used to give a Note (as opposed to a paper or letter)
% in PMB therefore use commands \notes#1 for numbered Note
% instead of \note
%

%
\catcode`\@=12
%
% Parameter values for `Preprint' style
%
\def\pptstyle{\ppttrue\headsize{17}{24}%
\textsize{12}{16}%
\smallsize{10}{12}%
\hsize=37.2pc\vsize=56pc
\textind=20pt\secindent=6pc}
%
% Parameter values for `Journal' style
%

%
% Parameter values for `Eleven point' style
%

%
% Parameter values for `Large size' style
%

%
\parindent=\textind
%
%\endinput
%
%%%%%%%%%%%%%%%%%%%%%%%%%%

%
% My own version
%  \jobname.xrf  changed to \jobname.aux
%
\catcode`@=11
\newwrite\auxfile
\newwrite\xreffile
\newif\ifxrefwarning \xrefwarningtrue
\newif\ifauxfile
\newif\ifxreffile
\def\testforxref{\begingroup
    \immediate\openin\xreffile = \jobname.aux\space
    \ifeof\xreffile\global\xreffilefalse
    \else\global\xreffiletrue\fi
    \immediate\closein\xreffile
    \endgroup}
\def\testforaux{\begingroup
    \immediate\openin\auxfile = \jobname.aux\space
    \ifeof\auxfile\global\auxfilefalse
    \else\global\auxfiletrue\fi
    \immediate\closein\auxfile
    \endgroup}
\def\openreffile{\immediate\openout\auxfile = \jobname.aux}%
\def\readreffile{%
    \testforxref
    \testforaux
    \ifxreffile
       \begingroup
         \@setletters
         \input \jobname.aux
       \endgroup
    \else
\message{No cross-reference file existed, some labels may be undefined}%
    \fi\openreffile}%
\def\@setletters{%
    \catcode`_=11 \catcode`+=11
    \catcode`-=11 \catcode`@=11
    \catcode`0=11 \catcode`1=11
    \catcode`2=11 \catcode`3=11
    \catcode`4=11 \catcode`5=11
    \catcode`6=11 \catcode`7=11
    \catcode`8=11 \catcode`9=11
    \catcode`(=11 \catcode`)=11
    \catcode`:=11 \catcode`'=11
    \catcode`&=11 \catcode`;=11
    \catcode`.=11}%
\gdef\el@b{\eqlabel}
\gdef\sl@b{\seclabel}
\gdef\tl@b{\tablabel}
\gdef\fl@b{\figlabel}
\def\l@belno{\ifx\labeltype\el@b
   \let\labelno=\en\def\@label{\eqlabel}%
   \else\let\labelno=\ignorespaces
   \ifx\labeltype\sl@b \def\@label{\seclabel}%
   \else\ifx\labeltype\tl@b \def\@label{\tablabel}%
   \else\ifx\labeltype\fl@b \def\@label{\figlabel}%
   \else\def\@label{\seclabel}\fi\fi\fi\fi}
\def\label#1{\l@belno\expandafter\xdef\csname #1@\endcsname{\@label}%
    \immediate\write\auxfile{\string
    \gdef\expandafter\string\csname @#1\endcsname{\@label}}%
    \labelno}%
\def\ref#1{%
    \expandafter \ifx \csname @#1\endcsname\relax
    \message{Undefined label `#1'.}%
    \expandafter\xdef\csname @#1\endcsname{(??)}\fi
    \csname @#1\endcsname}%
\readreffile
%
%%%%%%%%%%%
%
\def\bibitem#1{\global\advance\refno by 1%
    \immediate\write\auxfile{\string
    \gdef\expandafter\string\csname #1@\endcsname{\the\refno}}%
    \rf{[\the\refno]}}%
\def\cite#1{\hbox{[\splitarg{#1}]}}%
\def\splitarg#1{\@pt#1,\@ptend}%
\def\@pt#1,#2\@ptend{\ifempty{#1}\else
    \@pttwo #1\@pttwoend
    \ifempty{#2}\else\sp@cer\@pt#2\@ptend\fi\fi}%
\def\@pttwo#1\@pttwoend{\expandafter
    \ifx \csname#1@\endcsname\@pttwoend\else
    \@ifundefined{#1}{{\bf ?}%
    \message{Undefined citation `#1' on page
    \the\pageno}}{\csname#1@\endcsname}\fi}%
\def\@pttwoend{@@@@@}%
\def\sp@cer{,\nobreak\thinspace}%
\def\ifempty#1{\@ifempty #1\@xx\@xxx}%
\def\@ifempty#1#2\@xxx{\ifx #1\@xx}%
\def\@xx{@@@@}%
\def\@xxx{@@@@}%
\long\def\@ifundefined#1#2#3{\expandafter\ifx\csname
  #1@\endcsname\relax#2\else#3\fi}%
\catcode`@=12
%\endinput
%%%%%%%%%%%%%%%%%%%%%%%%%%%%%%%%%%%%%%%%%%%%%%%%%%%%%%%%

%%%%%%%%%%%%%%%%%%%%%%%%%%%%%%%%%%%%%%%%%%%%%%%%%%%%%%%%%%%%%%%%%%%%%
%%%                                                               %%%
%%%             CHPT  Review  for Rep. Prog. Phys.                %%%
%%%               Draft  8      10 February 1994                  %%%
%%%                                                               %%%
%%%%%%%%%%%%%%%%%%%%%%%%%%%%%%%%%%%%%%%%%%%%%%%%%%%%%%%%%%%%%%%%%%%%%
%%%
%%%
%\input iopppt
%\input xref         % to enable cross referencing to be used
%\input iopptmac     % extra macros for verbatim listings
                    % used in this article
%
% \endtable has been redefined as tables are inserted within the
% text rather than grouped at the end of the article as
% usual for preprints
%
\def\endtable{\endinsert}
\pptstyle          % Replace by \jnlstyle for journal style output
%\jnlstyle
%

\eqnobysec

%
%   My commands
%
\def\lrder{\buildrel \leftrightarrow \over \partial}

\def\cL{{\cal L}}
\def\cD{{\cal D}}
\def\cM{{\cal M}}

\def\cP{{\cal P}}

\def\dg{\dagger}
\def\toG{\buildrel G \over \longrightarrow}
\def\dfrac{\displaystyle \frac}

%
%

%\title{Chiral Perturbation Theory}[Chiral Perturbation Theory]
\review{Chiral Perturbation Theory}[Chiral Perturbation Theory]

\author{A Pich}

\address{\ Departament de F\'{\ii}sica Te\`orica and Institut de
F\'{\ii}sica Corpuscular, Universitat de
Val\`encia -- CSIC, E-46100 Burjassot, Val\`encia, Spain}

\abs
 An introduction to the basic ideas and methods of
Chiral Perturbation Theory is presented.
Several phenomenological applications of the
effective Lagrangian technique to strong, electromagnetic
and weak interactions are discussed.
\endabs

%\pacs{11.30.Hv; 11.30.Rd; 11.40.Fy; 12.38.Aw; 14.40.Aq}

\submitted

\date

%%%%%%%%%%%%%%%%%%%%%%%%%%%%%%%%%%%%%%%%%%%%%%%%

\section{Introduction}
\label{sec:introduction}
Quantum Chromodynamics (QCD)
is nowadays the established theory of the strong interactions.
Owing to its asymptotic-free nature
(Gross and Wilczek 1973, Politzer 1973),
perturbation theory
can be applied at short distances; the resulting predictions
have achieved a remarkable success,
explaining a wide range of phenomena where large momentum
transfers are involved.
In the low-energy domain, however,
the growing of the running QCD coupling and the associated
confinement of quarks
and gluons make very difficult to perform a thorough analysis
of the  QCD dynamics in terms of these fundamental degrees
of freedom.
A description in terms of the hadronic asymptotic states seems
more adequate; unfortunately, given the richness of the
hadronic spectrum, this is also a formidable task.

At very low energies, a great simplification of the
strong-interaction dynamics occurs.
Below the resonance region ($E<M_\rho$), the hadronic
spectrum only contains an octet of very light pseudoscalar
particles ($\pi$, $K$, $\eta$),
whose interactions can be easily understood with
global symmetry considerations.
This has allowed the development of a
powerful theoretical framework, Chiral Perturbation Theory
(ChPT), to systematically analyze the low-energy implications
of the QCD symmetries.
This formalism is based on two key ingredients:
the chiral symmetry properties of QCD
and the concept of effective field theory.

The pseudoscalar octet can be identified with the multiplet
of (approximately) massless Goldstone bosons associated with
the spontaneous breakdown of chiral symmetry.
Goldstone particles obey low-energy theorems, which result
in the known predictions of Current Algebra and PCAC
(Adler and Dashen 1968, de Alfaro \etal 1973).
Moreover, since there is mass gap separating the light
Goldstone states from the rest of the hadronic spectrum,
one can build an effective field theory,
incorporating the right symmetry requirements,
with Goldstone
particles as
the only dynamical degrees of freedom
(Weinberg 1967a, Cronin 1967, Schwinger 1967, Wess and Zumino 1967,
Dashen and Weinstein 1969, Gasiorowicz and Geffen 1969).
This leads to a great simplification of Current Algebra
calculations and, what is more important, allows for
a systematic investigation
of higher-order corrections in
the perturbative field-theory sense
(Weinberg 1979, Gasser and Leutwyler 1984, 1985).

Effective field theories
are the appropriate theoretical tool to describe
low-energy physics, where {\it low} is defined with respect to some
energy scale $\Lambda$.
They only take explicitly into account
the relevant
degrees of freedom, i.e. those states with $m<<\Lambda$, while the
heavier excitations with $M>>\Lambda$ are integrated out from
the action.
One gets in this way a string of non-renormalizable interactions
among the light states, which can be organized as an expansion
in powers of energy/$\Lambda$.
The information on  the heavier degrees of freedom is then
contained in the couplings of the resulting low-energy Lagrangian.
Although effective field theories contain an
infinite number of terms, renormalizability
is not an issue since, at a given order in the energy expansion,
the low-energy theory is specified by a finite number of couplings;
this allows for an order-by-order renormalization.

A simple example of effective field theory
is provided by QED at very low energies,
$E_\gamma <<m_e$.
%$\omega<<m_e$, where $\omega$ denotes the photon energy.
In this limit, one can describe the light-by-light scattering
using an  effective Lagrangian in terms of the electromagnetic
field only.
Gauge, Lorentz and Parity invariance
constrain the possible structures present
in the effective Lagrangian:
$$
\fl
{\cL}_{\rm eff} = -{1\over 4} F^{\mu\nu} F_{\mu\nu}
   + {a\over m_e^4} \, (F^{\mu\nu} F_{\mu\nu})^2
   + {b\over m_e^4} \, F^{\mu\nu} F_{\nu\sigma}
     F^{\sigma\rho} F_{\rho\mu}
   + \Or (F^6/m_e^8) \, .
\label{eq:QED}
$$
In the low-energy regime,
all the information on the original QED dynamics
is embodied in the values of the two low-energy couplings $a$ and $b$.
The values of these constants can be computed,
by explicitly integrating out the electron field from the original
QED generating
functional (or equivalently, by computing the relevant light-by-light
box diagrams). One then gets the well-known result
(Euler 1936, Euler and Heisenberg 1936):
$$
a = -{\alpha^2\over 36}\, , \qquad\qquad\qquad
b = {7 \alpha^2\over 90} \, .
\label{eq:EH}
$$
The important point to realize is that, even in the absence of an
explicit computation of the couplings $a$ and $b$, the Lagrangian
\ref{eq:QED} contains non-trivial information,
which is a consequence
of the imposed symmetries. The dominant contributions to the
amplitudes for different
low-energy photon reactions like
$\gamma\gamma\to 2\gamma, 4\gamma, \ldots$
can be directly obtained from
${\cL}_{\rm eff}$.
Moreover, the order of magnitude of the constants $a$, $b$ can also be
easily estimated through a na\"{\ii}ve counting of powers of the
electromagnetic coupling and combinatorial and loop
[$1/(16 \pi^2)$] factors.

The previous example is somehow academic, since perturbation theory
in powers of $\alpha$ works extremely well in QED.
 However, the effective
Lagrangian~\ref{eq:QED} would be valid even if the fine structure
constant were big;
the only difference would then be that we would not be
able to perturbatively compute the couplings $a$ and $b$.

In QCD,
due to confinement, the
quark and gluon fields are not asymptotic states.
Moreover, we do not know
how to derive the hadronic
interactions directly from the fundamental QCD Lagrangian.
However, we
do know the symmetry properties of the strong interactions;
therefore, we can write an effective field theory
in terms of the hadronic asymptotic states, and
parametrize the unknown dynamical information in a few couplings.

The theoretical basis of effective field theories can be formulated
(Weinberg 1979, Leutwyler 1994a) as a
{\bf theorem:}
{\it
for a given set of asymptotic states, perturbation theory with the
most general Lagrangian containing all terms allowed by the assumed
symmetries will yield the most general S-matrix elements consistent
with analyticity, perturbative unitarity,
cluster decomposition
and the assumed symmetries.}

In the following, I will present an overview of ChPT.
The chiral symmetry of the QCD Lagrangian is discussed
in section~\ref{sec:symmetry}.
The ChPT formalism is presented in
sections~\ref{sec:lo} and \ref{sec:p4},
where
the lowest-order and next-to-leading-order
terms in the chiral expansion
are analyzed.
Section~\ref{sec:phenomenology} contains a few selected
phenomenological applications.
The role of the lowest-mass resonances on the Goldstone interactions
is studied in section~\ref{sec:resonances} and
the relation between the effective Lagrangian and
the underlying fundamental QCD theory is discussed
in section~\ref{sec:couplings}, which summarizes
recent attempts to calculate the chiral couplings.
The effective realization of the non-leptonic
$\Delta S=1$ interactions
%is described in section~\ref{sec:weak},
and a brief overview of
the application of the chiral techniques to non-leptonic
$K$ decays is given in
section~\ref{sec:weak}.
%sections~\ref{sec:kpp}, \ref{sec:radiative} and
%\ref{sec:anomalous}.

Section~\ref{sec:baryons} presents the ChPT formalism in the
baryon sector. Some issues concerning the $U(1)_A$ anomaly and
the strong-CP problem are analyzed in section~\ref{sec:strongCP}.
The broad range of application of the ChPT techniques is finally
illustrated in sections~\ref{sec:light_Higgs} and
\ref{ref:electroweak}, which briefly discuss
the low-energy interactions of an hypothetical light Higgs boson and
the Goldstone dynamics associated with
the Standard Model electroweak symmetry breaking.
A few summarizing comments are collected in section~\ref{sec:summary}.

This report has grown out of a previous set of lectures (Pich 1994);
therefore, rather than giving an exhaustive and updated
summary of the field, it attempts to provide
a more pedagogical introduction.
I have made extensive use
of excellent reviews (Bijnens 1993a, Ecker 1993, 1995, Gasser 1990,
Leutwyler 1991, 1994c, Mei{\ss}ner 1993, de Rafael 1995) and books
(Donoghue, Golowich and Holstein 1992, Georgi 1984, Mei{\ss}ner 1992)
already existing in the literature.
Further details on particular subjects can be found in those
references.

\section{Chiral symmetry}
\label{sec:symmetry}

   In the absence of quark masses, the QCD Lagrangian
[$q = \hbox{\rm column}(u,d,\ldots)$]
$$
{\cL}_{QCD}^0 = -{1\over 4}\, G^a_{\mu\nu} G^{\mu\nu}_a
 + i \bar q_L \gamma^\mu D_\mu q_L  + i \bar q_R \gamma^\mu D_\mu q_R
\label{eq:LQCD}
$$
is invariant under
independent {\it global} $G\equiv SU(N_f)_L\otimes SU(N_f)_R$
transformations\footnote{$^{\dagger}$}{
%%%%%
Actually, the Lagrangian \ref{eq:LQCD}
has a larger $U(N_f)_L\otimes U(N_f)_R$ global symmetry. However, the
$U(1)_A$ part is broken by quantum effects [$U(1)_A$ anomaly],
while the
quark-number symmetry $U(1)_V$ is trivially realized
in the meson sector.}
%%%%%
of the left- and
right-handed quarks in flavour space:
$$
q_L \, \toG \, g_L \, q_L \, , \qquad\qquad
q_R \, \toG \, g_R \, q_R \, , \qquad\qquad
g_{L,R} \in SU(N_f)_{L,R} \, .
\label{eq:qrot}
$$
The Noether currents associated with the chiral group $G$ are
[$\lambda_a$ are Gell-Mann's matrices with
$\hbox{\rm Tr}(\lambda_a\lambda_b) = 2 \delta_{ab}$]:
$$
J^{a\mu}_X = \bar q_X \gamma^\mu {\lambda_a\over 2} q_X ,
\qquad\qquad (X = L,R;\quad a = 1,\,\ldots,\, 8) .
\label{eq:noether_currents}
$$
The corresponding Noether charges
$Q^a_X = \int d^3x \, J^{a0}_X(x)$ satisfy the familiar
commutation relations
$$
[Q_X^a,Q_Y^b] = i \delta_{XY} f_{abc} Q^c_X ,
\label{eq:conmutation_relations}
$$
which were the starting point of the Current Algebra
methods of the sixties.

This chiral symmetry, which should be
approximately good in the light quark
sector ($u$,$d$,$s$), is however not seen in
the hadronic spectrum. Although hadrons can be nicely classified in
$SU(3)_V$ representations,
degenerate multiplets with opposite parity do not exist.
Moreover, the octet of pseudoscalar mesons happens to be much
lighter than all the other hadronic states.
  To be consistent with this experimental  fact,
the ground state of the
theory (the vacuum) should not be symmetric under the chiral group.
The $SU(3)_L \otimes SU(3)_R$ symmetry
spontaneously breaks down to
$SU(3)_{L+R}$
and, according to Goldstone's (1961) theorem,
an octet of pseudoscalar massless
bosons appears in the theory.

More specifically,
let us consider a Noether charge $Q$,
and assume the existence of an
operator $O$ that satisfies
$$
\langle 0 | [Q,O] | 0 \rangle \not= 0 \, ;
\label{eq:order}
$$
this is clearly only possible if $Q|0\rangle\not= 0$.
Goldstone's theorem then tells us that there exists a
massless state $|G\rangle$
such that
$$
\langle 0|J^0|G\rangle \, \langle G|O|0\rangle\not= 0 \, .
\label{eq:Goldstone_theorem}
$$
The quantum numbers of the Goldstone boson are
dictated by those of $J^0$ and $O$.
The quantity in the left-hand side of equation~\ref{eq:order}
is called the order parameter of the spontaneous symmetry breakdown.

Since there are eight broken axial generators of the chiral
group, $Q^a_A = Q^a_R - Q^a_L$,
there should be eight pseudoscalar Goldstone states
$|G^a\rangle$, which we can identify with
the eight lightest hadronic states
($\pi^+$, $\pi^-$, $\pi^0$, $\eta$, $K^+$, $K^-$, $K^0$
and $\bar{K}^0$);
their small masses being generated by the quark-mass matrix,
which explicitly
breaks the global symmetry of the QCD Lagrangian.
The corresponding $O^a$ must be pseudoscalar operators. The simplest
possibility are $O^a = \bar q \gamma_5 \lambda_a q$, which satisfy
$$
\langle 0|[Q^a_A, \bar q \gamma_5 \lambda_b q] |0\rangle=
-{1\over 2} \,\langle 0|\bar q \{\lambda_a,\lambda_b\} q |0\rangle =
-{2\over 3} \,\delta_{ab} \,\langle 0|\bar q q |0\rangle \, .
\label{eq:vev_relation}
$$
The quark condensate
$$
\langle 0|\bar u u |0\rangle =
\langle 0|\bar d d |0\rangle =
\langle 0|\bar s s |0\rangle \not = 0
\label{eq:quark_condensate}
$$
is then
the natural order parameter of
Spontaneous Chiral Symmetry Breaking (SCSB).

\section{Effective chiral Lagrangian at lowest order}
\label{sec:lo}

The Goldstone nature of the pseudoscalar mesons implies strong
constraints on their interactions, which can be most easily analyzed
on the basis of an effective Lagrangian.
Since there is a mass gap separating the pseudoscalar octet from the
rest of the hadronic spectrum, we can build an effective field
theory containing only the Goldstone modes.
Our basic
assumption is the pattern of SCSB:
$$
G \equiv SU(3)_L\otimes SU(3)_R
\buildrel{\rm SCSB}\over\longrightarrow H \equiv SU(3)_V \, .
\label{eq:scsb}
$$

Let us denote $\phi^a$ ($a=1,\ldots,8$)  the coordinates
describing the Goldstone fields in the coset space $G/H$, and choose
a coset representative
$\bar\xi(\phi)\equiv(\xi_L(\phi),\xi_R(\phi))\in G$.
The change of the Goldstone coordinates under
a chiral transformation
$g\equiv(g_L,g_R)\in G$ is given by
$$
\xi_L(\phi)\, \toG\, g_L\,\xi_L(\phi)\, h^\dagger(\phi,g) \, ,
\qquad\quad
\xi_R(\phi)\, \toG\, g_R\,\xi_R(\phi)\, h^\dagger(\phi,g) \, ,
\label{eq:h_def}
$$
where $h(\phi,g)\in H$ is a compensating
transformation which is needed to return to the
given choice of coset representative $\bar\xi$;
in general, $h$ depends both on $\phi$ and $g$.
Since the same transformation $h(\phi,g)$ occurs in the
left and right sectors
(the two chiral sectors can be related by a parity transformation,
which obviously leaves $H$ invariant),
we can get rid of it by
combining the two chiral relations in \ref{eq:h_def}
into the simpler form
$$
U(\phi)\, \toG\, g_R\, U(\phi)\, g_L^\dagger \, ,
\qquad\qquad
%\label{eq:utransf}\cr
U(\phi)\,\equiv\,\xi_R(\phi)\,\xi_L^\dagger(\phi) \, .
\label{eq:u_def}
$$
Moreover, without lost of generality, we can take a canonical
choice of coset representative such that
$\xi_R(\phi) = \xi_L^\dagger(\phi) \equiv u(\phi)$.
%\equiv\exp{\left\{i\Phi/(\sqrt{2}f)\right\}}$.
The $3\times 3$ unitary matrix
$$
U(\phi)\, = \, u(\phi)^2\, =\,
\exp{\left\{i\sqrt{2}\Phi/f\right\}}
\label{eq:u_parametrization}
$$
gives a very convenient parametrization of the
Goldstone fields
$$
\Phi (x) \equiv {\vec{\lambda}\over\sqrt 2} \, \vec{\phi}
 = \,
\pmatrix{{1\over\sqrt 2}\pi^0 \, +
\, {1\over\sqrt 6}\eta_8
 & \pi^+ & K^+ \cr
\pi^- & - {1\over\sqrt 2}\pi^0 \, + \, {1\over\sqrt 6}\eta_8
 & K^0 \cr K^- & \bar K^0 & - {2 \over\sqrt 6}\eta_8 }.
\label{eq:phi_matrix}
$$
Notice that
$U(\phi)$ transforms linearly under the chiral group,
%[equation~\ref{eq:utransf}],
but the induced transformation on the Goldstone fields
$\vec{\phi}$ is highly non-linear.

To get a  low-energy effective Lagrangian realization of QCD,
for the light-quark sector ($u$, $d$, $s$),
we should write the most general Lagrangian involving the matrix
$U(\phi)$, which is consistent with chiral symmetry.
The Lagrangian can be organized in terms of increasing powers of
momentum or, equivalently, in terms of an increasing number of
derivatives
(parity conservation requires an even number of derivatives):
$$
\cL_{\rm eff}(U) \, = \, \sum_n \cL_{2n} \, .
\label{eq:l_series}
$$
 In the low-energy domain we are interested in, the
terms with a minimum number of derivatives will dominate.

Due to the unitarity of the $U$ matrix, $U U^\dagger = I$, at least
two derivatives are required to generate a non-trivial interaction.
To lowest order, the effective chiral Lagrangian is uniquely
given by the term
$$
\cL_2 = {f^2\over 4}
\langle \partial_\mu U^\dagger \partial^\mu U \rangle \, ,
\label{eq:l2}
$$
where $\langle A\rangle$ denotes the trace of the matrix $A$.

Expanding $U(\phi)$ in a power series in $\Phi$, one obtains the
Goldstone kinetic terms plus a tower of interactions involving
an increasing number of pseudoscalars.
The requirement that the kinetic terms are properly normalized
fixes the global coefficient $f^2/4$ in equation~\ref{eq:l2}.
All interactions among the Goldstones can then be predicted in terms
of the single coupling $f$:
$$
{\cal L}_2 \, = \, {1\over 2} \,\langle\partial_\mu\Phi
\partial^\mu\Phi\rangle
\, + \, {1\over 12 f^2} \,\langle (\Phi\lrder_\mu\Phi) \,
(\Phi\buildrel \leftrightarrow \over {\partial^\mu}\Phi)
\rangle \, + \, \Or (\Phi^6/f^4) \, .
\label{eq:l2_expanded}
$$

To compute the $\pi\pi$ scattering amplitude, for instance, is now
a trivial perturbative exercise. One gets the well-known
(Weinberg 1966) result  [$t\equiv (p_+' - p_+)^2$]
$$
T(\pi^+\pi^0\to\pi^+\pi^0) = {t\over f^2} \, .
\label{eq:WE1}
$$
Similar results can be obtained for
$\pi\pi\to 4\pi, 6\pi, 8\pi, \ldots $
The non-linearity of the effective Lagrangian relates
amplitudes with different numbers of Goldstone bosons, allowing
for absolute predictions in terms of $f$.

The effective field theory
technique becomes much more powerful if one introduces couplings
to external classical fields.
Let us consider an extended QCD Lagrangian, with quark
couplings to external Hermitian matrix-valued fields
$v_\mu$, $a_\mu$, $s$, $p$:
$$
\cL_{QCD} = \cL^0_{QCD} +
\bar q \gamma^\mu (v_\mu + \gamma_5 a_\mu ) q -
\bar q (s - i \gamma_5 p) q \, .
\label{eq:extendedqcd}
$$
The external fields will allow us to compute the effective realization
of general Green functions of quark currents in a very straightforward
way. Moreover, they can be used to incorporate the
electromagnetic and semileptonic weak interactions, and the
explicit breaking of chiral symmetry through the quark masses:
$$
r_\mu \,\equiv\, v_\mu + a_\mu \, = \, e Q A_\mu + \ldots
\cr
\ell_\mu \, \equiv \, v_\mu - a_\mu \, =
\,  e Q A_\mu + {e\over\sqrt{2}\sin{\theta_W}}
(W_\mu^\dagger T_+ + {\rm h.c.}) + \ldots
\label{eq:breaking}\cr
s \, = \, {\cal M} + \ldots
$$
Here, $Q$ and ${\cal M}$
denote the quark-charge and quark-mass matrices, respectively,
$$
Q = {1\over 3}\, \hbox{\rm diag}(2,-1,-1)\, , \qquad\qquad
{\cal M} = \hbox{\rm diag}(m_u,m_d,m_s) \, ,
\label{eq:q_m_matrices}
$$
and $T_+$ is a $3\times 3$ matrix containing the relevant
Cabibbo--Kobayashi--Maskawa factors
$$
T_+ \, = \, \pmatrix{
0 & V_{ud} & V_{us} \cr 0 & 0 & 0 \cr 0 & 0 & 0
}.
\label{eq:t_matrix}
$$

The Lagrangian \ref{eq:extendedqcd}
is invariant under the following set of {\it local}
$SU(3)_L\otimes SU(3)_R$ transformations:
$$
q_L  \,\longrightarrow\,  g_L \, q_L \, , \qquad
q_R  \,\longrightarrow\,  g_R \, q_R \, , \qquad
s + i p  \,\longrightarrow\,  g_R \, (s + i p) \, g_L^\dagger \, ,
\cr
\ell_\mu \,\longrightarrow\,  g_L \, \ell_\mu \, g_L^\dagger \, + \,
i g_L \partial_\mu g_L^\dagger \, ,
\qquad
r_\mu  \,\longrightarrow\,  g_R \, r_\mu \, g_R^\dagger \, + \,
i g_R \partial_\mu g_R^\dagger \, .
\label{eq:symmetry}
$$
We can use this symmetry to build a generalized effective
Lagrangian for the Goldstone bosons, in the presence of external
sources. Note that to respect the local invariance, the gauge fields
$v_\mu$, $a_\mu$ can only appear through the covariant derivatives
$$
D_\mu U = \partial_\mu U - i r_\mu U + i U \ell_\mu \, ,
\qquad
D_\mu U^\dagger = \partial_\mu U^\dagger  + i U^\dagger r_\mu
- i \ell_\mu U^\dagger \, ,
\label{eq:covariant_derivative}
$$
and through the field strength tensors
$$
F^{\mu\nu}_L =
\partial^\mu \ell^\nu - \partial^\nu \ell^\mu
- i [ \ell^\mu , \ell^\nu ] \, ,
\qquad
F^{\mu\nu}_R =
\partial^\mu r^\nu - \partial^\nu r^\mu - i [ r^\mu , r^\nu ] \, .
\label{eq:field_strength}
$$
At lowest order in momenta, the more general effective Lagrangian
consistent with Lorentz invariance and (local) chiral symmetry
is of the form (Gasser and Leutwyler 1985)
$$
{\cal L}_2 = {f^2\over 4}\,
\langle D_\mu U^\dagger D^\mu U \, + \, U^\dagger\chi  \,
+  \,\chi^\dagger U
\rangle ,
\label{eq:lowestorder}
$$
where
$$
\chi \, = \, 2 B_0 \, (s + i p) ,
\label{eq:chi}
$$
and $B_0$ is a constant, which, like $f$, is not fixed by
symmetry requirements alone.

Once special directions in flavour space, like the ones in
equation~\ref{eq:breaking}, are selected for the external fields,
chiral symmetry is of course explicitly broken.
The important point is that \ref{eq:lowestorder} then breaks the
symmetry in exactly the same way as the fundamental short-distance
Lagrangian \ref{eq:extendedqcd} does.

The power of the external field technique becomes obvious when
computing the chiral Noether currents.
The Green functions are obtained as functional
derivatives of the generating functional
$Z[v,a,s,p]$, defined via the path-integral formula
$$
\fl
\exp{\{i Z\}}  =  \int  {\cal D}q \,{\cal D} \bar q
\,{\cal D}G_\mu \,
\exp{\left\{i \int d^4x\, {\cal L}_{QCD}\right\}}  =
\int  {\cal D}U \,
\exp{\left\{i \int d^4x\, {\cal L}_{\rm eff}\right\}} .
\label{eq:generatingfunctional}
$$
At lowest order in momenta, the generating functional reduces to the
classical action $S_2 = \int d^4x \,{\cal L}_2$;
therefore, the currents can be trivially
computed by taking the appropriate derivatives
with respect to the external fields:
$$
\fl
\eqalign{
J^\mu_L \doteq {\delta S_2\over \delta \ell_\mu} \, = \, &
 {i\over 2} f^2 D_\mu U^\dagger U =
\hphantom{-}{f\over\sqrt{2}} D_\mu \Phi -
{i\over 2} \, \left(\Phi
\buildrel \leftrightarrow \over {D^\mu}\Phi\right) +
\Or (\Phi^3/f) , \cr
J^\mu_R \doteq {\delta S_2\over \delta r_\mu} \, = \, &
 {i\over 2} f^2 D_\mu U U^\dagger =
-{f\over\sqrt{2}} D_\mu \Phi -
{i\over 2} \, \left(\Phi
\buildrel \leftrightarrow \over {D^\mu}\Phi\right) +
\Or (\Phi^3/f) .}
\label{eq:l_r_currents}
$$

The physical meaning of the chiral coupling $f$ is now obvious;
at $\Or (p^2)$, $f$ equals the pion decay constant,
$f = f_\pi = 92.4$ MeV, defined as
$$
\langle 0 | (J^\mu_A)^{12} | \pi^+\rangle
 \equiv i \sqrt{2} f_\pi p^\mu .
\label{f_pi}
$$
Similarly, by taking derivatives with respect to the external scalar
and pseudoscalar
sources,
$$
\eqalign{
\bar q^j_L q^i_R
\doteq -{\delta S_2\over \delta (s-ip)^{ji}} \, = &\,
-{f^2\over 2} B_0 \, U(\phi)^{ij} ,
\cr
\bar q^j_R q^i_L
\doteq -{\delta S_2\over \delta (s+ip)^{ji}} \, = &\,
-{f^2\over 2} B_0 \, U(\phi)^{\dagger ij} ,}
\label{eq:s_p_currents}
$$
we learn that the constant $B_0$ is related to the
quark condensate:
$$
\langle 0 | \bar q^j q^i|0\rangle = -f^2 B_0 \delta^{ij} .
\label{eq:b0}
$$
The Goldstone bosons, parametrized by the matrix $U(\phi)$,
correspond to the zero-energy excitations over this
vacuum condensate.

Taking $s = {\cal M}$ and $p=0$,
the $\chi$ term in equation~\ref{eq:lowestorder}
gives rise to a quadratic pseudoscalar-mass term plus
additional interactions proportional to the quark masses.
Expanding in powers of $\Phi$
(and dropping an irrelevant constant), one has:
$$
{f^2\over 4} 2 B_0 \,\langle {\cal M} (U + U^\dagger) \rangle
=  B_0 \left\{ - \langle {\cal M}\Phi^2\rangle
+ {1\over 6 f^2} \langle {\cal M} \Phi^4\rangle
+ \Or (\Phi^6/f^4) \right\} .
\label{eq:massterm}
$$
The explicit evaluation of the trace in the quadratic
mass term provides
the relation between the physical meson masses and the quark masses:
$$
M_{\pi^\pm}^2 \, = \,  2 \hat{m} B_0 \, , \qquad\qquad
M_{\pi^0}^2 \, = \,  2 \hat{m} B_0 - \varepsilon +
\Or (\varepsilon^2) \, , \cr
M_{K^\pm}^2 \, = \,  (m_u + m_s) B_0 \, , \qquad\qquad
M_{K^0}^2 \, = \,  (m_d + m_s) B_0 \, ,
\label{eq:masses} \cr
M_{\eta_8}^2 \, = \,
{2\over 3} (\hat{m} + 2 m_s)  B_0 + \varepsilon +
\Or (\varepsilon^2) \, ,
$$
where\footnote{$^{\ddagger}$}{
%%%%%%%%%%%%%
The $\Or (\varepsilon)$ corrections to $M_{\pi^0}^2$
and $M_{\eta_8}^2$
originate from a small mixing term between the
$\pi^0$ and $\eta_8$ fields:
$ \quad
- B_0 \langle {\cal M}\Phi^2\rangle \longrightarrow
- (B_0/\sqrt{3})\, (m_u - m_d)\, \pi^0\eta_8 \, .
\quad $
The diagonalization of the quadratic $\pi^0$, $\eta_8$
mass matrix, gives
the mass eigenstates,
$\pi^0 = \cos{\delta} \,\phi^3 + \sin{\delta} \,\phi^8$
and
$\eta_8 = -\sin{\delta} \,\phi^3 + \cos{\delta} \,\phi^8$,
where
$
\tan{(2\delta)} = \sqrt{3} (m_d-m_u)/\left( 2 (m_s-\hat{m})\right) .
$
}
%%%%%%%%%%%%%
$$
\hat m = {1\over 2} (m_u + m_d) \, , \qquad\qquad
\varepsilon = {B_0\over 4} {(m_u - m_d)^2\over  (m_s - \hat m)} \, .
\label{eq:mhat}
$$

Chiral symmetry relates the magnitude of the meson and quark masses
to the size of the quark condensate.
Using the result \ref{eq:b0}, one gets from
the first equation in \ref{eq:masses}
the well-known relation
(Gell-Mann, Oakes and Renner 1968)
$$
f^2_\pi M_\pi^2 = -\hat m \,\langle 0|\bar u u + \bar d d|0\rangle\, .
\label{eq:gmor}
$$

Taking out the common $B_0$ factor, equations \ref{eq:masses} imply
the old Current Algebra mass ratios
(Gell-Mann, Oakes and Renner 1968, Weinberg 1977),
$$
{M^2_{\pi^\pm}\over 2 \hat m} = {M^2_{K^+}\over (m_u+m_s)} =
{M_{K^0}\over (m_d+m_s)}
\approx {3 M^2_{\eta_8}\over (2 \hat m + 4 m_s)} \, ,
\label{eq:mratios}
$$
and
(up to $\Or (m_u-m_d)$ corrections)
the Gell-Mann (1962)--Okubo (1962) mass relation,
$$
3 M^2_{\eta_8} = 4 M_K^2 - M_\pi^2 \, .
\label{eq:gmo}
$$
Note that the chiral Lagrangian
automatically implies the successful quadratic
Gell-Mann--Okubo mass relation, and not a linear one.
Since $B_0 m_q\propto M^2_\phi$, the external field $\chi$
is counted as $\Or (p^2)$ in the chiral expansion.

Although chiral symmetry alone cannot fix the absolute values
of the quark masses, it gives information about quark-mass
ratios. Neglecting the tiny $\Or (\varepsilon)$ effects,
one gets the relations
$$
{m_d - m_u \over m_d + m_u} \, = \,
{(M_{K^0}^2 - M_{K^+}^2) - (M_{\pi^0}^2 - M_{\pi^+}^2)
\over M_{\pi^0}^2}
\, = \, 0.29 \, ,
\label{eq:ratio1}\cr
{m_s -\hat m\over 2 \hat m} \, = \,
{M_{K^0}^2 - M_{\pi^0}^2\over M_{\pi^0}^2}
\, = \, 12.6 \, .
\label{eq:ratio2}
$$
In equation~\ref{eq:ratio1} we have subtracted the pion square-mass
difference, to take into account the electromagnetic contribution
to the pseudoscalar-meson self-energies;
in the chiral limit ($m_u=m_d=m_s=0$),
this contribution is proportional
to the square of the meson charge and it is the same for
$K^+$ and $\pi^+$
(Dashen 1969).
The mass formulae \ref{eq:ratio1} and \ref{eq:ratio2}
imply the quark-mass ratios advocated by Weinberg (1977):
$$
m_u : m_d : m_s = 0.55 : 1 : 20.3 \, .
\label{eq:Weinbergratios}
$$
Quark-mass corrections are therefore dominated by $m_s$, which is
large compared with $m_u$, $m_d$.
Notice that the difference $m_d-m_u$ is not small compared with
the individual up- and down-quark masses; in spite of that,
isospin turns out
to be a very good symmetry, because
isospin-breaking effects are governed by the small ratio
$(m_d-m_u)/m_s$.

The $\Phi^4$ interactions in equation~\ref{eq:massterm}
introduce mass corrections to the $\pi\pi$ scattering amplitude
\ref{eq:WE1},
$$
T(\pi^+\pi^0\to\pi^+\pi^0) = {t - M_\pi^2\over f_\pi^2}\, ,
\label{eq:WE2}
$$
in perfect agreement with the Current Algebra result
(Weinberg 1966).
Since $f=f_\pi$ is fixed from pion decay, this result
is now an absolute prediction of chiral symmetry.

The lowest-order chiral Lagrangian \ref{eq:lowestorder} encodes
in a very compact way all the Current Algebra results obtained in
the sixties
(Adler and Dashen 1968, de Alfaro \etal 1973).
The nice feature of the chiral approach is its elegant
simplicity. Moreover, as we will see in the next section,
the effective field theory method
allows us to estimate higher-order corrections in a systematic way.

\goodbreak
\section{ChPT at $\Or (p^4)$}
\label{sec:p4}

At next-to-leading order in momenta, $\Or (p^4)$, the
computation of the generating functional $Z[v,a,s,p]$ involves
three different ingredients:
\item{1.} The most general effective chiral Lagrangian of
$\Or (p^4)$, ${\cal L}_4$, to be considered at tree level.
\item{2.} One-loop graphs associated with the lowest-order
Lagrangian ${\cal L}_2$.
\item{3.} The Wess--Zumino (1971)--Witten (1983) functional
to account for the chiral anomaly.

\subsection{$\Or (p^4)$ Lagrangian}

At $\Or (p^4)$, the most general\footnote{$^{\S}$}{
%%%%%%%%%%
Since we will only need ${\cal L}_4$ at tree level,
the general expression of this Lagrangian has been simplified,
using the $\Or (p^2)$ equations of motion obeyed by $U$.
Moreover, a $3\times 3$ matrix relation has been used to reduce the
number of independent terms.
For the two-flavour case, not all of these terms are independent
(Gasser and Leutwyler 1984, 1985).}
Lagrangian, invariant under
parity, charge conjugation and
the local chiral transformations \ref{eq:symmetry},
is given by (Gasser and Leutwyler 1985)
$$
\fl\eqalign{
{\cal L}_4  \, = \, &
L_1 \,\langle D_\mu U^\dagger D^\mu U\rangle^2 \, + \,
L_2 \,\langle D_\mu U^\dagger D_\nu U\rangle\,
   \langle D^\mu U^\dagger D^\nu U\rangle
\cr  &
+~L_3 \,\langle D_\mu U^\dagger D^\mu U D_\nu U^\dagger
D^\nu U\rangle\,
+ \, L_4 \,\langle D_\mu U^\dagger D^\mu U\rangle\,
   \langle U^\dagger\chi +  \chi^\dagger U \rangle
\cr  &
+~L_5 \,\langle D_\mu U^\dagger D^\mu U \left( U^\dagger\chi +
\chi^\dagger U
\right)\rangle\,
+ \, L_6 \,\langle U^\dagger\chi +  \chi^\dagger U \rangle^2
\cr  &
+~L_7 \,\langle U^\dagger\chi -  \chi^\dagger U \rangle^2\,
+ \, L_8 \,\langle\chi^\dagger U \chi^\dagger U
+ U^\dagger\chi U^\dagger\chi\rangle
\cr  &
-~i L_9 \,\langle F_R^{\mu\nu} D_\mu U D_\nu U^\dagger +
     F_L^{\mu\nu} D_\mu U^\dagger D_\nu U\rangle\,
+ \, L_{10} \,\langle U^\dagger F_R^{\mu\nu} U F_{L\mu\nu} \rangle
\cr  &
+~H_1 \,\langle F_{R\mu\nu} F_R^{\mu\nu} +
F_{L\mu\nu} F_L^{\mu\nu}\rangle\,
+ \, H_2 \,\langle \chi^\dagger\chi\rangle \, .}
\label{eq:l4}
$$

The terms proportional to $H_1$ and $H_2$ do not contain the
pseudoscalar fields and are therefore not directly measurable.
Thus, at $\Or (p^4)$ we need ten additional coupling constants
$L_i$
to determine the low-energy behaviour of the Green functions.
These constants  parametrize our
ignorance about the details of the underlying QCD dynamics.
In principle, all the chiral couplings are calculable functions
of $\Lambda_{QCD}$ and the heavy-quark masses. At the present time,
however, our main source of information about these couplings
is low-energy phenomenology.

\subsection{Chiral loops}

ChPT is a quantum field theory, perfectly defined through
equation~\ref{eq:generatingfunctional}. As such, we must take
into account quantum loops with Goldstone-boson propagators in the
internal lines.
The chiral loops generate non-polynomial contributions,
with logarithms and threshold factors, as required by unitarity.

The loop integrals are homogeneous functions
of the external momenta and
the pseudoscalar masses occurring in the propagators.
A simple dimensional counting shows that,
for a general connected diagram with $N_d$ vertices of
$\Or (p^d)$ ($d=2,4,\ldots$) and $L$ loops,
the overall chiral dimension is given by (Weinberg 1979)
$$
D = 2 L + 2 + \sum_d N_d \, (d-2) \, .
\label{eq:d_counting}
$$
Each loop  adds two powers of momenta;
this power suppression of loop diagrams is at the basis of low-energy
expansions, such as ChPT.
The leading $D=2$ contributions are obtained with $L=0$ and $d=2$,
i.e. only tree-level graphs with ${\cal L}_2$ insertions.
At $\Or (p^4)$, we  have tree-level contributions from
${\cal L}_4$ ($L=0$, $d=4$, $N_4=1$) and one-loop graphs with the
lowest-order Lagrangian ${\cal L}_2$ ($L=1$, $d=2$).

The Goldstone loops are divergent and need to be renormalized.
Although effective field theories
are non-renormalizable (i.e. an infinite number
of counter-terms is required),
order by order in the momentum expansion
they define a perfectly renormalizable theory.
If we use a regularization which preserves the symmetries of
the Lagrangian\footnote{$^\sharp$}{
%%%%%%%
A rather comprehensive analysis of different regularization schemes in
ChPT has been given by Espriu and Matias (1994).
%%%%%%%
}, such as dimensional regularization,
the counter-terms needed to renormalize the theory will be
necessarily symmetric.
Since by construction the full effective Lagrangian
contains all terms permitted by the symmetry,
the divergences can then be absorbed in a renormalization of the
coupling constants occurring in the Lagrangian.
At one loop (in ${\cal L}_2$), the ChPT divergences are $\Or (p^4)$
and are therefore renormalized by the low-energy couplings
in equation~\ref{eq:l4}:
$$
L_i = L_i^r(\mu) + \Gamma_i \lambda \, , \qquad\qquad
H_i = H_i^r(\mu) + \widetilde\Gamma_i \lambda \, ,
\label{eq:renormalization}
$$
where
$$
\lambda = {\mu^{d-4}\over 16 \pi^2} \left\{
{1\over d-4} -{1\over 2} \left[ \log{(4\pi)} + \Gamma'(1) + 1 \right]
\right\} .
\label{eq:divergence}
$$
The explicit calculation of the one-loop generating functional $Z_4$
(Gasser and Leutwyler 1985) gives:
$$
\fl
\Gamma_1 = {3\over 32}\, , \quad\; \Gamma_2 = \dfrac{3}{16}\, ,
\qquad  \Gamma_3 = 0 \, ,
\qquad \Gamma_4 = {1\over 8} \, , \qquad\;\,
\Gamma_5 = {3\over 8} \, , \qquad \Gamma_6 = \dfrac{11}{144} \, ,
\cr \fl
\Gamma_7 = 0 \, ,
\qquad \Gamma_8 = {5\over 48}\, , \qquad
\Gamma_9 = {1\over 4} \, ,\qquad\! \Gamma_{10} = -\dfrac{1}{4} \, ,
\quad\; \widetilde\Gamma_1 = -{1\over 8} \, , \quad\;
\widetilde\Gamma_2 = {5\over 24}\, .
\label{eq:d_factors}
$$
The renormalized couplings $L_i^r(\mu)$ depend on the arbitrary
scale of dimensional regularization $\mu$.
This scale dependence is of course
cancelled by that of the loop amplitude, in any
measurable quantity.

A typical $\Or (p^4)$ amplitude will then consist of a non-polynomial
part, coming from the loop computation, plus a polynomial
in momenta and
pseudoscalar masses, which depends on the unknown constants $L_i$.
The non-polynomial part (the so-called chiral logarithms) is
completely predicted as a function
of the lowest-order coupling $f$ and
the Goldstone masses.

This chiral structure can be easily understood in terms
of dispersion relations.
Given the lowest-order Lagrangian ${\cal L}_2$, the non-trivial
analytic behaviour associated with  some physical intermediate state
is calculable without the introduction of new arbitrary
chiral coefficients.
Analyticity then allows us to reconstruct the full amplitude, through
a dispersive integral, up to a subtraction polynomial.
ChPT generates (perturbatively) the correct dispersion integrals and
organizes the subtraction polynomials in a derivative expansion.

ChPT is an expansion in powers of momenta over some typical hadronic
scale, usually called the scale of chiral symmetry breaking
$\Lambda_\chi$.
The variation of the loop contribution under a rescaling of
$\mu$, by say $e$, provides a natural order-of-magnitude
estimate\footnote{$^*$}{
%%%%%
Since the loop amplitude increases with
the number of possible Goldstone mesons in the internal lines,
this estimate results in a slight dependence of $\Lambda_\chi$ on
the number of light-quark flavours $N_f$
(Soldate and Sundrum 1990, Chivukula \etal 1993):
$\Lambda_\chi\sim 4\pi f_\pi/\sqrt{N_f}$.}
%%%%%
of $\Lambda_\chi$
(Weinberg 1979, Manohar and Georgi 1984):
$\Lambda_\chi\sim 4\pi f_\pi\sim 1.2\,{\rm GeV}$.

\subsection{The chiral anomaly}

Although the QCD Lagrangian \ref{eq:extendedqcd}
is invariant under local
chiral transformations, this is no longer true for the associated
generating functional.
The anomalies of the fermionic determinant break chiral symmetry
at the quantum level
(Adler 1969, Bardeen 1969, Bell and Jackiw 1969).
The fermionic determinant can always be defined with the convention
that $Z[v,a,s,p]$ is invariant under vector transformations.
Under an infinitesimal chiral transformation
$$
g_{L,R} = 1 + i \alpha \mp i \beta + \ldots
\label{eq:inf}
$$
the anomalous change of the generating functional
is then given by (Bardeen 1969):
$$
\delta Z[v,a,s,p]  \, = \,
-{N_C\over 16\pi^2} \, \int d^4x \,
\langle \beta(x) \,\Omega(x)\rangle \, ,
\label{eq:anomaly}\cr
\fl
\Omega(x)  = \varepsilon^{\mu\nu\sigma\rho} \!
 \left[
v_{\mu\nu} v_{\sigma\rho}
+ {4\over 3} \,\nabla_\mu a_\nu \nabla_\sigma a_\rho
+ {2\over 3} i \,\{ v_{\mu\nu},a_\sigma a_\rho\}
%\right.\cr
%\qquad\qquad\qquad\left.\hbox{}
+ {8\over 3} i \, a_\sigma v_{\mu\nu} a_\rho
+ {4\over 3} \, a_\mu a_\nu a_\sigma a_\rho \right] \! ,
\label{eq:anomaly_b}\cr
v_{\mu\nu} \, = \,
\partial_\mu v_\nu - \partial_\nu v_\mu - i \, [v_\mu,v_\nu] \, ,
\qquad\quad
\nabla_\mu a_\nu  \, = \,
\partial_\mu a_\nu - i \, [v_\mu,a_\nu] \, .
\label{eq:anomaly_c}
$$
($N_C =3$ is the number of colours, and $\varepsilon_{0123}=1$).
Note that $\Omega(x)$ only depends on the external fields $v_\mu$
and $a_\mu$.
This anomalous variation of $Z$ is an $\Or (p^4)$
effect, in the chiral counting.

So far, we have been imposing chiral symmetry to construct the
effective ChPT Lagrangian.
Since chiral symmetry is explicitly violated
by the anomaly
at the fundamental
QCD level,
we need to add a functional $Z_A$ with the property that its
change under a chiral gauge transformation reproduces
\ref{eq:anomaly}.
Such a functional was first constructed by Wess and Zumino (1971),
and reformulated in a nice geometrical way by Witten (1983).
It has the explicit form:
$$
\fl
S[U,\ell,r]_{WZW} = -\,\dfrac{i N_C}{240 \pi^2}
\int d\sigma^{ijklm} \left\langle \Sigma^L_i
\Sigma^L_j \Sigma^L_k \Sigma^L_l \Sigma^L_m \right\rangle
\cr \quad\!
 -\,\dfrac{i N_C}{48 \pi^2} \int d^4 x\,
\varepsilon_{\mu \nu \alpha \beta}\left( W (U,\ell,r)^{\mu \nu
\alpha \beta} - W ({\bf 1},\ell,r)^{\mu \nu \alpha \beta} \right) ,
\label{eq:WZW}
$$
$$
\fl
W (U,\ell,r)_{\mu \nu \alpha \beta}  =
\left\langle U \ell_{\mu} \ell_{\nu} \ell_{\alpha}U^{\dg} r_{\beta}
+ \frac{1}{4} U \ell_{\mu} U^{\dg} r_{\nu} U \ell_\alpha U^{\dg}
r_{\beta}
+ i U \partial_{\mu} \ell_{\nu} \ell_{\alpha} U^{\dg} r_{\beta}
\right.\cr
 +~ i \partial_{\mu} r_{\nu} U \ell_{\alpha} U^{\dg} r_{\beta}
- i \Sigma^L_{\mu} \ell_{\nu} U^{\dg} r_{\alpha} U \ell_{\beta}
+ \Sigma^L_{\mu} U^{\dg} \partial_{\nu} r_{\alpha} U \ell_\beta
\cr
 -~ \Sigma^L_{\mu} \Sigma^L_{\nu} U^{\dg} r_{\alpha} U \ell_{\beta}
+ \Sigma^L_{\mu} \ell_{\nu} \partial_{\alpha} \ell_{\beta}
+ \Sigma^L_{\mu} \partial_{\nu} \ell_{\alpha} \ell_{\beta}
 - i \Sigma^L_{\mu} \ell_{\nu} \ell_{\alpha} \ell_{\beta}
\cr \left.
+~\frac{1}{2} \Sigma^L_{\mu} \ell_{\nu} \Sigma^L_{\alpha} \ell_{\beta}
- i \Sigma^L_{\mu} \Sigma^L_{\nu} \Sigma^L_{\alpha} \ell_{\beta}
\right\rangle
%\cr
 - \left( L \leftrightarrow R \right) , \label{eq:WZW2}
$$
where
$$
\Sigma^L_\mu = U^{\dg} \partial_\mu U \, , \qquad\qquad
\Sigma^R_\mu = U \partial_\mu U^{\dg} \, ,
\label{eq:sima_l_r}
$$
and
$\left( L \leftrightarrow R \right)$ stands for the interchanges
$U \leftrightarrow U^\dg $, $\ell_\mu \leftrightarrow r_\mu $ and
$\Sigma^L_\mu \leftrightarrow \Sigma^R_\mu $.
The integration in the first term of equation~\ref{eq:WZW} is over a
five-dimensional manifold whose boundary is four-dimensional
Minkowski
space. The integrand is a surface term;
therefore both the first and the
second terms of $S_{WZW}$ are $\Or (p^4)$, according to the chiral
counting rules.

Since anomalies have a short-distance origin, their effect is
completely calculable. The translation from the fundamental
quark--gluon level to the effective chiral level is unaffected by
hadronization problems.
In spite of its considerable complexity, the anomalous action
\ref{eq:WZW} has no free parameters.

The anomaly functional gives rise to interactions that break
the intrinsic parity.
It is responsible for the $\pi^0\to 2\gamma$,
$\eta\to 2 \gamma$ decays, and the $\gamma 3\pi$,
$\gamma\pi^+\pi^-\eta$
interactions;
a detailed analysis of these processes has been given by
Bijnens (1993a).
The five-dimensional surface term generates interactions among five
or more Goldstone bosons.

\section{Low-energy phenomenology at $\Or (p^4)$}
\label{sec:phenomenology}

At lowest order in momenta,
the predictive power of the chiral Lagrangian was
really impressive; with only two low-energy couplings, it was
possible to describe all Green functions associated
with the pseudoscalar-meson interactions.
The symmetry constraints become less powerful at
higher orders. Ten additional constants appear in the
$\cL_4$ Lagrangian, and many more\footnote{$^\|$}{
%%%%%%%%%
According to a recent analysis (Fearing and Scherer 1994),
$\cL_6$ involves 111 (32) independent terms of even (odd)
intrinsic parity.}
%%%%%%%%%
would be needed
at $\Or (p^6)$.
Higher-order terms in the chiral expansion
are much more sensitive
to the non-trivial aspects of the underlying QCD dynamics.

With $p \lap M_K \, (M_\pi)$,
we expect $\Or (p^4)$
corrections to the lowest-order amplitudes at the level
of $p^2/\Lambda_\chi^2 \lap 20\% \, (2\% )$.
We need to include those corrections if we aim to increase
the accuracy of the ChPT predictions beyond this level.
Although the number of free constants in $\cL_4$ looks
quite big, only a few of them
contribute to a given  observable.
In the absence of external fields, for instance,
the Lagrangian reduces to the first three terms; elastic
$\pi\pi$ and $\pi K$ scatterings are then sensitive to
$L_{1,2,3}$.
The two-derivative couplings $L_{4,5}$ generate mass corrections
to the meson decay constants (and
mass-dependent wave-function renormalizations).
Pseudoscalar masses are affected by the non-derivative
terms $L_{6,7,8}$;
$L_9$ is mainly responsible for the charged-meson
electromagnetic radius
and $L_{10}$, finally, only contributes to amplitudes with at least
two external vector or axial-vector fields, like
the radiative semileptonic decay $\pi\to e\nu\gamma$.

Table~\ref{tab:Lcouplings}
(Bijnens, Ecker and Gasser 1994)
summarizes the present status of the phenomenological determination
of the constants $L_i$
(Gasser and Leutwyler 1985, Bijnens and Cornet 1988, Bijnens 1990,
Riggenbach \etal 1991, Bijnens, Colangelo and Gasser 1994).
The quoted numbers correspond to the
renormalized couplings, at a scale $\mu = M_\rho$.
The values of these couplings at any other
renormalization scale can be
trivially obtained, through the logarithmic running implied by
\ref{eq:renormalization}:
$$
L_i^r(\mu_2) \, = \, L_i^r(\mu_1) \, + \, {\Gamma_i\over (4\pi)^2}
\,\log{\left({\mu_1\over\mu_2}\right)} .
\label{eq:l_running}
$$

%%%%%%%%%%%%%%%%%%%%%%%%%%% TABLE %%%%%%%%%%%%%%%%%
\midinsert
\table{Phenomenological values of the
renormalized couplings $L_i^r(M_\rho)$.
The last column shows the source used to extract this information.}
%\vspace{0.2cm}
\label{tab:Lcouplings}
\align\C{#}&&\C{#}\cr
\br
$i$ & $L_i^r(M_\rho) \times 10^3$ & Source \cr
\mr
1 & $\hphantom{-}0.4\pm0.3$ & $K_{e4}$, $\pi\pi\to\pi\pi$
\cr
2 & $\hphantom{-}1.4\pm0.3$ & $K_{e4}$, $\pi\pi\to\pi\pi$
\cr
3 & $-3.5\pm1.1$ & $K_{e4}$, $\pi\pi\to\pi\pi$
\cr
4 & $-0.3\pm0.5$ &  Zweig rule
\cr
5 & $\hphantom{-}1.4\pm0.5$ & $F_K : F_\pi$
\cr
6 & $-0.2\pm0.3$ & Zweig rule
\cr
7 & $-0.4\pm0.2$ & Gell-Mann--Okubo, $L_5$, $L_8$
\cr
8 & $\hphantom{-}0.9\pm0.3$ & $M_{K^0} - M_{K^+}$, $L_5$,
$(m_s - \hat{m}) : (m_d-m_u)$
\cr
9 & $\hphantom{-}6.9\pm0.7$ & $\langle r^2\rangle^\pi_V$
\cr
10 & $-5.5\pm0.7$ & $\pi\to e\nu\gamma$
\cr\br
\endalign
\endtable
%%%%%%%%%%%%%%%%%%%%%%%%%%%%%%%%%%%%%%%%%%%%%%%%%%%

Comparing the Lagrangians $\cL_2$ and $\cL_4$,
one can make an estimate
of the expected size of the couplings $L_i$ in terms of the scale of
SCSB. Taking
$\Lambda_\chi \sim 4 \pi f_\pi \sim 1.2\, {\rm GeV}$,
one would get
$$
L_i \sim {f_\pi^2/4 \over \Lambda_\chi^2} \sim {1\over 4 (4 \pi)^2}
\sim 2\times 10^{-3} ,
\label{eq:l_size}
$$
in reasonable agreement with the phenomenological values quoted in
table~\ref{tab:Lcouplings}.
This indicates a good convergence of the momentum expansion
below the  resonance region, i.e. $p < M_\rho$.

The chiral Lagrangian allows us to make a good book-keeping of
phenomenological information with a few couplings.
Once these couplings have been fixed,
we can predict many other quantities. In addition,
the information contained in table~\ref{tab:Lcouplings}
is very useful to easily test different QCD-inspired models.
Given any particular model aiming to correctly describe QCD at low
energies, we no longer need to make an extensive
phenomenological analysis to test its reliability; it suffices
to calculate the low-energy couplings predicted by the model,
and compare them with the values in table~\ref{tab:Lcouplings}.

An exhaustive description of the chiral phenomenology
at $\Or (p^4)$  is beyond the
scope of these review.
Instead, I will just present a few examples to
illustrate both the power and limitations of the ChPT techniques.

\subsection{Decay constants}

In the isospin limit ($m_u = m_d = \hat m$),
the $\Or (p^4)$ calculation of the meson-decay constants
gives (Gasser and Leutwyler 1985):
$$
\fl\eqalign{
f_\pi \, = \, & f \left\{ 1 - 2\mu_\pi - \mu_K +
    {4 M_\pi^2\over f^2} L_5^r(\mu)
    + {8 M_K^2 + 4 M_\pi^2 \over f^2} L_4^r(\mu)
    \right\} ,
\cr
f_K \, = \, & f \left\{ 1 - {3\over 4}\mu_\pi - {3\over 2}\mu_K
    - {3\over 4}\mu_{\eta_8}
    + {4 M_K^2\over f^2} L_5^r(\mu)
    + {8 M_K^2 + 4 M_\pi^2 \over f^2} L_4^r(\mu)
    \right\} ,
\cr
f_{\eta_8} \, = \, & f \left\{ 1 - 3\mu_K +
    {4 M_{\eta_8}^2\over f^2} L_5^r(\mu)
    + {8 M_K^2 + 4 M_\pi^2 \over f^2} L_4^r(\mu)
    \right\} , }
\label{eq:f_meson}
$$
where
$$
\mu_P \equiv {M_P^2\over 32 \pi^2 f^2} \,
\log{\left( {M_P^2\over\mu^2}\right)} .
\label{eq:mu_p}
$$
The result depends on two $\Or (p^4)$ couplings, $L_4$ and $L_5$.
The $L_4$ term generates a universal shift of all meson-decay
constants,
$\delta f^2 = 16 L_4 B_0 \langle\cM\rangle$,
which can be eliminated taking ratios.
{}From the experimental value (Leutwyler and Roos 1984)
$$
{f_K\over f_\pi} = 1.22\pm 0.01 \, ,
\label{eq:f_k_pi_ratio}
$$
one can then fix $L_5(\mu)$; this gives the result quoted in
table~\ref{tab:Lcouplings}.
Moreover, one gets the absolute prediction
(Gasser and Leutwyler 1985)
$$
{f_{\eta_8}\over f_\pi} = 1.3 \pm 0.05 \, .
\label{eq:f_eta_pi_ratio}
$$
Taking into account isospin violations, one can also predict
(Gasser and Leutwyler 1985) a tiny
difference between $f_{K^\pm}$ and $f_{K^0}$, proportional to
$m_d-m_u$.

\subsection{Electromagnetic form factors}

At $\Or (p^2)$ the electromagnetic coupling of the Goldstone bosons
is just the minimal one, obtained through the covariant derivative.
The next-order corrections generate a momentum-dependent
form factor:
$$
F^{\phi^\pm}_V(q^2) = 1 + {1\over 6} \,
\langle r^2 \rangle^{\phi^\pm}_V \, q^2 + \ldots \quad ;
\quad
F^{\phi^0}_V(q^2) =  {1\over 6} \,
\langle r^2 \rangle^{\phi^0}_V \, q^2 + \ldots
\label{eq:ff}
$$
The meson electromagnetic radius
$\langle r^2 \rangle^\phi_V$
gets local contributions from the $L_9$ term,
plus logarithmic loop corrections (Gasser and Leutwyler 1985):
$$
\eqalign{
\langle r^2 \rangle^{\pi^\pm}_V \, =\, & {12 L^r_9(\mu)\over f^2}
    - {1\over 32 \pi^2 f^2} \left\{
   2 \log{\left({M_\pi^2\over\mu^2}\right)}
    + \log{\left({M_K^2\over\mu^2}\right)} + 3 \right\} ,
\cr
\langle r^2 \rangle^{K^0}_V \, =\, & - {1\over 16 \pi^2 f^2}
\,\log{\left({M_K\over M_\pi}\right) } ,
\cr
\langle r^2 \rangle^{K^\pm}_V \, =\, & \langle r^2 \rangle^{\pi^\pm}_V
   + \langle r^2 \rangle^{K^0}_V . }
\label{eq:radius}
$$

Since neutral bosons do not couple to the photon at tree level,
$\langle r^2 \rangle^{K^0}_V$
only gets a loop contribution, which is moreover finite
(there cannot be any divergence because there
exists no counter-term to renormalize it).
The predicted value,
$\langle r^2 \rangle^{K^0}_V = -0.04\pm 0.03 \, {\rm fm}^2$, is in
perfect agreement with the experimental determination
(Molzon \etal 1978)
$\langle r^2 \rangle^{K^0}_V = -0.054\pm 0.026 \, {\rm fm}^2$.

The measured electromagnetic pion radius,
$\langle r^2 \rangle^{\pi^\pm}_V = 0.439\pm 0.008 \, {\rm fm}^2$
(Amendolia \etal 1986),
is used as input to estimate the coupling $L_9$.
This observable provides  a good example of the importance of
higher-order local terms in the chiral expansion (Leutwyler 1989).
If one tries to ignore the $L_9$ contribution, using instead some
{\it physical} cut-off $p_{\rm max}$ to regularize the  loops,
one needs $p_{\rm max}\sim 60 \, {\rm GeV}$,
in order to reproduce the experimental value; this is clearly
nonsense.
The pion charge radius is dominated by the $L^r_9(\mu)$
contribution, for any reasonable value of $\mu$.

The measured $K^+$ charge radius (Dally \etal 1982),
$\langle r^2 \rangle^{K^\pm}_V = 0.28\pm 0.07 \, {\rm fm}^2$,
has a larger experimental uncertainty.
Within present errors, it is in agreement with the parameter-free
relation in equation~\ref{eq:radius}.

\subsection{$K_{l3}$ decays}

The semileptonic decays $K^+\to\pi^0 l^+ \nu_l$ and
$K^0\to\pi^- l^+ \nu_l$ are governed by the corresponding
hadronic matrix
elements of the vector current [$t\equiv (P_K-P_\pi)^2$],
$$
\langle \pi| \bar s\gamma^\mu u |K\rangle = C_{K\pi} \,\left[
\left( P_K + P_\pi\right)^\mu \, f_+^{K\pi}(t) \, + \,
\left( P_K - P_\pi\right)^\mu \, f_-^{K\pi}(t) \right] ,
\label{eq:vector_matrix}
$$
where $C_{K^+\pi^0} = 1/\sqrt{2}$, $C_{K^0\pi^-} = 1$.
At lowest order, the two form factors reduce to trivial constants:
$f_+^{K\pi}(t) = 1$ and $f_-^{K\pi}(t) = 0$.
There is however a sizeable correction to $f_+^{K^+\pi^0}(t)$,
due to $\pi^0\eta$ mixing, which is
proportional to $(m_d-m_u)$,
$$
f_+^{K^+\pi^0}(0) \, = \,
1 + {3\over 4} \, {m_d-m_u\over m_s - \hat m}
\, = \, 1.017 \, .
\label{eq:fp_kp_p0}
$$
This number should be compared with the experimental ratio
$$
{f_+^{K^+\pi^0}(0)\over f_+^{K^0\pi^-}(0)} \, = \,
1.028 \pm 0.010 \, .
\label{eq:expratio}
$$
The $\Or (p^4)$ corrections to $f_+^{K\pi}(0)$ can be expressed in
a parameter-free manner in terms of the physical meson masses
(Gasser and Leutwyler 1985).
Including those contributions,
one gets the more precise values
$$
 f_+^{K^0\pi^-}(0) = 0.977 \, , \qquad \qquad
{f_+^{K^+\pi^0}(0)\over f_+^{K^0\pi^-}(0)} = 1.022 \, ,
\label{eq:fp_predictions}
$$
which are in perfect agreement with the experimental result
\ref{eq:expratio}.
The accurate ChPT calculation of these quantities allows us to
extract (Leutwyler and Roos 1984) the most precise determination
of the Cabibbo--Kobayashi--Maskawa matrix element $V_{us}$:
$$
|V_{us}| = 0.2196 \pm 0.0023 \, .
\label{eq:v_us}
$$

At $\Or (p^4)$, the form factors get momentum-dependent contributions.
Since $L_9$ is
the only unknown chiral coupling occurring in $f_+^{K\pi}(t)$ at this
order, the slope
$\lambda_+$
of this form factor can be fully predicted:
$$
\lambda_+ \equiv {1\over 6 }\,\langle r^2\rangle^{K\pi}_V\, M_\pi^2
= 0.031\pm 0.003 \, .
\label{eq:slope}
$$
This number is in excellent agreement with the experimental
determinations (Particle Data Group 1994),
$\lambda_+ = 0.0300\pm 0.0016$ ($K^0_{e3}$) and
$\lambda_+ = 0.0286\pm 0.0022$ ($K^\pm_{e3}$).

Instead of $f_-^{K\pi}(t)$, it is usual to parametrize the
experimental results in terms of the so-called
scalar form factor
$$
f_0^{K\pi}(t) = f_+^{K\pi}(t) +{t\over M_K^2 - M_\pi^2} f_-^{K\pi}(t)
\, .
\label{eq:scalar_ff}
$$
The slope of this form factor is determined by the constant $L_5$,
which in turn is fixed by $f_K/f_\pi$.
One gets the result (Gasser and Leutwyler 1985):
$$
\lambda_0 \equiv {1\over 6 }\,\langle r^2\rangle^{K\pi}_S\, M_\pi^2
= 0.017\pm 0.004 \, .
\label{eq:slope2}
$$
The experimental situation concerning the
value of this slope is far from clear;
while an older high-statistics measurement
(Donaldson \etal 1974),
$\lambda_0 = 0.019\pm 0.004$, confirmed the theoretical expectations,
more recent experiments find higher values, which disagree with this
result. Cho \etal (1980), for instance, report
$\lambda_0 = 0.046\pm 0.006$,
which differs from \ref{eq:slope2} by more than 4 standard
deviations.
The Particle Data Group (1994) quotes a world average
$\lambda_0 = 0.025\pm 0.006$.

\subsection{Meson and quark masses}

The relations \ref{eq:masses} get modified at $\Or (p^4)$.
The additional contributions depend on the low-energy constants
$L_4$, $L_5$, $L_6$, $L_7$ and $L_8$.
It is possible, however, to obtain one relation between the
quark and meson masses, which does not contain any of the $\Or (p^4)$
couplings.
The dimensionless ratios
$$
Q_1 \equiv {M_K^2 \over M_\pi^2} \, , \qquad\qquad
Q_2 \equiv {(M_{K^0}^2 - M_{K^+}^2)_{\rm QCD}
%- (M_{\pi^0}^2 - M_{\pi^+}^2)
    \over M_K^2 - M_{\pi}^2}\,  ,
\label{eq:q1q2_def}
$$
get  the same $\Or (p^4)$ correction (Gasser and Leutwyler 1985):
$$
Q_1 = {m_s + \hat m \over 2 \hat m} \, \{ 1 + \Delta_M\} ,
\qquad\qquad
Q_2 = {m_d - m_u \over m_s - \hat m} \, \{ 1 + \Delta_M\} ,
\label{eq:q1q2}
$$
where
$$
\Delta_M = - \mu_\pi + \mu_{\eta_8} + {8\over f^2}\,
(M_K^2 - M_\pi^2)\, \left[ 2 L_8^r(\mu) - L_5^r(\mu)\right] .
\label{eq:delta_m}
$$
Therefore, at this order, the ratio $Q_1/Q_2$ is just given
by the corresponding
ratio of quark masses,
$$
Q^2 \equiv {Q_1\over Q_2} =
{m_s^2 - \hat m^2 \over m_d^2 - m_u^2} \, .
\label{eq:Q2}
$$
To a good approximation, equation~\ref{eq:Q2}
can be written as an ellipse,
$$
\left({m_u\over m_d}\right)^2 + {1\over Q^2}\,
\left({m_s\over m_d}\right)^2 = 1 \, ,
\label{eq:ellipse}
$$
which constrains the quark-mass ratios.
The meson masses in \ref{eq:q1q2_def} refer to pure QCD;
using the Dashen (1969) theorem
$(\Delta M^2_K - \Delta M^2_\pi)_{\rm em}\equiv
(M^2_{K^+} - M^2_{K^0} - M^2_{\pi^+} + M^2_{\pi^0})_{\rm em} = 0$
to correct for the electromagnetic contributions,
the observed values of the meson masses give $Q = 24$.

Obviously, the quark-mass ratios \ref{eq:Weinbergratios},
obtained at $\Or (p^2)$, satisfy this elliptic constraint.
At $\Or (p^4)$, however, it is not possible to make a separate
determination of $m_u/m_d$ and $m_s/m_d$ without having additional
information on some of the $L_i$ couplings.

A useful quantity is the deviation of the Gell-Mann--Okubo relation,
$$
\Delta_{\rm GMO} \equiv {4 M_K^2 - 3 M_{\eta_8}^2 - M_\pi^2
    \over M_{\eta_8}^2 - M_\pi^2} \, .
\label{eq:delta_gmo}
$$
Neglecting the mass difference $m_d-m_u$, one gets
(Gasser and Leutwyler 1985)
$$
\eqalign{
\Delta_{\rm GMO} \, = \, &
{-2 \,(4 M_K^2 \mu_K - 3 M_{\eta_8}^2 \mu_{\eta_8} - M_\pi^2 \mu_\pi)
    \over M_{\eta_8}^2 - M_\pi^2}
\cr  &
   -{6\over f^2} \, (M_{\eta_8}^2 - M_\pi^2)
   \, \left[ 12 L_7^r(\mu) + 6 L_8^r(\mu) - L_5^r(\mu)\right] .}
\label{eq:dGMO}
$$
Experimentally, correcting the masses for electromagnetic effects,
$\Delta_{\rm GMO} = 0.21$. Since $L_5$ is already known, this allows
the combination $2 L_7 + L_8$ to be fixed .

In order to determine the individual quark-mass ratios
from equations~\ref{eq:q1q2}, we
would need to fix
the constant $L_8$. However, there is no way to find an observable
that isolates this coupling.
The reason is an accidental symmetry of the Lagrangian
$\cL_2 + \cL_4$, which remains invariant under
the following simultaneous change (Kaplan and Manohar 1986)
of the quark-mass matrix
and some of the chiral couplings:
$$
\cM' \, = \, \alpha \,\cM + \beta\, (\cM^\dagger)^{-1} \, \det\cM\, ,
\qquad B_0' \, = \, B_0 / \alpha\, ,
\cr
L'_6 \, = \, L_6 - \zeta \, , \qquad
L'_7 \, = \, L_7 - \zeta \, , \qquad
L'_8 \, = \, L_8 + 2 \zeta \, ,
\label{eq:kmsymmetry}
$$
where $\alpha$ and $\beta$ are arbitrary constants, and
$\zeta = \beta f^2 / (32\alpha B_0)$.
The only information on the quark-mass matrix $\cM$ that we used
to construct the effective Lagrangian was that it transforms as
$\cM\to g_R \cM g_L^\dagger$.
The matrix $\cM'$ transforms in the same manner;
therefore, symmetry alone does not allow us to distinguish between
$\cM$ and $\cM'$.
Since only the product $B_0 \cM$ appears in the Lagrangian,
$\alpha$ merely changes the value of the constant $B_0$.
The term proportional to $\beta$ is a correction of $\Or (\cM^2)$;
when inserted in $\cL_2$, it generates a contribution to
$\cL_4$, which is reabsorbed by the redefinition
of the $\Or (p^4)$ couplings.
All chiral predictions will be invariant under the transformation
\ref{eq:kmsymmetry}; therefore it is not possible to
separately determine the values of the quark masses and the
constants $B_0$, $L_6$, $L_7$ and $L_8$.
We can only fix those combinations of chiral couplings and masses
that remain invariant under \ref{eq:kmsymmetry}.

Notice that  \ref{eq:kmsymmetry}
is certainly not a symmetry of the underlying
QCD Lagrangian.
The accidental symmetry arises in the effective theory
because we are not making use of the explicit form of the
QCD Lagrangian; only its symmetry properties under chiral rotations
have been taken into account.
For instance, the matrix elements of the scalar and pseudoscalar
currents involve the physical quark masses and are not
invariant under \ref{eq:kmsymmetry};
if we had a low-energy probe of those currents
(such as a very light Higgs particle), we could directly fix $L_8$
in exactly the same way as we have determined $L_5$ using the
weak interactions to test the axial-current matrix elements
(Leutwyler 1994b).

We can resolve the ambiguity by obtaining
one additional information from outside the pseudoscalar-meson
chiral Lagrangian framework.
For instance, by analyzing the
isospin breaking in the baryon mass spectrum and the $\rho$-$\omega$
mixing (Gasser and Leutwyler 1982), it is possible to fix the ratio
$$
R\,\equiv\, {m_s - \hat m\over m_d - m_u}\, =\, 43.7\pm 2.7 \, .
\label{eq:r}
$$
Inserting this number in \ref{eq:Q2}, one gets
(Gasser and Leutwyler 1985)
$$
{m_s \over \hat m} = 25.7 \pm 2.6 \, ,
\qquad\qquad
{m_d - m_u \over 2 \hat m} \, = \, 0.28\pm 0.03 \, .
\label{eq:ms_m_ratio}
$$
Moreover, one can now determine  $L_8$ from
\ref{eq:q1q2}, and  therefore fix $L_7$ with
equation~\ref{eq:dGMO};
one gets then the values quoted in table~\ref{tab:Lcouplings}.

The error in \ref{eq:ms_m_ratio} includes an educated guess of the
uncertainties associated with higher-order corrections and
electromagnetic effects.
It has been pointed out recently that the Dashen theorem
receives large $\Or (e^2\cM)$ corrections which tend to increase the
electromagnetic contribution to the kaon mass difference.
The 1-loop logarithmic corrections are known to be sizeable
(Langacker and Pagels 1973, Maltman and Kotchan 1990, Urech 1995,
Neufeld and Rupertsberger 1995),
but the numerical result depends on the scale used
to evaluate the logarithms.
The magnitude of the non-logarithmic contribution has been recently
estimated  by two groups; although they use
a rather different framework, they get similar results:
$(\Delta M^2_K - \Delta M^2_\pi)_{\rm em} =
(1.0\pm 0.1)\times 10^{-3} \, {\rm GeV}^2$
(Donoghue, Holstein and Wyler 1992)
and
$(1.3\pm 0.4)\times 10^{-3} \, {\rm GeV}^2$ (Bijnens 1993b).
A lower number is obtained if one assumes that the
$L_7$ coupling is dominated by the $\eta'$ contribution
(see section \ref{sec:resonances}); using the measured $\eta$--$\eta'$
mixing angle and equation \ref{eq:r}, one gets then from $Q_2$:
$(\Delta M^2_K - \Delta M^2_\pi)_{\rm em} =
-(0.1\pm 1.0)\times 10^{-3} \, {\rm GeV}^2$
(Leutwyler 1990, 1994b, Urech 1995).
In view of the present uncertainties, we can take the conservative
range
$(\Delta M^2_K - \Delta M^2_\pi)_{\rm em} =
(0.75\pm 0.75)\times 10^{-3} \, {\rm GeV}^2$, which implies
$Q=22.7\pm 1.4$. The corresponding quark mass ratios are:
$$
{m_s \over \hat m} = 22.6 \pm 3.3 \, ,
\qquad\qquad
{m_d - m_u \over 2 \hat m} \, = \, 0.25\pm 0.04 \, .
\label{eq:ms_m_ratio_2}
$$

\section{The role of resonances in ChPT}
\label{sec:resonances}

It seems rather natural to expect that the lowest-mass resonances,
such as $\rho$ mesons, should have an important impact on the physics
of the pseudoscalar bosons. In particular, the low-energy
singularities due to the exchange of those resonances should generate
sizeable contributions to the chiral couplings. This can be easily
understood, making a Taylor expansion of the
$\rho$ propagator:
$$ {1\over p^2 - M_\rho^2} \, = \, {-1\over M_\rho^2} \,
\left\{ 1 + {p^2\over M_\rho^2} + \ldots \right\},
\qquad\qquad (p^2 < M_\rho^2) .
\label{eq:rho_propagator}
$$ Below the $\rho$-mass scale, the singularity associated with the
pole of the resonance propagator is replaced by the corresponding
momentum expansion. The exchange of virtual $\rho$ mesons should
result in derivative Goldstone couplings proportional to powers of
$1/M_\rho^2$.

A systematic analysis of the role of resonances in the ChPT Lagrangian
has been performed by Ecker \etal (1989a) [see also Donoghue \etal
1989]. One writes first a general chiral-invariant Lagrangian
$\cL(U,V,A,S,P)$, describing the couplings of meson resonances
of the type $V(1^{--})$, $A(1^{++})$, $S(0^{++})$ and
$P(0^{-+})$ to
the Goldstone bosons, at lowest-order in derivatives. The coupling
constants of this Lagrangian are phenomenologically extracted from
physics at the resonance-mass scale. One has then an effective chiral
theory defined in the intermediate-energy region. The
generating functional \ref{eq:generatingfunctional} is given in this
theory by the path-integral formula
$$
\fl\exp{\{i Z\}} \, = \,
\int \, {\cal D}U(\phi)\, \cD V \,\cD A \,\cD S \,\cD P
\, \exp{\left\{ i \int d^4x \,\cL(U,V,A,S,P) \right\}} .
\label{eq:pi_relation}
$$
The integration of the resonance fields  results in a low-energy
theory with  only Goldstone bosons, i.e. the usual ChPT Lagrangian. At
lowest-order, this integration can be explicitly performed by
expanding around the classical solution for the resonance fields.

The formal procedure to introduce higher-mass states in the
chiral Lagrangian was first discussed by Coleman \etal (1969) and
Callan \etal (1969).
The wanted ingredient for a non-linear representation of the chiral
group is the compensating $SU(3)_V$ transformation $h(\phi,g)$ which
appears under the action of $G$ on the
coset representative $u(\phi)$
[see equations \ref{eq:h_def} to \ref{eq:u_parametrization}]:
$$
u(\phi)\,\toG\, g_R\,u(\phi)\, h^\dagger(\phi,g)\, = \,
h(\phi,g)\,u(\phi)\, g_L^\dagger \, .
\label{eq:h_def_2}
$$

In practice, we shall only be
interested in resonances transforming as octets or singlets under
$SU(3)_V$. Denoting  the resonance multiplets generically by
$R= \vec{\lambda}\vec{R}/\sqrt{2}$  (octet) and $R_1$ (singlet), the
non-linear realization of $G$ is given by
$$
R\,\toG\, h(\phi,g) \, R \,
h(\phi,g)^\dagger
\, ,
\qquad\qquad R_1\,\toG\, R_1 \, .
\label{eq:R_transformation}
$$
Since the action of $G$ on the octet field $R$
is local, we are led to define a covariant derivative
$$
\nabla_\mu R \,=\, \partial_\mu R + [\Gamma_\mu,R] \, ,
\label{eq:d_covariant}
$$
with
$$
\Gamma_\mu \,=\,
\frac{1}{2} \left\{ u^\dagger (\partial_\mu - ir_\mu) u +
  u(\partial_\mu - i\ell_\mu) u^\dagger \right\}
\label{eq:connection}
$$
ensuring the proper transformation
$$
\nabla_\mu R \,\toG\, h(\phi,g)\,
\nabla_\mu R \,\, h(\phi,g)^\dagger\, .
\label{eq:dc_transf}
$$
Without external fields, $\Gamma_\mu$ is the usual natural connection
on  coset space.

To determine the resonance-exchange contributions to the effective
chiral  Lagrangian, we need the lowest-order couplings
%of meson resonances
to the pseudoscalar Goldstones
which are linear in the resonance fields.
It is useful to define objects transforming as $SU(3)_V$ octets:
$$
u_\mu \,\equiv\, i u^\dagger D_\mu U u^\dagger\, =\, u_\mu^\dagger\, ,
\cr
\chi_\pm \,\equiv\, u^\dagger \chi u^\dagger \pm u \chi^\dagger u\, ,
\label{eq:octet_objects}\cr
f^{\mu\nu}_\pm \, = \, u F_L^{\mu\nu} u^\dagger \pm
u^\dagger   F_R^{\mu\nu} u \, .
$$
Invoking $P$ and $C$ invariance,
the relevant lowest-order Lagrangian can be written as
(Ecker \etal 1989a)
$$
\cL_{\rm R} \, =\,
\sum_{R=V,A,S,P} \left\{\cL_{\rm Kin}(R) + \cL_2(R)\right\}
\, ,
\label{eq:res_Lagrangian}
$$
with kinetic terms\footnote{$^\star$}{
%%%%%%%
The vector and axial-vector mesons are described
in terms of antisymmetric tensor fields $V_{\mu\nu}$
and $A_{\mu\nu}$
(Gasser and Leutwyler 1984, Ecker \etal 1989a)
instead of the more familiar vector
fields.}
%This formulation is especially convenient when considering
%interactions  with external gauge fields such as the electromagnetic
%field.}
%%%%%%%%%
$$
\fl
\cL_{\rm Kin}(R \! = \! V,A) =
    - {1\over 2}\, \langle \nabla^\lambda R_{\lambda\mu}
\nabla_\nu R^{\nu\mu} -{M^2_R\over 2} \, R_{\mu\nu} R^{\mu\nu}\rangle
-{1\over 2}\, \partial^\lambda R_{1,\lambda\mu}
\partial_\nu R_1^{\nu\mu} +
{M^2_{R_1}\over 4} \, R_{1,\mu\nu} R_1^{\mu\nu} ,
%\label{eq:kin_v}
\cr\fl
\cL_{\rm Kin}(R \! = \! S,P) =  {1\over 2} \,
\langle \nabla^\mu R
\nabla_\mu R - M^2_R R^2\rangle
+ {1\over 2}  \partial^\mu R_1 \partial_\mu R_1 -
 {M^2_{R_1}\over 2} R_1^2\, ,
\label{eq:kin_s}
$$
where $M_R$, $M_{R_1}$ are the corresponding masses in the chiral
limit.  The interactions ${\cal L}_2(R)$ read
$$
\fl\eqalignno{
\cL_2[V(1^{--})] =& \, {F_V\over 2\sqrt{2}} \,
     \langle V_{\mu\nu} f_+^{\mu\nu}\rangle +
    {iG_V\over \sqrt{2}} \, \langle V_{\mu\nu} u^\mu u^\nu\rangle
\, ,
\label{eq:V_int}\cr
\cL_2[A(1^{++})] =& \, {F_A\over 2\sqrt{2}} \,
    \langle A_{\mu\nu} f_-^{\mu\nu} \rangle \, ,
\label{eq:A_int}\cr
\cL_2[S(0^{++})]=& \, c_d \, \langle S u_\mu
u^\mu\rangle + c_m \, \langle S \chi_+ \rangle +
   \tilde c_d \, S_1 \, \langle u_\mu u^\mu \rangle +
    \tilde c_m \, S_1 \, \langle \chi_+\rangle \, ,
\label{eq:S_int}\cr
\cL_2[P(0^{-+})] =& \, id_m \, \langle P \chi_-
\rangle +  i \tilde d_m \, P_1 \langle \chi_-\rangle \, .
 \label{eq:P_int}}
$$
All coupling constants are real.
The octet fields are written in the usual matrix notation
$$
\fl
V_{\mu\nu} \,=\, {\vec{\lambda}\over\sqrt{2}}\vec{V}_{\mu\nu}\, = \,
\pmatrix{
{1\over\sqrt{2}}\rho^0_{\mu\nu} + {1\over \sqrt{6}}\omega_{8,\mu\nu}
& \rho^+_{\mu\nu} & K^{*+}_{\mu\nu}
\cr
\rho^-_{\mu\nu} & - {1\over\sqrt{2}}\rho^0_{\mu\nu} +
{1\over \sqrt{6}}\omega_{8,\mu\nu} & K^{*0}_{\mu\nu}
\cr
K^{*-}_{\mu\nu} & \overline{K}^{*0}_{\mu\nu}
& -{2\over \sqrt{6}}\omega_{8,\mu\nu}
} \, ,
\label{eq:v_multiplet}
$$
and similarly for the other octets. We observe that for $V$ and $A$
only  octets can couple whereas both octets and singlets appear for
$S$ and $P$  (always to lowest order $p^2$).

{}From the measured
decay rates for $\rho^0\to e^+e^-$ and $\rho\to 2\pi$,
one can determine the vector couplings
$|F_V|= 154$ MeV and $|G_V| = 69$ MeV.
Since the pions are far from being soft, chiral corrections
to $G_V$ are expected to be important.
We can estimate the size of these corrections from
the electromagnetic form factor of the pion,
which is known to be well-reproduced by vector-meson
dominance (VMD):
$$
F^{\pi^\pm}_V(t) \approx {M_\rho^2 \over M_\rho^2 -t } \, ,
\label{eq:vmd}
$$
i.e.
$\langle r^2\rangle^{\pi^\pm}_V \approx 6/M_\rho^2 = 0.4 \,{\rm fm}^2$,
to be compared with the measured value
$\langle r^2 \rangle^{\pi^\pm}_V = 0.439\pm 0.008 \, {\rm fm}^2$.
The exchange of a $\rho$ meson between the $G_V$
and $F_V$ vertices, generates a contribution to
the electromagnetic pion radius (Ecker \etal 1989a):
$\langle r^2\rangle^{\pi^\pm}_V = 6 F_V G_V / (f^2 M_V^2)$.
Taking, $M_V=M_\rho$,
the success of the na\"{\ii}ve  VMD formula \ref{eq:vmd} requires
$G_V F_V>0$ and $|G_V|\approx |f_\pi^2/F_V| = 55$ MeV.
Including also the contribution from chiral loops
(Gasser and Leutwyler 1985), reduces this estimate to
$|G_V|= 53$ MeV, which is the value we shall adopt.
The axial parameters can be fixed using the old Weinberg
(1967b) sum rules:
$F_A^2 = F_V^2 - f_\pi^2 = (123 \, {\rm MeV})^2$ and $M_A^2 = M_V^2
F_V^2/ F_A^2 = (968 \, {\rm MeV})^2$.

$V$ exchange generates contributions to
$L_1$, $L_2$, $L_3$, $L_9$ and $L_{10}$, while
$A$ exchange only contributes to $L_{10}$ (Ecker \etal 1989a):
$$
\eqalign{ & L_1^V = {G_V^2\over 8 M_V^2}\, , \qquad L_2^V = 2 L_1^V ,
\qquad L_3^V = -6 L_1^V , \cr & L_9^V = {F_V G_V\over 2 M_V^2}\, , \qquad
L_{10}^{V+A} = - {F_V^2\over 4 M_V^2} + {F_A^2\over 4 M_A^2} \, .}
\label{eq:vmd_results}
$$
The resulting values of the $L_i$ couplings (Ecker \etal 1989a)
are summarized in
table~\ref{tab:vmd}, which compares the different resonance-exchange
contributions with the phenomenologically determined values of
$L_i^r(M_\rho)$.
The results shown in the table
clearly establish a chiral version of vector (and axial-vector) meson
dominance: whenever they can contribute at all, $V$ and $A$ exchange
seem to completely dominate the relevant coupling constants.

%%%%%%%%%%%%%%%%%%%%%%%%%%% TABLE %%%%%%%%%%%%%%%%%
\midinsert
\table{$V$, $A$, $S$, $S_1$ and $\eta_1$ contributions to the
 coupling constants $L_i^r$ in units of $10^{-3}$.
 The last column shows the results obtained using the relations
 \ref{eq:vmdpred}.}[w]
\label{tab:vmd}
\align\C{#}&&\C{#}\cr
\br i & $L_i^r(M_\rho)$ & $\hphantom{.0}V$ & $A\,$ & $\,S$ &
      $S_1$ & $\eta_1$ & Total & Total$^{c)}$
\cr\mr 1 & $\hphantom{-}0.4\pm0.3$ &
      $\hphantom{-1}0.6$ & $0\hphantom{.0}$ & $-0.2$ &
      $0.2^{b)}$ & $0$ & $\hphantom{-}0.6$ & $\hphantom{-}0.9$
\cr 2 & $\hphantom{-}1.4\pm0.3$ &
      $\hphantom{-1}1.2$ & $0\hphantom{.0}$ & $0$ &
      $0\hphantom{.2^{b)}}$ & $0$ & $\hphantom{-}1.2$
      & $\hphantom{-}1.8$
\cr 3 & $-3.5\pm1.1$ &
      $\,-3.6$ & $0\hphantom{.0}$ & $\hphantom{-}0.6$ &
      $0\hphantom{.2^{b)}}$ & $0$ & $-3.0$ & $-4.9$
\cr 4 & $-0.3\pm0.5$ &
      $\hphantom{-1}0\hphantom{.0}$ & $0\hphantom{.0}$ & $-0.5$ &
      $0.5^{b)}$ & $0$ & $\hphantom{-}0.0$ & $\hphantom{-}0.0$
\cr 5 & $\hphantom{-}1.4\pm0.5$ &
      $\hphantom{-1}0\hphantom{.0}$ & $0\hphantom{.0}$ &
      $\hphantom{-1}1.4^{a)}$ &
      $0\hphantom{.2^{b)}}$ & $0$ & $\hphantom{-}1.4$
      & $\hphantom{-}1.4$
\cr 6 & $-0.2\pm0.3$ &
      $\hphantom{-1}0\hphantom{.0}$ & $0\hphantom{.0}$ & $-0.3$ &
      $0.3^{b)}$ & $0$ & $\hphantom{-}0.0$ & $\hphantom{-}0.0$
\cr 7 & $-0.4\pm0.2$ &
      $\hphantom{-1}0\hphantom{.0}$ & $0\hphantom{.0}$ & $0$ &
      $0\hphantom{.2^{b)}}$ & $-0.3$ & $-0.3$ & $-0.3$
\cr 8 & $\hphantom{-}0.9\pm0.3$ &
      $\hphantom{-1}0\hphantom{.0}$ & $0\hphantom{.0}$ &
      $\hphantom{-1}0.9^{a)}$ &
      $0\hphantom{.2^{b)}}$ & $0$ & $\hphantom{-}0.9$
      & $\hphantom{-}0.9$
\cr 9 & $\hphantom{-}6.9\pm0.7$ &
      $\hphantom{-11}6.9^{a)}$ & $0\hphantom{.0}$ & $0$ &
      $0\hphantom{.2^{b)}}$ & $0$ & $\hphantom{-}6.9$
      & $\hphantom{-}7.3$
\cr 10 & $-5.5\pm0.7$ &
      $-10.0$ & $4.0$ & $0$ &
      $0\hphantom{.2^{b)}}$ & $0$ & $-6.0$ & $-5.5$
\cr\br
\endalign
\tabnote{$\qquad\qquad\quad$ $^{a)}$ Input. $\qquad$
$^{b)}$ Large-$N_C$ estimate. $\qquad$
$^{c)}$ With \ref{eq:vmdpred}}
\endtable
%%%%%%%%%%%%%%%%%%%%%%%%%%%%%%%%%%%%%%%%%%%%%%%%%%%%%%%%%%%%%%%%

There are different phenomenologically successful models in the
literature for $V$ and $A$ resonances [tensor-field description
(Gasser and Leutwyler 1984, Ecker \etal 1989a), massive Yang--Mills
(Mei{\ss}ner 1988), hidden gauge formulation (Bando \etal 1988),
etc.]. It can be shown (Ecker \etal 1989b)   that all models are
equivalent  (i.e. they give the same contributions to the $L_i$),
provided
they incorporate the appropriate QCD constraints at high energies.
Moreover, with additional QCD-inspired assumptions of high-energy
behaviour, such as an unsubtracted dispersion relation for the pion
electromagnetic form factor, all $V$ and $A$ couplings can be
expressed in terms of
$f_\pi$ and $M_V$ only (Ecker \etal 1989b):
$$
F_V = \sqrt{2} f_\pi\, , \qquad G_V= f_\pi/\sqrt{2}\, , \qquad
F_A=f_\pi\, , \qquad M_A=\sqrt{2} M_V\, .
\label{eq:f_relations}
$$
In that case, one has
$$
L_1^V = L_2^V/2 = - L_3^V/6 = L_9^V/8 = -L_{10}^{V+A}/6 =
f_\pi^2/(16 M_V^2) \, .
\label{eq:vmdpred}
$$
The last column in table~\ref{tab:vmd} shows the predicted
numerical values of the $L_i$ couplings, using the relations
\ref{eq:vmdpred}.

The exchange of scalar resonances
generates the contributions (Ecker \etal 1989a):
$$
\eqalign{
& L_1^{S+S_1} = -{c_d^2\over 6 M_S^2} +
     {\tilde{c}_d^2\over 2 M_{S_1}^2} \, ,
\qquad
L_3^S = {c_d^2\over 2 M_S^2} \, ,
\cr
&L_4^{S+S_1} = -{c_d c_m\over 3 M_S^2} +
     {\tilde{c}_d \tilde{c}_m\over M_{S_1}^2} \, ,
\qquad\;
L_5^S = {c_d c_m\over M_S^2} \, ,
\cr
& L_6^{S+S_1} = -{c_m^2\over 6 M_S^2} +
     {\tilde{c}_m^2\over 2 M_{S_1}^2} \, ,
\qquad
L_8^S = {c_m^2\over 2 M_S^2} \, . }
\label{eq:s_exchange}
$$
Since the experimental information is quite scarce in the scalar
sector, one needs to assume that the couplings $L_5$ and $L_8$ are
due exclusively to scalar-octet exchange, to determine the
scalar-octet couplings $c_d$ and $c_m$.
Taking $M_S=M_{a_0}=983$ MeV,
the scalar-octet contributions to the other
$L_i$ ($i=1,3,4,6$) are then fixed. Moreover, one can then predict
$\Gamma(a_0\to\eta\pi)=59$ MeV,
in good agreement with the experimental value
$\Gamma(a_0\to\eta\pi)\approx \Gamma(a_0) = (57\pm 11)$ MeV.
The $S_1$-exchange contributions can be expressed in terms of
the octet parameters using large-$N_C$ arguments. For $N_C=\infty$,
$M_{S_1}=M_S$, $|\tilde{c_d}|=|c_d|/\sqrt{3}$ and
$|\tilde{c_m}|=|c_m|/\sqrt{3}$ (Ecker \etal 1989a); therefore,
octet- and singlet-scalar exchange cancel in $L_1$, $L_4$ and $L_6$.
Although the results in table~\ref{tab:vmd} cannot be considered as a
proof for scalar dominance, they provide at least a convincing
demonstration of its consistency.

Neglecting the higher-mass
$0^{-+}$ resonances, the only  remaining meson-exchange is the one
associated with the $\eta_1$, which generates a sizeable contribution
to $L_7$ (Gasser and Leutwyler 1985, Ecker \etal 1989a):
$$
L_7^{\eta_1} = - {\tilde{d}_m^2\over 2 M_{\eta_1}^2} \, .
\label{eq:etas_contrib}
$$
The magnitude of this contribution can be calculated from the
quark-mass expansion of $M_\eta^2$ and $M_{\eta'}^2$, which fixes
the $\eta_1$
parameters in the large $N_c$ limit (Ecker \etal 1989a):
$M_{\eta_1}=804$ MeV, $|\tilde{d}_m|=20$ MeV.
The final result for $L_7$ is in close agreement
with its phenomenological value.

The combined resonance contributions appear to saturate the $L_i^r$
almost entirely (Ecker \etal 1989a). Within the uncertainties of the
approach, there is no need for invoking any additional contributions.
Although the comparison has been made for $\mu=M_\rho$, a similar
conclusion would apply for any value of $\mu$ in the low-lying
resonance region between 0.5 and 1 GeV.

\section{Short-distance estimates of ChPT parameters}
\label{sec:couplings}

All chiral couplings are in principle calculable from QCD.
They are functions of $\Lambda_{QCD}$ and the heavy-quark
masses $m_c$, $m_b$, $m_t$. Unfortunately, we are not able
at present to make such a first-principle computation.
Although the integral over the quark fields in
\ref{eq:generatingfunctional}
can be done explicitly, we do not know how to
perform analytically the remaining integration over the
gluon fields.
A perturbative evaluation of the gluonic contribution would
obviously fail in reproducing the correct dynamics of SCSB.
A possible way out is to parametrize phenomenologically the
SCSB and make a weak gluon-field expansion around the
resulting physical vacuum.

The simplest parametrization (Espriu \etal 1990)
is obtained by adding to the QCD Lagrangian the
chiral invariant term
$$
\Delta\cL_{\rm QCD} = - M_Q \left( \bar q_R U q_L +
    \bar q_L U^\dagger q_R \right) ,
\label{eq:ERTmodel}
$$
which serves to introduce the $U$ field, and a mass parameter
$M_Q$, which regulates the infra-red behaviour of the low-energy
effective action. In the presence of this term
the operator $\bar q q$ acquires a vacuum expectation value;
therefore, \ref{eq:ERTmodel} is an effective way to
generate the order parameter due to SCSB.
Making a chiral rotation of the quark fields,
$Q_L \equiv u(\phi) q_L$, $Q_R \equiv u(\phi)^\dagger q_R$,
with $U=u^2$,
the interaction \ref{eq:ERTmodel} reduces to a
mass-term for the {\it dressed} quarks $Q$; the parameter
$M_Q$ can then be interpreted as a
{\it constituent-quark mass}.

The derivation of the low-energy effective chiral Lagrangian
within this framework has been extensively discussed by Espriu \etal
(1990).
In the chiral and large-$N_C$ limits,
and including the leading gluonic
contributions, one gets:
$$
\eqalign{
& 8 L_1 = 4 L_2 = L_9 = {N_C\over 48\pi^2}
  \left[ 1 + \Or\!\left(1/M_Q^6\right)\right] , \cr
& L_3 = L_{10} = -{N_C\over 96\pi^2}
   \left[ 1 + {\pi^2\over 5 N_C}
   {\langle{\alpha_s\over\pi}GG\rangle\over M_Q^4} +
\Or\!\left(1/M_Q^6\right)\right] . }
\label{eq:ERTresult}
$$
Due to dimensional reasons, the leading contributions
to the $\Or (p^4)$ couplings only depend on
$N_C$ and geometrical factors.
It is remarkable  that $L_1$, $L_2$ and $L_9$
do not get any gluonic correction at this order; this
result is independent of the way SCSB has been
parametrized
($M_Q$ can be taken to be infinite).
Table~\ref{tab:ERT90} compares the predictions obtained with
only the leading term in \ref{eq:ERTresult}
(i.e. neglecting the gluonic correction) with the
phenomenological determination of the
$L_i$ couplings.
The numerical agreement is quite impressive;
both the order of magnitude and the sign are correctly
reproduced (notice that this is just a free-quark result!).
Moreover, the gluonic corrections shift the values of
$L_3$ and $L_{10}$ in the right direction, making them
more negative.

%%%%%%%%%%%%%%%%%%%%%%%%%%% TABLE %%%%%%%%%%%%%%%%%
\midinsert
\table{Leading-order ($\alpha_s=0$) predictions for the
$L_i$'s, within the QCD-inspired model \ref{eq:ERTmodel}.
The phenomenological values are shown in the second row
for comparison.
All numbers are given in units of $10^{-3}$.}
\label{tab:ERT90}
\align\C{#}&&\C{#}\cr
\br
& $L_1$ & $L_2$ & $L_3$ & $L_9$ & $L_{10}$
\cr\mr
$L_i^{\rm th}(\alpha_s=0)$ & 0.79 & 1.58 & $-3.17$ & 6.33 & $-3.17$
\cr
$L_i^r(M_\rho)$ & $0.4\pm0.3$ & $1.4\pm0.3$ & $-3.5\pm1.1$ &
$6.9\pm0.7$ & $-5.5\pm0.7$
\cr\br
\endalign
\endtable
%%%%%%%%%%%%%%%%%%%%%%%%%%%%%%%%%%%%%%%%%%%%%%%%%%%%%%%%%%%%%%%%

The results \ref{eq:ERTresult}
obey almost all relations in \ref{eq:vmdpred}.
Comparing the predictions for $L_{1,2,9}$ in the VMD approach
of equation~\ref{eq:vmdpred} with the QCD-inspired
ones in \ref{eq:ERTresult},
one gets a quite good estimate of the $\rho$ mass:
$$
M_V = 2\sqrt{2}\pi f = 821 \, {\rm MeV} .
\label{eq:rho_mass}
$$

Is it quite easy to prove that the interaction \ref{eq:ERTmodel}
is equivalent to the mean-field approximation of the
Nambu--Jona-Lasinio (1961) model,
where SCSB is triggered by four-quark operators.
It has been conjectured recently (Bijnens, Bruno and de Rafael 1993)
that integrating out the quark and gluon fields of QCD,
down to some intermediate scale
$\Lambda_\chi$, gives rise to an extended Nambu--Jona-Lasinio
Lagrangian.
By introducing collective fields (to be identified later with
the Goldstone fields and
$S$, $V$, $A$ resonances) the model can be transformed into a
Lagrangian bilinear in the quark fields, which can therefore
be integrated out.
One then gets an effective Lagrangian,
describing the couplings of the pseudoscalar bosons to
vector, axial-vector and scalar resonances.
Extending the analysis beyond the mean-field approximation,
Bijnens, Bruno and de Rafael (1993) obtain predictions for
20 measurable quantities, including the $L_i$'s, in terms of only
4 parameters. The quality of the fits is quite impressive.
Since the model contains all resonances that are known to saturate
the $L_i$ couplings, it is not surprising that one gets an
improvement of the
mean-field-approximation results, specially for
the constants $L_5$ and $L_8$, which are sensitive to
scalar exchange.
What is more important, this analysis clarifies a
potential problem of double-counting:
in certain limits the model approaches either the pure
quark-loop predictions \ref{eq:ERTresult} or
the VMD results \ref{eq:vmdpred},
but in general it interpolates between these two cases.

\section{$\Delta S=1$ non-leptonic weak interactions}
\label{sec:weak}

The Standard Model predicts strangeness-changing transitions with
$\Delta S=1$ via $W$-exchange between two weak charged currents.
At low energies ($E<<M_W$), the heavy fields $W$, $Z$, $t$, $b$, $c$
can be integrated out; using standard operator-product-expansion
techniques, the non-leptonic
$\Delta S=1$ weak interactions are described by an
effective Hamiltonian (Gilman and Wise 1979)
$$
{\cal H}^{\Delta S = 1}_{\rm eff}
 \, = \, {G_F \over \sqrt{2}} V_{ud}^{\hphantom{*}} V_{us}^* \,
\sum_i C_i(\mu) \, Q_i \, + \, {\rm h.c.} \, ,
\label{eq:ds_hamiltonian}
$$
which is a sum of local
four-quark operators, constructed with the light
($u, d, s$) quark fields
only,
$$
\fl\eqalign{
Q_1  \,\equiv \, 4 \, (\bar s_L \gamma^\mu d_L)
\, (\bar u_L \gamma_\mu u_L) , \qquad \qquad\,\;
&Q_2 \, \equiv \, 4 \, (\bar s_L \gamma^\mu u_L)
\, (\bar u_L \gamma_\mu d_L) ,
\cr
Q_3  \,\equiv \, 4 \, (\bar s_L \gamma^\mu d_L) \, \sum_{q=u,d,s}
(\bar q_L \gamma_\mu q_L) ,\qquad\,
&Q_4 \, \equiv \, 4 \, \sum_{q=u,d,s} (\bar s_L \gamma^\mu q_L) \,
(\bar q_L \gamma_\mu d_L) ,\quad\qquad
\cr
Q_5 \,\equiv\, 4 \, (\bar s_L \gamma^\mu d_L) \sum_{q=u,d,s} \,
(\bar q_R \gamma_\mu q_R) , \qquad
&Q_6 \,\equiv \, -8 \, \sum_{q=u,d,s} (\bar s_L q_R)
\, (\bar q_R d_L) ,}
\label{eq:four_quark_operators}
$$
modulated by Wilson coefficients $C_i(\mu)$,
which are functions of the heavy
$W$, $t$, $b$, $c$ masses and an overall renormalization scale $\mu$.
Only five of these operators are independent, since
$Q_4 = - Q_1 + Q_2 + Q_3$.
{}From the point of view of chiral $SU(3)_L \otimes SU(3)_R$ and isospin
quantum numbers, $Q_- \equiv Q_2 - Q_1$ and $Q_i \; (i=3,4,5,6)$
transform as
($8_L,1_R$) and induce $|\Delta I| = \frac12$ transitions, while
$Q_1 + 2/3 Q_2 - 1/3 Q_3$ transforms like ($27_L,1_R$) and
induces both
$|\Delta I| = \frac12$ and $|\Delta I| = \frac32$ transitions.

The effect of $\Delta S=1$ non-leptonic weak interactions can be
incorporated in the low-energy chiral theory (Cronin 1967),
as a perturbation
to the strong effective Lagrangian $\cL_{\rm eff}(U)$.
At lowest order in the number of derivatives,
the most general effective bosonic Lagrangian, with the
same $SU(3)_L\otimes SU(3)_R$ transformation properties
as the short-distance Hamiltonian
\ref{eq:ds_hamiltonian},
contains two terms\footnote{$^{\P}$}{
One can build an additional octet term with the external $\chi$ field,
$\langle\lambda\left(U^\dagger\chi + \chi^\dagger U\right)\rangle$;
however, this term does not contribute to on-shell amplitudes.}:
$$
\fl
\cL_2^{\Delta S=1} =
-{G_F\over\sqrt{2}} V_{ud}^{\hphantom{*}} V_{us}^*\,\left\{
g_8 \, \langle \lambda L_\mu L^\mu\rangle +
g_{27} \left( L_{\mu 23} L^\mu_{11} + \frac{2}{3} L_{\mu 21}
L^\mu_{13}\right) + {\rm h.c.} \right\} ,
\label{eq:L2w}
$$
where
$$
\lambda =  (\lambda_6 - i \lambda_7) /2 \, , \qquad\qquad
L_\mu = i f^2 U^\dg D_\mu U \, .
\label{eq:lambda_def}
$$
The chiral couplings $g_8$ and $g_{27}$ measure the strength
of the two parts in the effective Hamiltonian \ref{eq:ds_hamiltonian}
transforming as $(8_L,1_R)$ and $(27_L,1_R)$, respectively,
under chiral rotations.
Their values can be extracted from $K\to 2\pi$ decays
(Pich \etal 1986):
$$
|g_8| \approx 5.1 \, , \qquad\qquad
g_{27} / g_8 \approx 1/18 \, .
\label{eq:g8g27}
$$
The huge difference between these two couplings shows the
well-known enhancement of the octet $|\Delta I|=\frac12$ transitions.

Using the effective Lagrangian \ref{eq:L2w},
the calculation of hadronic weak decays becomes a straightforward
perturbative problem. The highly non-trivial QCD dynamics
has been parametrized in terms of the two chiral couplings.
Of course, the interesting problem that remains to be solved is
to compute $g_8$ and $g_{27}$ from the underlying QCD theory
and, therefore, to gain a dynamical understanding of the so-called
$|\Delta I|=\frac12$ rule.
Although this is a very difficult task, considerable progress
has been achieved recently
(Pich and de Rafael 1991a, Jamin and Pich 1994).
Applying the QCD-inspired model of equation~\ref{eq:ERTmodel}
to the weak sector, a quite successful estimate of these
two couplings has been obtained;
a very detailed description of
this calculation, and a comparison with other
approaches, has been given by Pich and de Rafael (1991a).

Once the couplings $g_8$ and $g_{27}$ have been phenomenologically
fixed to the values in \ref{eq:g8g27},
other decays like $K\to 3\pi$ or $K\to 2\pi\gamma$ can be easily
predicted at $\Or (p^2)$.
As in the strong sector,
one reproduces in this way the successful soft-pion
relations of Current Algebra.
However, the data are already accurate enough
for the next-order corrections to be sizeable.
Moreover, many transitions do not occur at $\Or (p^2)$.
For instance, due to a mismatch between the
minimum number of powers of momenta
required by gauge invariance and the powers of momenta that
the lowest-order
effective Lagrangian can provide,
the amplitude for any non-leptonic radiative
$K$-decay with at most one pion in the final state
($K \rightarrow \gamma
\gamma  , K \rightarrow \gamma l^+ l^- ,
K \rightarrow \pi \gamma \gamma ,
K \rightarrow \pi l^+ l^-$, ...)
vanishes to lowest order in ChPT
(Ecker, Pich and de Rafael 1987a, 1987b, 1988).
These decays are then sensitive to the non-trivial
quantum field theory aspects of ChPT.

Unfortunately, at $\Or (p^4)$ there is a very large number of
possible terms,
satisfying the appropriate $(8_L,1_R)$ and $(27_L,1_R)$
transformation properties (Kambor \etal 1990).
Using the $\Or (p^2)$ equations of motion obeyed by $U$ to reduce the
number of terms, 35 independent structures
(plus 2 contact terms involving external fields only)
remain in the octet
sector alone (Kambor \etal 1990, Ecker 1990,
Esposito-Far\`ese 1991).
Restricting the attention to those terms that contribute to
non-leptonic amplitudes where the only external gauge fields
are photons, still leaves 22 relevant octet terms
(Ecker, Kambor and Wyler 1993).
Clearly, the predictive power of a completely general chiral
analysis, using  only symmetry constraints, is rather limited.
Nevertheless, as we are going to see,
it is still possible to make predictions.

Due to the complicated interplay of electroweak and strong
interactions, the low-energy constants of the weak non-leptonic
chiral Lagrangian encode a much richer information than
in the pure strong sector.
These chiral couplings contain both long- and short-distance
contributions, and some of them (like $g_8$) have in addition
a CP-violating imaginary part.
Genuine short-distance physics, such as the electroweak penguin
operators, have their corresponding effective realization
in the chiral Lagrangian.
Moreover, there are four $\Or (p^4)$ terms containing an
$\varepsilon_{\mu\nu\alpha\beta}$ tensor,
which get a direct (probably dominant) contribution from
the chiral anomaly
(Ecker, Neufeld and Pich 1992, Bijnens, Ecker and Pich 1992).

In recent years, there have been several attempts to estimate
these low-energy
couplings using different approximations, such as
factorization (Fajfer and G\'erard 1989, Cheng 1990,
Pich and de Rafael 1991a),
weak-deformation model (Ecker, Pich and de Rafael 1990),
effective-action approach
(Pich and de Rafael 1991a, Bruno and Prades 1993),
or resonance exchange
(Isidori and Pugliese 1992, Ecker, Kambor and Wyler 1993).
Although more work in this direction is certainly needed,
a qualitative picture of the size of the different couplings
is already emerging.

\subsection{$K\to 2\pi, 3\pi$ decays}
\label{subsec:kpp}

Imposing isospin and Bose symmetries, and keeping terms up to
$\Or (p^4)$, a general parametrization (Devlin and Dickey 1979)
of the $K\to 3\pi$ amplitudes involves ten measurable parameters:
$\alpha_i$, $\beta_i$, $\zeta_i$, $\xi_i$, $\gamma_3$ and
$\xi'_3$, where $i=1,3$ refers to the $\Delta I =\frac12, \frac32$
pieces.
At $\Or (p^2)$, the quadratic slope parameters
$\zeta_i$, $\xi_i$ and $\xi'_3$ vanish; therefore the
lowest-order Lagrangian \ref{eq:L2w} predicts five
$K\to 3\pi$ parameters in terms of the two couplings
$g_8$ and $g_{27}$, extracted from $K\to 2 \pi$.
These predictions give the right qualitative pattern,
but there are sizeable differences with the
measured amplitudes.
Moreover, non-zero values for some of the slope parameters
have been clearly established experimentally.

The agreement is substantially improved at $\Or (p^4)$
(Kambor \etal 1991).
In spite of the large number of unknown couplings in the general
effective $\Delta S=1$ Lagrangian,
only 7 combinations of these weak chiral constants
are relevant for describing the $K\to 2\pi$ and $K\to 3 \pi$
amplitudes (Kambor \etal 1992).
Therefore, one has 7 parameters for 12 observables, which
results in 5 relations.
The extent to which these relations are satisfied provides
a non-trivial test of chiral symmetry at the four-derivative level.
The results of such a test (Kambor \etal 1992) are shown in
table~\ref{tab:KDHMW92}, where the 5 conditions have been
formulated as predictions for the 5 slope parameters.
The comparison is very successful for the two
$\Delta I = \frac12$ parameters, but
the data are not good enough to say anything conclusive about
the other three $\Delta I = \frac32$ predictions.
%moreover, a possible discrepancy
%in the value of $\xi_3$ would not be very significative, because
%this parameter is expected to be rather sensitive to
%electromagnetic effects, which have been omitted in the analysis.

%%%%%%%%%%%%%%%%%%%%%%%%%%% TABLE %%%%%%%%%%%%%%%%%
\midinsert
\table{Predicted and measured values of the quadratic
slope parameters in the $K\to 3\pi$ amplitudes
(Kambor \etal 1992).
All values are given in units of $10^{-8}$.}
\label{tab:KDHMW92}
\align\C{#}&&\C{#}\cr
\br
Parameter & Experimental value & Prediction
\cr\mr
$\zeta_1$ & $-0.47\pm0.15$ & $-0.47\pm0.18$ \cr
$\xi_1$ & $-1.51\pm0.30$ & $-1.58\pm0.19$ \cr
$\zeta_3$ & $-0.21\pm0.08$ & $-0.011\pm0.006$ \cr
$\xi_3$ & $-0.12\pm0.17$ & $\hphantom{-}0.092\pm0.030$ \cr
$\xi'_3$ & $-0.21\pm0.51$ & $-0.033\pm0.077$ \cr\br
\endalign
\endtable
%%%%%%%%%%%%%%%%%%%%%%%%%%%%%%%%%%%%%%%%%%%%%%%%%%%

The $\Or (p^4)$ analysis of these decays has also clarified the
role of long-distance effects ($\pi\pi$ rescattering)
in the dynamical enhancement of $\Delta I = \frac12$ amplitudes.
The $\Or (p^4)$ corrections give indeed a sizeable
constructive contribution, which results (Kambor \etal 1991)
in a fitted value for $|g_8|$ that is about $30\%$ smaller
than the lowest-order determination \ref{eq:g8g27}.
While this certainly goes in the right direction,
it also shows that the bulk of the enhancement mechanism
comes from a different source.

\subsection{Radiative $K$ Decays}
\label{subsec:radiative}

Owing to the constraints of electromagnetic gauge invariance,
radiative
$K$ decays with at most one pion in the final state do not occur
at $\Or (p^2)$.
Moreover, only a few terms of the octet $\Or (p^4)$ Lagrangian
are relevant for this kind of processes
(Ecker, Pich and de Rafael 1987a, 1987b, 1988):
$$
\fl\cL_4^{\Delta S=1,{\rm em}} \, \doteq \,
- {G_F\over\sqrt{2}} V_{ud}^{\hphantom{*}}
V_{us}^* \, g_8  \, \biggl\{
-{i e  \over  f^2} F^{\mu \nu}\,
   \left[ w_1 \,\langle Q \lambda L_\mu L_\nu\rangle
 + w_2 \,\langle Q L_\mu \lambda L_\nu\rangle\right]
\biggr.\cr  \biggl.
\qquad\qquad\qquad\;\;
 +~e^2 f^2  w_4 F^{\mu\nu} F_{\mu\nu}\,
 \langle\lambda Q U^\dagger Q U\rangle + {\rm h.c.} \biggr\}\, .
\label{eq:l4_rad}\cr
$$
The small number of unknown chiral couplings allows us to
derive useful relations among different processes and
to obtain definite predictions. The absence of a tree-level
$\Or (p^2)$ contribution makes the final results very sensitive to the
loop structure of the amplitudes.

\subsubsection{$K_S\to\gamma\gamma$}

%%%%%%%%%%%%%% FIGURE %%%%%%%%%%%%%%%%%%%%%
%\begin{figure}[h]
%\begin{center}
%\mbox{\epsfig{file=ksgg.eps,height=7.0cm}}
%\end{center}
%\caption{Feynman diagrams for $K_1^0\to\gamma^*\gamma^*$.}
%\label{fig:ksgg}
%\end{figure}
%%%%%%%%%%%%%%%%%%%% END FIGURE %%%%%%%%%%%%%%%%%%%%%%%%

The symmetry constraints do not allow any direct tree-level
$K_1^0\gamma\gamma$ coupling at $\Or (p^4)$
($K^0_{1,2}$ refer to the CP-even and CP-odd eigenstates,
respectively).
This decay proceeds then
through a loop of charged pions as shown in
figure~\ref{fig:ksgg} (there are
similar diagrams with charged kaons in the loop, but
their sum is proportional to
$M^2_{K^0} - M^2_{K^+}$ and therefore can be neglected).
%gives a zero contribution to the decay amplitude).
Since there are no possible counter-terms to renormalize
divergences, the one-loop amplitude is necessarily finite.
Although each of the four diagrams in figure~\ref{fig:ksgg}
is quadratically divergent, these divergences cancel  in the
sum.
The resulting prediction
(D'Ambrosio and Espriu 1986, Goity 1987)
is in very good agreement
with the experimental measurement (Burkhardt \etal 1987):
$$
{\rm Br}(K_S\to\gamma\gamma) \, = \,\cases{ \quad
2.0 \times 10^{-6}   &  (theory)
\cr
(2.4 \pm 1.2) \times 10^{-6} & (experiment) \cr } .
\label{eq:ksgg}
$$

\subsubsection{$K_{L,S}\to\mu^+\mu^-$}

There are well-known short-distance contributions
(electroweak penguins and box diagrams)
to the decay $K_L\to\mu^+\mu^-$.
However, this transition is dominated by long-distance
physics. The main contribution proceeds through a two-photon
intermediate state: $K_2^0\to\gamma^*\gamma^*\to\mu^+\mu^-$.
Contrary to $K_1^0\to\gamma\gamma$,
the prediction for the $K_2^0\to\gamma\gamma$ decay is
very uncertain, because the first non-zero contribution
occurs\footnote{$^\natural$}{
%%%%%%%%%%%%%
At $\Or (p^4)$, this decay proceeds through
a tree-level $K_2^0\to\pi^0,\eta$ transition, followed by
$\pi^0,\eta\to\gamma\gamma$ vertices.
Because of the Gell-Mann--Okubo relation,
the sum of the $\pi^0$ and $\eta$ contributions
cancels exactly to lowest order.
The decay amplitude is then very sensitive to $SU(3)$ breaking.}
%%%%%%%%%%%%%
at $\Or (p^6)$.
That makes very difficult any attempt to
predict the $K_{L}\to\mu^+\mu^-$ amplitude.

%%%%%%%%%%%%%%%%%%%%%% FIGURE %%%%%%%%%%%%%%%%%%%%%%%%%%%
%\begin{figure}[htb]
%\begin{center}
%\mbox{\epsfig{file=ksmm.eps,height=4.0cm,width=9cm}}
%\end{center}
%\caption{Feynman diagram for the $K_1^0\to \mu^+ \mu^-$ decay.
%The $K_1^0 \gamma^* \gamma^*$ vertex is generated through
%the one-loop
%diagrams shown in Fig.~\protect\ref{fig:ksgg}}
%\end{figure}
%%%%%%%%%%%%%%%%%%%% END FIGURE %%%%%%%%%%%%%%%%%%%%%%%%

The situation is completely different for the $K_S$ decay.
A straightforward chiral analysis (Ecker and Pich 1991)
shows that, at lowest order in momenta, the only allowed
tree-level $K^0\mu^+\mu^-$ coupling corresponds to the
CP-odd state $K_2^0$.
Therefore, the $K_1^0\to\mu^+\mu^-$ transition can only be
generated by a finite non-local two-loop contribution.
The calculation has been performed recently
(Ecker and Pich 1991), with the result:
$$
{\Gamma(K_S\to\mu^+\mu^-)\over\Gamma(K_S\to\gamma\gamma)}
= 2\times 10^{-6}, \qquad
{\Gamma(K_S\to e^+ e^-)\over\Gamma(K_S\to\gamma\gamma)}
= 8\times 10^{-9},
\label{eq:ksmm_ratios}
$$
well below the present experimental upper limits.
Although, in view of
the smallness of the predicted ratios,
this calculation seems quite academic, it has important
implications for CP-violation studies.

The longitudinal muon polarization $\cP_L$
in the decay $K_L\to\mu^+\mu^-$ is an interesting measure of
CP violation.
As for every CP-violating observable in the neutral kaon system,
there are in general two different kinds of contributions to $\cP_L$:
indirect CP violation through the small
$K_1^0$ admixture of the $K_L$
($\varepsilon$ effect), and direct CP violation in the
$K_2^0\to\mu^+\mu^-$
decay amplitude.

   In the Standard Model, the direct-CP-violating amplitude is
induced by Higgs exchange with an effective one-loop flavour-changing
$\bar s d H$ coupling (Botella and Lim 1986).
The present lower bound
on the Higgs mass, $M_H>58.4$ GeV ($95 \%$ C.L.), implies a
conservative upper limit
$|\cP_{L,{\rm Direct}}| < 10^{-4}$.
Much larger values, $\cP_L \sim \Or (10^{-2})$, appear quite naturally
in various extensions of the Standard Model
(Geng and Ng 1990, Mohapatra 1993).
It is worth emphasizing that $\cP_L$ is especially
sensitive to the presence of light scalars with CP-violating
Yukawa couplings. Thus, $\cP_L$ seems to be a good signature to look
for new physics beyond the Standard Model; for this to be the case,
however, it is very important to have a good quantitative
understanding of the Standard Model prediction to allow us to infer,
from a measurement of $\cP_L$, the existence of a new CP-violation
mechanism.

  The chiral calculation of the $K_1^0\to\mu^+\mu^-$ amplitude
allows us to make a reliable estimate\footnote{$^{**}$}{
Taking only the absorptive parts of the $K_{1,2}\to\mu^+\mu^-$
amplitudes into account,
a value $|\cP_{L,\varepsilon}| \approx 7\times 10^{-4}$ was
estimated previously (Herczeg 1983).
However, this is only one out of four contributions
to $\cP_L$ (Ecker and Pich 1991),
which
could all interfere constructively with unknown magnitudes.
}
of the contribution to $\cP_L$ due to $K^0$--$\bar K^0$ mixing
(Ecker and Pich 1991):
$$
1.9 < |\cP_{L,\varepsilon}| \times 10^3 \Biggl(
{2 \times 10^{-6} \over
{\rm Br}(K_S\to\gamma\gamma)} \Biggr)^{1/2} < 2.5 \, .
\label{eq:p_l}
$$
Taking into account
the present experimental errors in ${\rm Br}(K_S\to\gamma\gamma)$  and
the inherent theoretical uncertainties due to uncalculated
higher-order corrections,
one can conclude that experimental indications for
$|\cP_L|>5\times 10^{-3}$ would constitute clear evidence
for additional
mechanisms of CP violation beyond the Standard Model.

\subsubsection{$K_L\to\pi^0\gamma\gamma$}

Assuming CP conservation, the most general form of the amplitude for
$K_2^0\to\pi^0\gamma\gamma$ depends on two independent invariant
amplitudes $A$ and $B$ (Ecker, Pich and de Rafael 1988),
$$
\fl\eqalign{
{\cal A}[&K_L(p_K)\to\pi^0(p_0)\gamma(q_1)\gamma(q_2)] =
  \epsilon_\mu(q_1) \,\epsilon_\nu(q_2)
\, \Biggl\{
{A(y,z)\over M^2_K}\,
\Bigl(  q_2^\mu q_1^\nu - q_1\cdot q_2 \, g^{\mu\nu}\Bigr)
\Biggr. \cr
& \Biggl.
+ {2 B(y,z)\over M^4_K}\,
\Bigl(p_K\cdot q_1 \, q_2^\mu p_K^\nu
+ p_K\cdot q_2\, q_1^\nu p_K^\mu
- q_1\cdot q_2 \,  p_K^\mu p_K^\nu  -
p_K\cdot q_1\, p_K\cdot q_2 \, g^{\mu\nu}\Bigr)
 \Biggr\} , }
\label{eq:a_b_def}
$$
where $y\equiv|p_K\cdot(q_1-q_2)|/M_K^2$ and $z=(q_1+q_2)^2/M_K^2$.

%%%%%%%%%%%%%%%%%%%%%  Figures %%%%%%%%%%%%%%%%%%%%%%%%%%
%\begin{figure}[thb]
%\vfill
%\vspace*{-1.0cm}
%\begin{minipage}{.47\linewidth}
%\begin{minipage}[h]{1.0\textwidth}
%\begin{center}
%\mbox{\epsfig{file=NA31M.eps,height=10.9cm,width=5.65cm}}
%\vspace*{-1.6cm}
%\caption{$2\gamma$-invariant-mass distribution for
%$K_L\to\pi^0\gamma\gamma$:
%$\Or (p^4)$ (dotted curve),
%$\Or (p^6)$ with $a_V=0$ (dashed curve),
%$\Or (p^6)$ with $a_V=-0.9$ (full curve).
%The spectrum is normalized to the 50 unambiguous
%events of NA31 (without acceptance corrections).}
%\label{fig:spectrum}
%\end{center}
%\end{minipage}
%\end{minipage}
%\hspace{0.6cm}
%\begin{minipage}{.47\linewidth}
%\begin{minipage}[h]{1.0\textwidth}
%\begin{center}
%\vspace*{2.35 cm}
%\mbox{\epsfig{file=NA31.eps,height=6.95cm,width=6.8cm}}
%\vspace*{-0.50cm}
%\caption{Measured
%\protect\cite{ref:NA31_92}
%$2\gamma$-invariant-mass distribution for
%$K_L\to\pi^0\gamma\gamma$ (solid line).
%The dashed line shows the estimated background.
%The experimental acceptance is given by the crosses.
%The dotted line simulates the $\Or (p^4)$ ChPT prediction.}
%\label{fig:spectrum_NA31}
%\end{center}
%\end{minipage}
%\end{minipage}
%\vfill
%\end{figure}
%%%%%%%%%%%%%%%%%%%%% End figures %%%%%%%%%%%%%%%%%%%%%%%%%%

Only the amplitude $A(y,z)$ is non-vanishing to lowest non-trivial
order, $\Or (p^4)$, in ChPT.
Again, the symmetry constraints do not allow any
tree-level contribution from $\Or (p^4)$ terms in the Lagrangian.
The $A(y,z)$ amplitude is therefore determined by a
finite loop calculation (Ecker, Pich and de Rafael 1987b).
The relevant Feynman diagrams are analogous to the ones in
figure~\ref{fig:ksgg}, but with an additional $\pi^0$ line
emerging from the weak vertex;
charged kaon loops also give a small contribution in this case.
Due to the large absorptive $\pi^+\pi^-$ contribution,
the spectrum in the invariant mass of the two photons
is predicted
(Ecker, Pich and de Rafael 1987b, Cappiello and D'Ambrosio 1988)
to have a very characteristic behaviour
(dotted line in figure~\ref{fig:spectrum}),
peaked at high values of $m_{\gamma\gamma}$.
The agreement with the measured two-photon distribution
(Barr \etal 1992),
shown in figure~\ref{fig:spectrum_NA31},
is remarkably good.
However, the $\Or (p^4)$ prediction for the rate
(Ecker, Pich and de Rafael 1987b, Cappiello and D'Ambrosio 1988),
${\rm Br}(K_L \rightarrow \pi^0 \gamma \gamma) = 0.67\times 10^{-6}$,
is smaller than the experimental value:
$$
\fl{\rm Br}(K_L \rightarrow \pi^0 \gamma \gamma )
\, = \,\cases{
(1.7 \pm 0.3) \times 10^{-6}  & (Barr \etal 1992), \cr
(2.2 \pm 1.0) \times 10^{-6}  & (Papadimitriou \etal 1991). \cr }
\label{eq:br_klpgg}
$$

Since the effect of the amplitude $B(y,z)$ first appears at
$\Or (p^6)$, one should worry about the size of the next-order
corrections. A na\"{\ii}ve VMD estimate through the decay
chain
$K_L\to\pi^0,\eta,\eta'\to V \gamma\to\pi^0\gamma\gamma$
(Sehgal 1988, Morozumi and Iwasaki 1989, Flynn and Randall 1989,
Heiliger and Sehgal 1993)
results in a sizeable contribution to $B(y,z)$
(Ecker, Pich and de Rafael 1990),
$$
A(y,z)\big|_{\rm VMD} \, = \,  \tilde a_V
\left( 3 - z + {M^2_\pi\over M_K^2}\right) ,
\qquad
B(y,z)\big|_{\rm VMD} \, = \,  -2\tilde a_V \, ,
\label{eq:vmd_contribution}
\cr
\tilde a_V \,\equiv\,
-{G_F\over\sqrt{2}} V_{ud}^{\hphantom{*}} V_{us}^*\, g_8 \,
{M_K^2 \alpha\over\pi} \, a_V \, ,
\label{eq:av_def}
$$
with $a_V \approx 0.32$.
However, this type of calculation predicts a photon
spectrum peaked at low values of $m_{\gamma\gamma}$,
in strong disagreement with
experiment.
As first emphasized by Ecker, Pich and de Rafael (1990),
there are also so-called direct weak contributions
associated with $V$ exchange, which cannot be written as a strong
VMD amplitude with an external weak transition.
Model-dependent estimates of this direct contribution
(Ecker, Pich and de Rafael 1990)
suggest a strong cancellation with the
na\"{\ii}ve vector-meson-exchange effect, i.e.
$|a_V| < 0.32$;
but the final result is unfortunately quite uncertain.

A detailed calculation of the most important $\Or (p^6)$ corrections
has been performed recently (Cohen, Ecker and Pich 1993).
In addition to the VMD contribution, the unitarity corrections
associated with the two-pion intermediate state
(i.e. $K_L\to\pi^0\pi^+\pi^-\to\pi^0\gamma\gamma$) have been
included\footnote{$^{\dagger\dagger}$}{
%%%%%%%%%%%%
The charged-pion loop has also been computed by Cappiello \etal
(1993).}.
%%%%%%%%%%%
Figure~\ref{fig:spectrum} shows the resulting photon spectrum
for $a_V=0$ (dashed curve) and $a_V=-0.9$ (full curve).
The corresponding branching ratio is:
$$
{\rm Br}(K_L\to\pi^0\gamma\gamma) \, = \, \cases{
0.67\times 10^{-6} , & $\Or (p^4)$, \cr
0.83\times 10^{-6} , & $\Or (p^6), \, a_V=0\, $, \cr
1.60\times 10^{-6} , & $\Or (p^6), \, a_V=-0.9\, $. }
\label{eq:br_pred_p6}
$$
The unitarity corrections by themselves raise the rate only
moderately. Moreover, they produce an even more pronounced
peaking of the spectrum at large $m_{\gamma\gamma}$, which
tends to ruin the success of the $\Or (p^4)$ prediction.
The addition of the $V$ exchange contribution restores again
the agreement.
Both the experimental rate and the spectrum
can be simultaneously reproduced with  $a_V = -0.9$.
A more complete unitarization of the $\pi$--$\pi$ intermediate
states (Kambor and Holstein 1994),
including the experimental $\gamma\gamma\to\pi^0\pi^0$
amplitude, increases the $K_L\to\pi^0\gamma\gamma$ decay width
some 10\%, leading to a slightly smaller value of $|a_V|$.

\subsubsection{$K\to\pi l^+ l^-$}

In contrast to the previous processes,
the  $\Or (p^4)$  calculation of $K^+\to\pi^+ l^+ l^-$
and $K_S\to\pi^0 l^+ l^-$ involves a divergent loop,
which is renormalized by the $\Or (p^4)$ Lagrangian.
The decay amplitudes
can then be written (Ecker, Pich and de Rafael 1987a)
as the sum of a calculable loop
contribution plus an unknown combination of chiral couplings,
$$
\eqalign{
w_+\, = \, & -{1\over 3} (4\pi)^2 [w_1^r + 2 w_2^r - 12 L_9^r]
  -{1\over 3} \log{\left(M_K M_\pi/\mu^2\right)} ,
\cr
w_S \, = \, & -{1\over 3} (4\pi)^2 [w_1^r - w_2^r]
  -{1\over 3} \log{\left(M_K^2/\mu^2\right)} , }
\label{eq:wp_w0}
$$
where $w_+$, $w_S$
refer to the decay of the $K^+$ and
$K_S$ respectively.
These constants are expected to be of order 1 by
na\"{\ii}ve power-counting arguments.
The logarithms have been included
to compensate the
renormalization-scale dependence of the chiral couplings,
so that $w_+$, $w_S$
are observable quantities.
If the final amplitudes are required to transform as
octets, then $w_2 = 4 L_9$, implying
$w_S = w_+ + \log{\left(M_\pi/M_K\right)}/3$.
It should be emphasized
that this relation goes beyond the usual requirement
of chiral invariance.

The measured $K^+\to\pi^+ e^+ e^-$ decay rate determines
two possible solutions for $w_+$
(Ecker, Pich and de Rafael 1987a).
The two-fold ambiguity can be solved, looking to
the shape of the invariant-mass distribution of the final lepton
pair, which is regulated by the same parameter $w_+$.
A fit to the BNL--E777 data (Alliegro \etal 1992)
gives
$$
w_+ = 0.89^{+0.24}_{-0.14}\, ,
\label{eq:omega}
$$
in agreement with model-dependent
theoretical estimates (Ecker, Pich and de Rafael 1990,
Bruno and Prades 1993).
Once $w_+$ has been fixed, one can make
predictions (Ecker, Pich and de Rafael 1987a)
for the rates and Dalitz-plot distributions
of the related modes
$K^+\to\pi^+ \mu^+ \mu^-$,
$K_S\to\pi^0 e^+ e^-$ and $K_S\to\pi^0 \mu^+ \mu^-$.

\subsubsection{$K_L\to\pi^0 e^+ e^-$}

 The rare decay $K_L \rightarrow \pi^0 e^+ e^-$
is an interesting process
in looking for new CP-violating signatures.
If CP were an exact symmetry,
only the CP-even state $K_1^0$ could decay
via one-photon emission, while
the decay of the CP-odd state $K_2^0$ would proceed through a
two-photon intermediate state and, therefore,
its decay amplitude would be suppressed
by an additional power of $\alpha$.
When CP-violation is taken into account,
however, an $\Or (\alpha)$ $K_L \rightarrow \pi^0 e^+ e^-$ decay
amplitude is induced, both through the small
$K_1^0$ component of the $K_L$
($\varepsilon$ effect) and through direct CP-violation in the
$K_2^0 \rightarrow \pi^0 e^+ e^-$ transition.
The electromagnetic suppression of the CP-conserving amplitude then
makes it plausible that this decay is
dominated by the CP-violating contributions.

  The short-distance analysis of the product of
weak and electromagnetic
currents allows a reliable calculation of the direct CP-violating
$K_2^0 \rightarrow \pi^0 e^+ e^-$ amplitude.
The corresponding branching ratio
%induced by this amplitude
has been estimated
(Buras \etal 1994) to be around
$$
{\rm Br}(K_L \rightarrow \pi^0 e^+ e^-)\Big|_{\rm Direct}
\simeq 6 \times 10^{-12} ,
\label{eq:direct}
$$
the exact number depending on the values of
$m_t$ and the quark-mixing
angles.

The indirect CP-violating amplitude induced by the
$K_1^0$ component of
the $K_L$ is given by the $K_S \rightarrow \pi^0 e^+ e^-$ amplitude
times the CP-mixing parameter $\varepsilon$.
Using the octet relation between $w_+$ and $w_S$,
the determination of the parameter $\omega_+$ in
\ref{eq:omega}
implies
$$
{\rm Br}(K_L \rightarrow \pi^0 e^+ e^-)\Big|_{\rm Indirect} \le
           1.6 \times 10^{-12}.
\label{eq:indirect}
$$
Comparing this  value   with \ref{eq:direct},
we see that the direct
CP-violating contribution is expected to be bigger than the
indirect one. This is very different from the situation in
$K \rightarrow \pi \pi$, where the contribution due to mixing
completely dominates.

   The present experimental upper bound (Harris \etal 1993),
%(Barker \etal 1990, Ohl \etal 1990),
$$
{\rm Br}(K_L \rightarrow \pi^0 e^+ e^-)\Big|_{\rm Exp}
 < 4.3 \times 10^{-9} \qquad (90\% {\rm C.L.}) ,
\label{eq:klpee_exp}
$$
is still far away from the expected Standard Model signal,
but the prospects
for getting the needed sensitivity of around $10^{-12}$ in
the next few years are rather encouraging.
To be able to interpret a future experimental measurement of
this decay as a CP-violating signature,
it is first necessary, however,
to pin down the actual
size of the two-photon exchange CP-conserving amplitude.

Using the computed  $K_L\to\pi^0\gamma\gamma$ amplitude,
one can estimate the two-photon exchange contribution
to $K_L\to\pi^0e^+e^-$,
by taking the absorptive part due to the two-photon
discontinuity as an
educated guess of the actual size of the complete amplitude.
At $\Or (p^4)$, the $K_L\to\pi^0e^+e^-$ decay
amplitude is
strongly suppressed (it is proportional to $m_e$), owing to the
helicity structure of the $A(y,z)$ term
(Donoghue, Holstein and Valencia 1987,
Ecker, Pich and de Rafael 1988):
$$
{\rm Br}(K_L \rightarrow \pi^0 \gamma ^* \gamma ^* \rightarrow \pi^0
     e^+ e^-)\Big|_{\Or (p^4)} \,\sim\, 5 \times 10^{-15} .
\label{eq:klpee_a}
$$
This helicity suppression is, however, no longer true at
the next order in the chiral expansion.
The $\Or (p^6)$ estimate of the amplitude
$B(y,z)$ (Cohen, Ecker and Pich 1993) gives rise to
$$
\fl
{\rm Br}(K_L \rightarrow \pi^0 \gamma^* \gamma^* \rightarrow \pi^0 e^+ e^-)
 \Big|_{\Or (p^6 )} \,\sim\,
\cases{
0.3 \times 10^{-12}, & $a_V=0\, $, \cr
1.8  \times 10^{-12}, & $a_V=-0.9\, $. }
\label{eq:klpee_p6}
$$
Although the rate increases with $|a_V|$,
there is some destructive interference between the unitarity
corrections of $\Or (p^6)$ and the $V$-exchange contribution
(for $a_V=-0.9$).
To get a more accurate estimate,
it would be necessary to make a careful fit to the
$K_L \to\pi^0\gamma\gamma$
data, taking the experimental acceptance into account,
to extract the actual value of $a_V$.

Thus, the decay width seems to be dominated by the CP-violating
amplitude, but the CP-conserving contribution could also be
important. Notice that if both amplitudes were comparable
there would be a sizeable CP-violating energy asymmetry between the
$e^-$ and the $e^+$ distributions
(Sehgal 1988, Heiliger and Sehgal 1993, Donoghue and Gabbiani 1994).

\subsection{The chiral anomaly in non-leptonic $K$ decays}
\label{subsec:anomalous}

The chiral anomaly also  appears in the non-leptonic
weak interactions.
A systematic study of all non-leptonic $K$ decays where
the anomaly contributes at leading order, $\Or (p^4)$,
has been performed recently
(Ecker, Neufeld and Pich 1992).
Only radiative $K$ decays are sensitive to the
anomaly in the non-leptonic sector.

The manifestations of the anomaly can be grouped in
two different classes of anomalous amplitudes:
reducible and direct contributions.
The reducible amplitudes arise from the contraction of meson
lines between a weak non-leptonic $\Delta S=1$ vertex and the
Wess--Zumino--Witten  functional \ref{eq:WZW}.
In the octet limit, all reducible anomalous amplitudes of
$\Or (p^4)$ can be predicted in terms of the coupling $g_8$.
The direct anomalous contributions are generated through
the contraction of the $W$ boson field between
a strong Green function on one side and the
Wess--Zumino--Witten functional
on the other.
Their computation is not straightforward, because of the
presence of strongly interacting fields on both
sides of the $W$.
Nevertheless, due to the non-renormalization theorem
of the chiral anomaly (Adler and Bardeen 1969),
the bosonized form of the direct anomalous amplitudes
can be fully predicted (Bijnens, Ecker and Pich 1992).
In spite of its anomalous origin, this contribution
is chiral-invariant. The anomaly turns out
to contribute to all possible octet terms of
$\cL_4^{\Delta S=1}$ proportional to the
$\varepsilon_{\mu\nu\alpha\beta}$ tensor.
Unfortunately, the coefficients of these terms
get also non-factorizable contributions
of non-anomalous origin, which cannot be computed
in a model-independent way. Therefore, the final
predictions can only be parametrized in terms of four
dimensionless chiral couplings, which are expected
to be positive and of order one.

The most frequent {\it anomalous} decays
$K^+\to\pi^+\pi^0\gamma$ and
$K_L\to\pi^+\pi^-\gamma$ share the remarkable feature that the
normally dominant bremsstrahlung amplitude is strongly suppressed,
making the experimental verification of the anomalous amplitude
substantially easier.
This suppression has different origins: $K^+\to\pi^+\pi^0$ proceeds
through the small 27-plet part of the non-leptonic weak interactions,
whereas $K_L\to\pi^+\pi^-$ is CP-violating.
The remaining non-leptonic $K$ decays with direct anomalous contributions
are either suppressed by phase space
[$K^+\to\pi^+\pi^0\pi^0\gamma(\gamma)$,
$K^+\to\pi^+\pi^+\pi^-\gamma(\gamma)$,
$K_L\to\pi^+\pi^-\pi^0\gamma$, $K_S\to\pi^+\pi^-\pi^0\gamma(\gamma)$]
or by the presence
of an
extra photon in the final state
[$K^+\to\pi^+\pi^0\gamma\gamma$, $K_L\to\pi^+\pi^-\gamma\gamma$].

\section{Baryons}
\label{sec:baryons}

The inclusion of baryons in the low-energy effective field theory
follows the same procedure used in section~\ref{sec:resonances}
to incorporate the mesonic resonances.
The octet of baryon fields is collected in a $3 \times 3$ matrix
$$
B(x) \, \equiv \, \pmatrix{{1 \over \sqrt{2}}\Sigma^0 +
{1 \over \sqrt{6}} \Lambda^0
& \Sigma^+ & p \cr \Sigma^- & -{1 \over \sqrt{2}}\Sigma^0 +
{1 \over \sqrt{6}} \Lambda^0 & n \cr \Xi^- & \Xi^0 & -{2 \over \sqrt{6}}
\Lambda^0 \cr } \, ,
\label{eq:B_def}
$$
which under $\, SU(3)_L \otimes SU(3)_R \, $ transforms
non-linearly
$$
B \, \toG \, h(\phi,g) \, B \, h^{\dagger}(\phi,g)
\, .
\label{eq:B_transformation}
$$

We look for the most general chiral-invariant
effective Lagrangian one can write in terms of the matrices
$B(x)$, $\overline B(x) \equiv  B(x)^{\dagger} \gamma_0$,
and $u(\phi)$.
We can easily write down the lowest-order
baryon-meson Lagrangian for Green functions with
at most two baryons:
$$
\fl
\cL_1^{(B)} \, = \,
\langle \overline B i\gamma^\mu \nabla_\mu B \rangle
 \, - \, M_B \langle\overline B B\rangle
\,
 + \, {D \over 2} \, \langle\overline B
\gamma^\mu \gamma_5 \{ u_{\mu}, B \} \rangle \, + \,
  {F \over 2} \, \langle\overline B \gamma^{\mu} \gamma_5 [
u_{\mu}, B ]\rangle
\, ,
\label{eq:L1}
$$
where
$$
\nabla_\mu B\,\equiv\,\partial_\mu B + [\Gamma_\mu,B] \, ,
\label{eq:nabla}
$$
with $\Gamma_\mu$ defined in equation \ref{eq:connection}.
The covariant derivative incorporates the correct
minimal coupling to the electromagnetic  field,
$\langle \overline B \gamma^\mu [v_\mu,B] \rangle \dot=
e A_\mu \langle \overline B \gamma^\mu [Q,B] \rangle$,
and interactions with the pseudoscalar mesons, such as the so-called
Weinberg term
$i\langle\overline B\gamma^\mu \left[\left[\Phi,\partial_\mu\Phi
 \right],B\right]\rangle/(4f^2)$.

$M_B$ is a common mass of the baryon octet,
due to the SCSB; it is the mass that the baryons would have
if the $u$, $d$ and $s$ quarks were exactly massless.
The fact that the baryon masses do not vanish
in the chiral limit and, moreover, are not small parameters
compared with $\Lambda_\chi$ complicates the chiral analysis
of the baryon sector.
Notice that from the point of view of chiral power counting
$\nabla_{\mu}B$ and $M_B B$ are
$\Or (1)$, but
$i \gamma^{\mu} \nabla_{\mu}B - M_B B$ is $\Or (p)$.

The last two terms in \ref{eq:L1} contain
the baryonic coupling to the external axial source $a_\mu$, and
$\overline B\Phi B$ interactions.
If one restricts the discussion to the two flavour sector ($u$, $d$),
only the sum of the $D$ and $F$ terms is relevant
[$\overline N\equiv (\bar p, \bar n)$,
$\Pi\equiv \vec\tau \vec\pi/\sqrt{2}$]:
$$
\fl\cL_1^{(N)}\, = \,
\langle \overline N i\gamma^{\mu} \nabla_\mu  N \rangle
 \, - \, M_N \langle\overline N N\rangle
\cr
+ (D+F) \, \left\langle\overline N \gamma^{\mu} \gamma_5
\left\{ a_\mu - {1\over\sqrt{2}f}\partial_\mu\Pi
+ {i\over\sqrt{2}f} [v_\mu,\Pi] + \Or (\Pi^2)
\right\} N  \right\rangle \, .
%\, + \, \Or (v_\mu,\Pi^2) .
\label{eq:L1_N}
$$
Thus, $D+F$ is the usual nucleon axial-vector coupling constant
measured in $n\to p e^-\bar\nu_e$ decay,
$$
D+F\, = \, g_A \, = \, 1.257\pm 0.003 \, ,
\label{eq:gA}
$$
and the strength of the $\pi N N$ interaction is fixed
in terms of $g_A$:
$$
\fl
T(N\to N\pi^i) \, = \, - i g_{\pi NN} \,
\bar u(p')\gamma_5\tau^i u(p) \, ,
\qquad
g_{\pi NN} \, = \, {g_A M_N\over f} \, = 12.8 \, .
\label{eq:GT}
$$
Equation \ref{eq:GT} is the well-known Goldberger--Treiman (1958)
relation, which is well satisfied by the measured
value $g_{\pi NN} \approx 13.3\pm 0.3$.
The same coupling $g_A$ determines many other interactions with
the pion fields, like the
Kroll--Ruderman (1954) $\overline N \pi\gamma N$ term.

With 3 light flavours, the baryon axial-vector currents get
different contributions from the
$D$ and $F$ terms. Therefore, using
semileptonic hyperon decays, one can make a separate
determination of the two couplings. A fit to the
experimental data, neglecting higher-order corrections,
gives (Luty and White 1993):
$$
D \, = \, 0.85\pm 0.06 \, , \qquad\qquad F \, =\, 0.52\pm 0.04 \, .
\label{eq:d_f_couplings}
$$

The baryon vector currents have of course their canonical form
at zero momentum transfer.
%, i.e.
%$\cL_1^{(B)}\doteq \langle\overline B\gamma^\mu [v_\mu,B]\rangle$.
Whereas the quark axial-vector currents are renormalized by the
strong interaction, giving rise to the $D\pm F$ factors,
the unbroken $SU(3)_V$ symmetry protects the vector currents.

Baryon mass splittings appear at higher orders
in the chiral expansion.
The possible lowest
$\Or ({\cM})$ interactions induced in the effective meson-baryon
Lagrangian are
$$
\cL_\cM^{(B)} \, = \,
- b_0 \, \langle\chi_+\rangle\,\langle\overline B B\rangle
\, - \, b_1 \, \langle\overline B \chi_+ B\rangle
\, - \, b_2 \, \langle\overline B B \chi_+ \rangle \, ,
\label{eq:L_B_M}
$$
where $b_0$, $b_1$ and $b_2$ are  coupling constants with dimensions
of an inverse mass.
The $b_0$ term gives an overall contribution to the common baryon mass
$M_B$, and therefore cannot be extracted from baryon mass splittings.
The other two couplings can be easily determined from the measured
masses, with the result (for $m_u = m_d = \hat m$)
$$
\fl
b_1 \, = \, {M_{\Xi} - M_{\Sigma} \over 4 (M_K^2 - M_{\pi}^2 ) }
\, = \, 0.14 \, {\rm GeV}^{-1}  , \qquad b_2 \, = \,
   {M_{N} - M_{\Sigma} \over 4 (M_K^2 - M_{\pi}^2 ) }
\, = \, -0.28 \, {\rm GeV}^{-1} .
\label{eq:b_1_2}
$$
The Lagrangian $\cL_\cM^{(B)}$ implies the
Gell-Mann (1962)--Okubo (1962) baryon mass relation
$$
M_\Sigma \, + \, 3 M_\Lambda \, = \, 2 (M_\Xi + M_N)  ,
\label{eq:B_GMO}
$$
which is experimentally satisfied to better than 1\%.

\subsection{Loops}

Goldstone loops generate non-analytic corrections to the
lowest-order results (Li and Pagels 1971,
Langacker and Pagels 1973, Pagels 1975).
Due to the different Lorentz structure of meson and baryon fields,
the baryon chiral expansion contains terms of $\Or (p^n)$ for each
positive integer $n$, unlike in the mesonic sector where the
expansion proceeds in steps of two powers of $p$.
This implies additional types of non-analyticity in the baryonic
amplitudes.
The baryon masses, for instance, get calculable corrections
of order $M_\pi^3$,
$$
\delta M_N \sim -{3 g_A^2 M_\pi^3\over 32\pi f^2} \, ,
\label{eq:m_cubic}
$$
i.e. a non-analytic (in the quark masses)
contribution proportional to $\cM^{3/2}$
(Langacker and Pagels 1973).

The structure of
the 1-loop generating functional has been analyzed in the
$SU(2)$ case by Gasser, Sainio and \v Svark (1988), and
later extended to $SU(3)$ by Krause (1990).
The presence of the large mass scale $M_B$ gives rise to a
very complicated power counting. In the meson sector, contributions
from n-loop graphs are suppressed by $(p^2)^n$ and, therefore,
there is
a one-to-one correspondence between the loop and small
momentum expansions.
If baryons are present, however, the loops can produce powers
of the heavy baryon mass instead of powers of the low external
momenta;
the chiral power of the loop contribution is then
reduced.
An amplitude with given chiral dimension may receive contributions
from diagrams with an arbitrary number of loops.
In particular, the coupling constants of the baryon Lagrangian
%$\cL_1^{(B)}$
get renormalized in every order of the loop expansion.
Thus, the evaluation of one-loop graphs associated with
$\cL_1^{(B)}$ produces divergences of $\Or (1)$ and $\Or (p)$ which
renormalize the lowest-order parameters $M_B$, $D$ and $F$;
they give in addition contributions of $\Or (p^2)$
and $\Or (p^3)$, which renormalize higher-order terms in the
chiral Lagrangian.
After appropriate mass and coupling constant renormalization,
higher loops start again to contribute at $\Or (p^2)$.

\subsection{Heavy baryon ChPT}

Since $M_B/\Lambda_\chi\sim \Or (1)$, higher-loop contributions
are not necessarily suppressed.
Thus, in the presence of baryons, the standard chiral expansion
is not only complicated but in addition its convergence is suspect.
The problem can be circumvented by taking the limit $p/M_B<<1$
and making and additional expansion in inverse powers of $M_B$.
Using heavy quark effective theory techniques
(Isgur and Wise 1989, Grinstein 1990,
Eichten and Hill 1990, Georgi 1990),
developed for the study of bottom physics,
Jenkins and Manohar (1991a, 1991b, 1992a) have
reformulated baryon ChPT in such a way as to transfer $M_B$
from the propagators to the vertices (as an inverse scale).

The heavy baryon Lagrangian describes the interactions of a heavy
static baryon with low-momentum pions.
The velocity of the baryon is nearly unchanged when it exchanges
some small momentum with the pion.
The baryon momentum can be written as
$$
p^\mu\, = \, M_B v^\mu \, + \, k^\mu ,
\label{eq:p_baryon}
$$
with $v^\mu$ the four-velocity satisfying $v^2=1$,
and
$k^\mu$ a small off-shell momentum ($k \cdot v<<\Lambda_\chi$).
The effective theory can be
formulated in a Lorentz covariant way by defining
velocity-dependent fields (Georgi 1990)
$$
B_v(x) \,\equiv\, \e^{i M_B v_\mu x^\mu}\, P_v^+\,B(x) ,
\qquad\qquad\qquad P_v^+\,\equiv\, {1+\slashchar{v} \over 2} .
\label{eq:b_v}
$$
The projection operator $P_v^+$ projects onto the particle
portion of the spinor, i.e.
$B_v$ is a two-component spinor.
In the baryon rest frame $v=(1,0,0,0)$ and $B_v$ corresponds
to the usual non-relativistic projection of the Dirac spinor
into the upper-component Pauli spinor.
The new baryon fields obey a modified Dirac equation,
$$
i \slashchar{\partial} B_v \, = \, 0 \, ,
\label{eq:dirac}
$$
which no longer contains a baryon mass term.
Derivatives acting on the field $B_v$ produce factors of $k$, rather
than $p$, so that higher derivative terms in the effective theory
are suppressed by powers of $k/\Lambda_\chi$, which is small.
Moreover, factors of $M_B$ cannot occur in any loop.
Thus, the heavy baryon Lagrangian has a consistent
power-counting expansion
(Weinberg 1990, Ecker 1994a):
the chiral dimension increases with the number of loops and the
lowest-order coupling constants are not renormalized by
higher-order loops.

All Lorentz tensors made from spinors can be written in terms of
$v^\mu$ and a spin operator $S_v^\mu$, that acts on the baryon fields,
defined through the properties (Jenkins and Manohar 1991a)
$$
v\cdot S_v = 0 \, , \quad S^2_v B_v = -{3\over 4} B_v \, ,
\label{eq:S_v_1}\cr
\left\{ S_v^\lambda,S_v^\sigma\right\}={1\over 2}
\left( v^\lambda v^\sigma - g^{\lambda\sigma}\right) ,
\quad
\left[ S_v^\lambda,S_v^\sigma\right] = i
\epsilon^{\lambda\sigma\alpha\beta} v_\alpha (S_v)_\beta \, .
\label{eq:S_v_2}
$$
In the baryon rest frame, $S_v$ reduces to the usual spin operator
$\vec\sigma/2$.

In the heavy baryon formulation,
the equivalent of the lowest-order Lagrangian
$\cL_1^{(B)}+\cL_\cM^{(B)}$  is given by
(Jenkins and Manohar 1991a)
$$
%\fl
\cL_v \, = \,
\langle \overline B_v iv^\mu \nabla_\mu B_v \rangle
\, + \, D \, \langle\overline B_v S_v^\mu\{ \xi_{\mu}, B_v \} \rangle
\, + \, F \, \langle\overline B_v S_v^\mu [ \xi_\mu, B ]\rangle
\cr\qquad
- b_0 \, \langle\chi_+\rangle\,\langle\overline B_v B_v\rangle
\, - \, b_1 \, \langle\overline B_v \chi_+ B_v\rangle
\, - \, b_2 \, \langle\overline B_v B_v \chi_+ \rangle \, .
\label{eq:L_B_v}
$$
The $1/M_B$ effects
(and the antibaryon spinor components)
in the original Dirac theory can be reproduced
in the effective theory by including higher-dimension operators
suppressed by powers of $1/M_B$.
Since $M_B\sim \Lambda_\chi$, the $1/M_B$ and $1/\Lambda_\chi$
expansions can be combined into a single derivative expansion.

A complete analysis of the heavy baryon generating functional
of $\Or (p^3)$ has been recently performed (Ecker 1994b) in the
$SU(2)$ case. The non-analytic pieces have been fully calculated.
Moreover, the $\Or (p^2)$ and some $\Or (p^3)$ couplings have been
either determined from phenomenology or estimated from
resonance exchange
(Bernard \etal 1992, 1994, Jenkins 1992a).
For some particular observables, like the nucleon electromagnetic
polarizabilities, $\Or (p^4)$ calculations already exist
(Bernard \etal 1993a, 1994).
A recent summary of chiral predictions
compared to experimental data has been given by
Mei{\ss}ner (1994).

The status of the three-flavour theory is less satisfactory.
While a complete one-loop analysis is still lacking,
rather large non-analytic
corrections associated with kaon loops have been found in several
observables
(Bijnens \etal 1985, Jenkins and Manohar 1991a, 1992b,
Jenkins 1992a) .
For instance, taking into account the non-analytic one-loop
contributions evaluated at a scale $\mu= M_\rho$,
the fit to the semileptonic hyperon data gives
(Luty and White 1993):
$$
D = 0.60\pm 0.03 \, , \qquad
F = 0.36 \pm 0.02 \, .
\label{eq:d_f_loop_fit}
$$
The difference with the lowest-order determination in equation
\ref{eq:d_f_couplings} is rather large.
Notice, however, that the contributions from local terms in the
chiral Lagrangian have been neglected;
thus, it is not clear how meaningful those results are.
In fact, a fit to the $\pi N$ $\sigma$-term
(Gasser \etal 1991) and to the baryon masses
reveals (Bernard \etal 1993b) that there are
large cancellations between the strange loops and the
counterterms.

Jenkins and Manohar (1991b) have advocated the inclusion
of the spin-${3\over 2}$ decuplet baryons as explicit
degrees of freedom in the low-energy Lagrangian.
Since the octet--decuplet mass splitting is not very large
($\Delta\approx 300$ MeV ),
one can expect significant contributions from these close--by
baryon states. They find that, in the limit $\Delta\to 0$,
decuplet loops tend to compensate
the large octet-loop contributions, improving the convergence of
the chiral expansion.
This approach has been, however, criticized by recent analyses
(Bernard \etal 1993b, Luty and White 1993), which show that
setting $\Delta=0$ gives a very poor approximation to the
decuplet contribution.
In the usual Lagrangian without decuplet fields, the decuplet
effects are already contained in the chiral couplings.
The Jenkins--Manohar approach is nothing else that a way to
make an estimate of the decuplet contribution to those couplings
(and sum some higher-order corrections).
However, their results are still incomplete because they have not
taken into account all possible terms in the Lagrangian, at the
considered order.
Clearly, a complete analysis of the three-flavour theory
is needed in order to clarify these issues.

\subsection{Non-leptonic hyperon decays}

Neglecting the small $(27_L,1_R)$ contribution, the lowest-order
$\Delta S=1$ non-leptonic effective Lagrangian
involving baryons contains two terms (Manohar and Georgi 1984,
Georgi 1984):
$$
\cL^{(B)}_{\Delta S=1} \, = \,
-{G_F\over\sqrt{2}} V_{ud}^{\phantom{*}} V_{us}^* \,\left\{
h_D \,\langle \overline B \{ u \lambda u^\dagger,B\}\rangle
+ h_F \,\langle \overline B [u \lambda u^\dagger,B]\rangle \right\}.
\label{eq:ds_baryon}
$$

The invariant matrix elements for non-leptonic hyperon decays
are conventionally parametrized as
$$
T(B_i\to B_j\pi) \, = \, \bar u_{B_j} \left[A^{(S)}_{ij} +
\gamma_5 A^{(P)}_{ij} \right] u_{B_i} \, ,
\label{eq:ds_amplitude}
$$
where $A^{(S)}_{ij}$ and $A^{(P)}_{ij}$ are the $S$- and
$P$-wave amplitudes, respectively.
The Lagrangian \ref{eq:ds_baryon} implies that the tree-level
$S$-wave amplitudes obey the Lee (1964)--Sugawara (1964) relation:
$$
\sqrt{3\over 2} A^{(S)}_{\Sigma^- n} + A^{(S)}_{\Lambda p}
+ 2 A^{(S)}_{\Xi^-\Lambda} \, = \, 0 \, ,
\label{eq:LS_relation}
$$
which is well satisfied experimentally.
A similar relation for the $P$-wave amplitudes does not exist,
since these involve pole diagrams in which the baryon changes
strangeness before or after pion emission.

Fitting the parameters $h_D$ and $h_F$ to the measured $S$-wave
amplitudes, one obtains a very bad description of the
$A^{(P)}_{ij}$ amplitudes. Therefore, higher-order corrections
are crucial in order to understand these decays.
In fact, keeping only the non-analytic contributions evaluated at
$\mu=1$ GeV, Bijnens \etal (1985) found very large
1-loop corrections,
which spoil the successful Lee-Sugawara relation for the $S$-waves,
and are even larger than the tree-level result
for some $P$-wave amplitudes.

The inclusion of the baryon decuplet
(Jenkins 1992b, Jenkins and Manohar 1992a) seems to
improve the convergence of the chiral expansion.
There is again a large cancellation between
the octet and decuplet loop contributions, reducing considerably
the total loop correction.
For the $S$-wave amplitudes one gets a very good
fit, with a quite small correction to the relation
\ref{eq:LS_relation}.
Although a satisfactory description of the $P$-wave amplitudes
is still not obtained, one finds that in this case
the tree-level result consists of two terms which tend to cancel
to a large extent, for the parameter values determined by the
$S$-wave fit;  normal-size chiral logarithmic corrections
are then of order one compared to the tree-level amplitudes.
Thus, the missing contributions from the relevant local terms
in the effective Lagrangian could easily explain the measured
amplitudes.

Strangeness-changing radiative hyperon decays have been also
analyzed within the chiral framework at the 1-loop level
(Jenkins \etal 1993, Neufeld 1993).

\section{Large-$N_C$ limit, $U(1)_A$ anomaly and strong CP violation}
\label{sec:strongCP}

In the large-$N_C$ limit the $U(1)_A$ anomaly
(Adler 1969, Adler and Bardeen 1969, Bell and Jackiw 1969)
is absent.
The massless QCD Lagrangian \ref{eq:LQCD} has then a larger
$U(3)_L\otimes U(3)_R$ chiral symmetry, and there are
nine Goldstone bosons associated with the SCSB to the diagonal
subgroup $U(3)_V$. These Goldstone excitations can be
conveniently collected in the $3\times 3$ unitary matrix
$$
\fl
\widetilde U(\phi) \,\equiv\,\langle 0|\widetilde U|0\rangle\,
U(\phi) \,\equiv\,
\langle 0|\widetilde U|0\rangle \,
\exp{\left\{ i\sqrt{2}\widetilde\Phi/f\right\}}
\, , \qquad\quad
\widetilde\Phi\,\equiv\, {\eta_1\over\sqrt{3}}\, I_3 \, + \,
{\vec\lambda\over\sqrt{2}}\,\vec\phi \, ,
\label{eq:U_tilde}
$$
where we have explicitly factor out from the $\widetilde U(\phi)$ matrix
its vacuum expectation value.
Under the chiral group, $\widetilde U(\phi)$ transforms as
$\widetilde U\to g_R \widetilde U g_L^\dagger\; $ ($g_{R,L}\in U(3)_{R,L}$).
To lowest order in the chiral expansion, the interactions of the nine
Goldstone bosons are described by the Lagrangian \ref{eq:lowestorder}
with $\widetilde U(\phi)$ instead of $U(\phi)$.
Notice that the $\eta_1$ kinetic term in
$\langle D_\mu\widetilde U D_\mu\widetilde U^\dagger\rangle$ decouples from
the $\phi$'s and the $\eta_1$ particle becomes stable in the
chiral limit (Di Vecchia \etal 1981).

To lowest non-trivial order in $1/N_C$, the chiral symmetry breaking
effect induced by the $U(1)_A$ anomaly can be taken into
account in the effective low-energy theory, through the term
(Di Vecchia and Veneziano 1980, Witten 1980, Rosenzweig \etal 1980)
$$
\cL_{U(1)_A} \, = \, - {f^2 \over 4} {a \over N_C} \,
\left\{ {i \over 2 } \left[\log{(\det{\widetilde U})} - \log{(\det{
\widetilde U^\dagger})}\right] \right\}^2  ,
\label{eq:anom_term}
$$
which breaks $U(3)_L \otimes U(3)_R$ but preserves
$SU(3)_L \otimes SU(3)_R \otimes U(1)_V$. The
parameter $\, a \,$ has dimensions of mass squared and, with the
factor
$1/N_C$ pulled out, is booked to be of $\Or (1)$ in the large-$N_C$
counting rules. Its value is not fixed by symmetry requirements alone;
it depends crucially on the dynamics of instantons. In the presence
of the term \ref{eq:anom_term},
the $\eta_1$-field becomes massive even in the
chiral limit: $M_{\eta_1^2} = 3 a/N_C + \Or (\cM)$.

Deeply related to the $U(1)_A$ anomaly is the possible presence
of an additional term in the QCD Lagrangian,
$$
\cL_\theta\, = \,\theta_0 \, {g^2 \over 64 \pi^2} \,
\epsilon_{\mu \nu \rho \sigma}\,
G^{\mu\nu}_a(x) G^{a,\rho\sigma}(x) \, ,
\label{eq:L_theta}
$$
with $\theta_0$, the so-called vacuum angle, a hitherto unknown
parameter.
This term violates {\sl P, T} and {\sl CP} and may lead
to observable effects in flavour conserving transitions.
Owing to the $U(1)_A$-anomaly, the $\theta_0$-vacuum is not invariant
under $U(1)_A$ transformations,
$g_R = g_L^{\dagger} = \e^{i\beta} I_3$
($N_f = 3$ is the number of light-quark flavours):
$$
\theta_0 \, \longrightarrow \, \theta_0 - 2N_f \beta \, .
\label{eq:theta_tranf}
$$

To simplify the discussion, let us fix the external scalar and
pseudoscalar fields in \ref{eq:extendedqcd}
to the values $s=\widetilde\cM$ and $p=0$, where
$\widetilde\cM$ denotes the full mass matrix emerging from the
Yukawa couplings of the light quarks in the electroweak sector.
In full generality, $\widetilde\cM$ is non-diagonal and non-Hermitian.
However, with the help of an appropriate $SU(3)_L \otimes SU(3)_R$
transformation one can always restrict $\widetilde\cM$ to the form
$$
\widetilde\cM = \exp{\left\{ {i \over 3} \arg(\det{\widetilde\cM})\right\}}
 \, \cM \, ,
\label{eq:M_rot}
$$
with $\cM$ the diagonal (and positive-definite) quark-mass matrix
\ref{eq:q_m_matrices}.
%In the absence of the $\theta_0$-vacuum term in the QCD-Lagrangian,
The phase
$\arg(\det{\widetilde\cM})$ could be reabsorbed by a simple $U(1)_A$
rotation
of the quark fields; however, because of the $U(1)_A$-anomaly,
the $U(1)_A$ rotation which eliminates
$\arg(\det{\widetilde\cM})$
from the mass term  generates a new $\theta_0$-vacuum
$$
\theta\,\equiv\,\theta_0 + \arg (\det{\widetilde\cM}) \, .
\label{eq:theta_eff}
$$
The combination $\theta$ remains invariant under arbitrary
$U(1)_A$ transformations.

In order to analyze the implications of the $\theta$ term on
the effective chiral theory, it is convenient to put the full
$\theta$-dependence on the quark-mass matrix.
Performing an appropriate chiral transformation, we can eliminate
the term \ref{eq:L_theta} from the QCD Lagrangian,
and write the mass matrix in the form
$\widetilde\cM = \exp{\left\{i\theta/3\right\}}\,\cM$.
In the absence of the anomaly term \ref{eq:anom_term}, the $\theta$
phase could
be reabsorbed by the $U(1)_A$ transformation
$$
\widetilde U \, \, \longrightarrow \, \,
\e^{i \theta / 6} \,\, \widetilde U \,\, \e^{i \theta / 6}
  \, .
\label{eq:U_theta_rot}
$$
In the presence of the $U(1)_A$ anomaly, and hence the term
\ref{eq:anom_term}, this transformation generates new physical
interactions:
$$
\cL_2 + \cL_{U(1)_A} \, = \,
{f^2 \over 4} \langle D_\mu\widetilde U D^\mu\widetilde U^\dagger\rangle
\, - \, V(\widetilde U) \, ,
\label{eq:eff_L}
\cr\fl
V(\widetilde U) \, = \, - {f^2 \over 4} \left\{
\langle \widetilde\chi^\dagger\widetilde U \, + \,
\widetilde U^\dagger\widetilde\chi\rangle
\, - \, {a \over N_C} \left\{ {i\over 2} \left[
\log{\left( {\det{\widetilde U}\over \det{\widetilde U^\dagger}}\right)}
\right]
 - \theta \right\}^2 \right\} ,
\label{eq:potential}
$$
where now the matrix
$\widetilde\chi = 2 \widetilde B_0 \cM \equiv
{\rm diag}(\widetilde\chi_u^2,\widetilde\chi_d^2,\widetilde\chi_s^2 )$
is real, positive and diagonal.

If the term proportional to $a/N_C$ were absent,
we could take without loss of generality
$\langle 0|\widetilde U|0\rangle = 1 $  and the diagonal entries
$\widetilde\chi_i^2$
would correspond to the Goldstone boson masses.
In the presence of the anomaly, however,
we should minimize the potential energy $V(\widetilde U)$
in order to fix $\langle 0|\widetilde U|0\rangle $.
With $\widetilde\chi$ diagonal, $\langle 0|\widetilde U|0\rangle$ can be
restricted to be diagonal as well and of the form
$$
\langle 0|\widetilde U|0\rangle \, = \, {\rm diag}(\e^{-i \varphi_u} ,
\e^{-i \varphi_d} , \e^{-i \varphi_s} ) \, .
\label{eq:phi_angles_def}
$$
The minimization conditions $\partial V/\partial\varphi_i = 0 \, $
restrict the $\varphi_i$'s to
satisfy the Dashen (1971)--Nuyts (1971) equations:
$$
\widetilde\chi_i^2 \, \sin{\varphi_i} \, = \, {a \over N_C}
\left(\theta - \sum_j\varphi_j \right)
\equiv \, {a \over N_C} \,\bar \theta
 \, , \qquad   (i=u,d,s) .
\label{eq:Dashen_Nuyts}
$$
The $\varphi_i$'s appearing in the effective Lagrangian can be
reabsorbed in  Hermitian matrices $\chi$ and $H$ defined by
$$
\langle 0|\widetilde U^\dagger |0\rangle \,\widetilde\chi \, \equiv \,
  \chi \, + \, i H  \, , \qquad
  \widetilde\chi^\dagger \,\langle 0|\widetilde U |0\rangle \, \equiv \,
 \chi \, - \, i H \, .
\label{eq:chi_H}
$$
Equations \ref{eq:Dashen_Nuyts} fix $H$ to be
proportional to the unit matrix:
$H  = {a \over N_C} \bar \theta \, I_3$.

The effective bosonic Lagrangian as a functional of $U(\phi)$,
with $\langle 0| U|0\rangle =1$, is then
(Pich and de Rafael 1991b):
$$
\fl
\cL_2 + \cL_{U(1)_A} \, = \, {f^2 \over 4} \Biggl\{
 \langle D_\mu U D^\mu U^\dagger + \chi (U+U^\dagger)\rangle \, - \,
 {a \over N_C} \left\{ \bar\theta^2 - {1 \over 4} \left[
\log{\left({\det U \over \det U^\dagger}\right)}\right]^2 \right\}
 \Biggr. \cr \Biggl.
   - i {a \over N_C} \bar\theta \left\{
\langle U-U^\dagger\rangle - \log{\left(
  {\det U \over \det U^\dagger} \right)} \right\}
  \Biggr\} \, .
\label{eq:final_L}
$$

The diagonalization of the quadratic piece of the Lagrangian gives
rise to the physical fields. In the isospin limit
($m_u=m_d=\hat m$), only the $\eta_1$ and $\eta_8$ mix:
$$
\eta \,  = \, \eta_8 \cos \theta_P \, - \, \eta_1
\sin \theta_P \, , \qquad
\eta' \,  = \, \eta_8 \sin \theta_P \, + \, \eta_1
\cos \theta_P \, .
\label{eq:eta_1_mixing}
$$
{}From the measured pseudoscalar-mass spectrum, one can get the values
of the mixing angle and the parameter $a$:
$\theta_P \approx -20^{\circ}$,
$a = M_\eta^2 + M_{\eta'}^2 - 2 M_K^2 = 0.726\, {\rm GeV}^2$.
Since
$\widetilde\chi_u^2 , \widetilde\chi_d^2  \ll \widetilde\chi_s^2 ,a/N_C \, $
and $\,\hat m \ll m_s$, equations \ref{eq:Dashen_Nuyts} imply
the approximate relation
$$
{a \over N_C} \bar\theta \, \approx \,
{\theta \over \sum_i \widetilde\chi_i^{-2} + N_c/a} \,
\approx \, {1 \over 2 } M_{\pi}^2 \theta \, .
\label{eq:theta_approx}
$$

The last term in \ref{eq:final_L} generates strong CP-violating
transitions between pseudoscalar particles. In particular it induces
the phase-space allowed decays
$\eta_{1,8} \to \pi\pi$.
Comparing the prediction
Br$(\eta\to\pi^+\pi^-)  = 1.8 \times 10^2 \,\theta^2$
with the present experimental upper bound,
Br$(\eta\to\pi^+\pi^-) < 1.5 \times 10^{-3}$,
one gets the limit
$|\theta|< 3 \times 10^{-3}$ (Pich and de Rafael 1991b).

\subsection{Baryon electric dipole moments}

It is completely straightforward to extend the previous analysis
to the baryon sector (Pich and de Rafael 1991b). One simply
writes the matrix $\widetilde U(\phi)$ in terms of the canonical
coset representative,
$$
\fl
\widetilde U(\phi) \, = \,
\tilde\xi_R(\phi) \,\tilde\xi_L^\dagger(\phi) \, ,
\qquad
\tilde\xi_L(\phi)\, =\,\langle 0|\tilde\xi_L|0\rangle\,
u(\phi)^\dagger\, ,
\qquad
\tilde\xi_R(\phi)\, =\,\langle 0|\tilde\xi_R|0\rangle\, u(\phi) \, ,
\label{eq:can_choice}
$$
with $\langle 0|u|0\rangle = 1$ and $U(\phi)=u(\phi)^2$.

{}From equation \ref{eq:chi_H} it follows that
$$
\langle 0|\tilde \xi_R^\dagger |0\rangle \,
 \widetilde\chi \,\langle 0|\tilde \xi_L |0\rangle \, = \,
\langle 0|\tilde\xi_L |0\rangle \,
\langle 0|\tilde\xi_R^\dagger |0\rangle \,\widetilde \chi \, = \,
 \chi + i H \, ,
\label{eq:baryon_H}
$$
where we have used the fact that $\widetilde \chi$ is diagonal and,
without loss of generality,
$\langle 0|\tilde\xi_L |0\rangle $ and
$\langle 0|\tilde\xi_R |0\rangle$
can be restricted to be diagonal as well.

The lowest-order baryon Lagrangian is directly obtained from
equation \ref{eq:L1}, making the obvious replacements
$$
u_\mu \,\longrightarrow\, \tilde\xi_\mu\,\equiv\,
i\,\left\{
\tilde\xi_R^\dagger \left(\partial_\mu - i r_\mu\right)
\tilde\xi_R \, - \,
\tilde\xi_L^\dagger \left(\partial_\mu - i l_\mu\right)
\tilde\xi_L \right\} \, ,
\label{eq:tilde_xi_mu}\cr
\Gamma_\mu \,\longrightarrow\, \widetilde\Gamma_\mu\,\equiv\,
{1\over 2}\,\left\{
\tilde\xi_R^\dagger \left(\partial_\mu - i r_\mu\right)
\tilde\xi_R \, + \,
\tilde\xi_L^\dagger \left(\partial_\mu - i l_\mu\right)
\tilde\xi_L \right\} \, ,
\label{eq:tilde_Gamma}
$$
and adding the additional singlet piece
$$
\Delta\cL_1^{(B)} \, = \, g_S\,\langle\tilde\xi_\mu\rangle\,
\langle\overline B\gamma^\mu\gamma_5 B\rangle \, .
\label{eq:singlet_L_B}
$$
The $\Or (\cM)$ interactions are given by the Lagrangian \ref{eq:L_B_M},
with the change
$$
\chi_+ \,\longrightarrow\, \widetilde\chi_+\,\equiv\,
\tilde\xi_R^\dagger\,\widetilde\chi\,\tilde\xi_L
\, + \,
\tilde\xi_L^\dagger\,\widetilde\chi^\dagger\,\tilde\xi_R \, .
\label{eq:tilde_chi}
$$

Equation \ref{eq:baryon_H} implies that
$$
\widetilde\chi_+\, = \, \chi_+ \, + \, i  {a\over N_C} \,\bar\theta
\, (U^\dagger - U) \, .
\label{eq:chi_change}
$$
Inserting this relation into $\cL_\cM^{(B)}$, leads to a
CP non-conserving meson--baryon interaction term modulated
by the coupling $a\bar\theta/N_C$:
$$
\fl
\cL_{\bar\theta}^{(B)}  =  -i {a\over N_C}\,\bar\theta\,\left\{
  b_0 \, \langle U^\dagger - U \rangle \, \langle\overline B B \rangle \,
+ \, b_1 \, \langle\overline B (U^\dagger -U) B \rangle \,
+ \, b_2 \, \langle\overline B B (U^\dagger -U)\rangle \right\}  .
\label{eq:L_theta_B}
$$

At the one-loop level, the Lagrangian $\cL_{\bar\theta}^{(B)}$
generates baryon electric dipole moments.
The logarithmic chiral contribution is fully
calculable. For the neutron electric dipole moment one obtains
(Pich and de Rafael 1991b):
$$
\fl
d_n^\gamma  =
{a \bar\theta\over N_c} \, {e \over 16\pi^2 f_\pi^2 } \,  \left\{
{M_{\Xi} - M_{\Sigma} \over m_K^2 - m_{\pi}^2 } \, (D + F) \,
 \log{\left( {M_N^2 \over m_{\pi}^2} \right)}
% \Bigr.\cr\Bigl. \hbox{}
+  {M_{\Sigma} - M_N \over m_K^2 - m_{\pi}^2 } \, (D - F) \,
 \log{\left({M_{\Sigma}^2 \over m_K^2}\right)}
\right\}
\cr\label{eq:nedm_formulae}
$$
in agreement with
the old Current Algebra result
(Crewther \etal 1979, Di Vecchia 1980).
Taking into account the uncertainty associated with the unknown
contribution from higher-order local terms in the baryon
Lagrangian, one finally gets
(Pich and de Rafael 1991b):
$$
d_n^\gamma \, = \,
(3.3 \pm 1.8 ) \times 10^{-16} \, \theta \, \, {\rm e \, cm} \, .
\label{eq:edm_th}
$$
{}From a comparison between this result and the experimental
(95\% CL) upper limit
$d_n^\gamma < 11 \times 10^{-26}$ e cm (Altarev \etal 1992),
one concludes that
$$
|\theta | \, < \, 7 \times 10^{-10} \, .
\label{eq:theta_bound}
$$

\section{Interactions of a light Higgs}
\label{sec:light_Higgs}

An hypothetical light Higgs particle provides a good example of the broad
range of application of the chiral techniques.
Its hadronic couplings are fixed by
low-energy theorems
which relate the $\phi\to\phi' h^0$ transition with
a zero-momentum Higgs to the corresponding $\phi\to\phi'$
coupling
(Gunion \etal 1990).
Although, within the Standard Model, the possibility
of a light Higgs boson is already excluded,
an extended scalar sector with additional degrees of
freedom could easily avoid the present experimental
limits.
%leaving the question of a light Higgs open to any
%speculation.

The quark--Higgs interaction can be written down in the general form
$$
{\cal L}_{h^0 \bar q q} \, = \, - {h^0\over v} \,
\left \{ k_d \, \bar d M_d d \, + \, k_u \, \bar u M_u u \right \},
\label{eq:yuk}
$$
where
$v = (\sqrt{2} G_F)^{-1/2} \approx 246 \, {\rm GeV}$,
$M_u$ and $M_d$ are the diagonal mass matrices for up-
and down-type quarks respectively, and the couplings $k_u$
and $k_d$ depend on the model considered.
In the Standard Model, $k_u=k_d=1$, while in the usual
two-Higgs-doublet models
(without tree-level flavour-changing neutral currents)
 $k_d = k_u = \cos{\alpha}/\sin{\beta}$ (model I)
or
$k_d = - \sin{\alpha}/\cos{\beta}$,
$k_u = \cos{\alpha}/\sin{\beta}$ (model~II),
where $\alpha$ and $\beta$ are functions of the
parameters of the scalar potential.

The Yukawa interactions of the light-quark flavours can be trivially
incorporated into the low-energy chiral Lagrangian
through the external scalar field $s$,
together with the light-quark-mass matrix $\cM$:
$$
s = \cM \left\{ 1 + {h^0\over v} (k_d A + k_u B) \right\},
\label{eq:s_matrix}
$$
where $A\equiv {\rm diag}(0,1,1)$ and
$B\equiv {\rm diag}(1,0,0)$.
It remains to compute the contribution from the heavy flavours
$c$, $b$, $t$.
Their Yukawa interactions induce a Higgs--gluon coupling through
heavy-quark loops (Shifman \etal 1978),
$$
\cL_{h^0GG} = {\alpha_s\over 12\pi} \, (n_d k_d + n_u k_u) \,
{h^0\over v}\, G^a_{\mu\nu} G^{\mu\nu}_a \, .
\label{eq:hgg}
$$
Here, $n_d=1$ and $n_u=2$ are the number of heavy quarks of
type down and up respectively.
The operator $G^a_{\mu\nu} G_a^{\mu\nu}$ can be related to the
trace of the energy-momentum tensor; in the
three light-flavour theory, one has
$$
\Theta^\mu_\mu = {\beta_1\alpha_s\over 4 \pi} \,
G^a_{\mu\nu} G_a^{\mu\nu}
\, + \, \bar q \cM q  \, ,
\label{eq:theta}
$$
where $\beta_1=-9/2$ is the first coefficient of the QCD $\beta$-function.
To obtain the low-energy representation of
$\cL_{h^0GG}$ it therefore suffices to replace
$\Theta^\mu_\mu$
and $\bar q \cM q$ by their corresponding expressions
in the effective chiral Lagrangian theory. One gets
(Chivukula \etal 1989, Leutwyler and Shifman 1990,
Prades and Pich 1990),
$$
\cL_{h^0GG}^{\rm eff} = \xi {h^0\over v} {f^2\over 2}\,
\left\{ \langle D_\mu U^\dagger D^\mu U \rangle
+ 3 B_0 \langle U^\dagger \cM + \cM U \rangle \right\} .
\label{eq:hgg_eff}
$$
The information on the heavy quarks, which survives in the
low-energy limit, is contained in the coefficient
$\xi \equiv - (n_d k_d + n_u k_u)/(3\beta_1) = 2 (k_d + 2 k_u)/27$.

Using the chiral formalism, the present experimental
constraints on a very light neutral scalar have
been investigated,
% by Pich \etal (1992),
in the context of two-Higgs-doublet
models.
A Higgs in the mass range $2 m_\mu < M_{h^0} < 2 M_\pi$
can be excluded (within model II), analyzing
the decay $\eta\to\pi^0 h^0$ (Prades and Pich 1990).
A more general analysis (Pich \etal 1992), using the
light-Higgs production channels
$Z\to Z^*h^0$, $\eta'\to\eta h^0$, $\eta\to\pi^0 h^0$
and $\pi\to e\nu h^0$,
allows us to exclude a large area in the parameter space
($\alpha,\beta,M_{h^0}$) of both models
(I and II) for $M_{h^0} < 2 m_\mu$.

\section{Effective theory at the electroweak scale}
\label{ref:electroweak}

In spite of the spectacular success of the Standard Model, we
still do not really understand the dynamics
underlying the electroweak symmetry breaking
$SU(2)_L\otimes U(1)_Y\to U(1)_{\rm QED}$.
The Higgs mechanism provides a renormalizable way
to generate the $W$ and $Z$ masses and, therefore, their
longitudinal degrees of freedom.
However, an experimental verification of this mechanism
is still lacking.

The scalar sector of the Standard Model Lagrangian
can be written in the form
$$
\cL(\Phi) = {1\over 2} \langle D^\mu\Sigma^\dagger D_\mu\Sigma\rangle
- {\lambda\over 16} \left(\langle\Sigma^\dagger\Sigma\rangle
- v^2\right)^2 ,
\label{eq:l_sm}
$$
where
$$
\Sigma \equiv \pmatrix{
\Phi^0 & -\Phi^+ \cr \Phi^- & \Phi^{0*}\cr }
\label{eq:sigma_matrix}
$$
and $D_\mu\Sigma$ is the usual gauge-covariant derivative
$$
D_\mu\Sigma \equiv \partial_\mu\Sigma
+ i g {\vec\tau\over 2} {\buildrel \rightarrow\over W}_\mu \Sigma
- i g' \Sigma {\tau_3\over 2} B_\mu  \, .
\label{eq:d_sigma}
$$
In the limit where the coupling $g'$ is neglected,
$\cL(\Phi)$ is invariant
under global $G\equiv SU(2)_L\otimes SU(2)_C$ transformations
($SU(2)_C$ is the so-called custodial-symmetry group),
$$
\Sigma \, \toG \,
g_L \,\Sigma\, g_C^\dagger , \qquad\qquad
g_{L,C}  \in SU(2)_{L,C} \, .
\label{eq:sigma_transf}
$$

Performing a polar decomposition,
$$
\fl
\Sigma(x) \, = \, {1\over\sqrt{2}}
\left( v + H(x) \right) \, U(\phi(x)) \, , \qquad
U(\phi(x)) \, = \, \exp{\left( i \vec{\tau}
\vec{\phi}(x) / v \right) } \, ,
\label{eq:polar}
$$
in terms of the Higgs field $H$ and the Goldstones
$\vec{\phi}$,
and taking the limit $\lambda>>1$ (heavy Higgs),
we can rewrite $\cL(\Phi)$ in the standard chiral form
(Appelquist and Bernard 1980):
$$
\cL(\Phi) = {v^2\over 4}
\langle D_\mu U^\dagger D^\mu U \rangle  +
\Or\!\!\left({H\over v}\right).
\label{eq:sm_goldstones}
$$
In the unitary gauge $U=1$, this $\Or (p^2)$ Lagrangian
reduces to the usual bilinear gauge-mass term.

Equation \ref{eq:sm_goldstones}
is the universal model-independent interaction of the
Goldstone bosons induced by the assumed pattern of SCSB,
$SU(2)_L\otimes SU(2)_C\longrightarrow SU(2)_{L+C}$.
The scattering of electroweak Goldstone bosons
(or equivalently longitudinal gauge bosons)
is then described by the same formulae as
the scattering of pions, changing $f$ by $v$
(Cornwall \etal 1974, Lee \etal 1977, Chanowitz and Gaillard 1985).
To the extent that the present data are still not sensitive to
the virtual Higgs effects, we have only tested up to now
the symmetry properties of the scalar sector encoded in
equation~\ref{eq:sm_goldstones}.

In order to really prove the particular scalar dynamics
of the Standard Model, we need to test the
model-dependent part involving the Higgs field $H$.
If the Higgs turns out to be too heavy to be directly
produced (or if it does not exist at all!),
one could still investigate the higher-order effects
by applying
the standard chiral-expansion techniques in a completely
straightforward way
(Appelquist 1980, Appelquist and Bernard 1980, Longhitano 1980).
The Standard Model gives definite predictions for the
corresponding chiral couplings of the $\Or (p^4)$ Lagrangian,
which could be tested in future
experiments\footnote{$^{\ddagger\ddagger}$}{
%%%%%%%%%%%%%
There is a large number of publications devoted to this subject:
Dobado and Herrero 1989, Donoghue and Ramirez 1990,
Holdom and Terning 1990, Dawson and Valencia 1991,
Dobado \etal 1991, Georgi 1991, Golden and Randall 1991,
Holdom 1991, Espriu and Herrero 1992,
De Rujula \etal 1992, Dobado and Pel\'aez 1994,
Herrero and Ruiz-Morales 1994, Espriu and Matias 1995.
Many other relevant references can be traced back from these
papers.}.
%%%%%%%%%%%%
It remains to be seen if the experimental determination
of the higher-order electroweak chiral couplings will confirm
the renormalizable Standard Model Lagrangian,
or will constitute an evidence of new physics

\section{Summary}
\label{sec:summary}

ChPT is a powerful tool to study the low-energy interactions
of the pseudoscalar-meson octet.
This effective Lagrangian framework incorporates all the
constraints implied by the chiral symmetry of the
underlying Lagrangian at the quark--gluon level,
allowing for a clear distinction between genuine aspects of
the Standard Model and additional assumptions of variable
credibility, usually related to the problem of long-distance
dynamics.
The low-energy amplitudes of the Standard Model are calculable in
ChPT, except for some coupling constants which are not restricted by
chiral symmetry. These constants reflect our lack of understanding
of the QCD confinement mechanism and must be determined experimentally
for the time being. Further progress in QCD can only improve
our knowledge of these chiral constants,
but it cannot modify the low-energy structure of the amplitudes.

ChPT provides a convenient language to improve our understanding of
the long-distance dynamics. Once the chiral couplings are
experimentally known, one can test different dynamical models,
by comparing the predictions that they give for those couplings with
their phenomenologically determined values.
The final goal would be, of course, to derive the low-energy
chiral constants
from the Standard Model Lagrangian itself.
Although this is a very difficult problem,
the recent attempts done in this direction look quite promising.

It is important to emphasize that:
\item{1.}    ChPT is not a model.
The effective Lagrangian generates the most general S-matrix elements
consistent with analyticity, perturbative unitarity, cluster decomposition and
the assumed symmetries.
Therefore,  ChPT is the effective theory of the Standard Model at
low energies.
\item{2.}  The experimental verification of the ChPT predictions
does not provide a test of the detailed dynamics of the
Standard Model; only the implications of the underlying symmetries
are being proved.
Any other model with identical chiral-symmetry properties would give
rise to the same effective Lagrangian, but with different
values for the low-energy couplings.
\item{3.} The dynamical information on the underlying fundamental
Lagrangian
is encoded in the chiral couplings.
In order to actually test the non-trivial low-energy dynamics
of the Standard Model, one needs first to know the
Standard Model predictions for the chiral couplings.

 In this report I have presented the basic formalism of ChPT
and some selected phenomenological applications.
The ChPT techniques can be applied in many more situations:
any system involving Goldstone bosons can be studied
in a similar way.

\ack
I am grateful to G. Ecker, E. Oset and E. de Rafael for useful discussions
and comments.
This work has been supported in part by CICYT (Spain) under grant
AEN-93-0234.

%%%%%%%%%%%%%%%%%%%%%%%%%%%
\references

\refjl{Adler S L 1969}{\PR}{177}{2426--38}

\refjl{Adler S L and Bardeen W A 1969}{\PR}{182}{1517--36}

\refbk{Adler S and Dashen R F 1968}{Current Algebras}
{(Benjamin, New York)}

\refbk{de Alfaro V, Fubini S, Furlan G and Rossetti C 1973}
{Currents in Hadron Physics}{(North-Holland, Amsterdam)}

\refjl{Alliegro C \etal 1992}{\PRL}{68}{278--81} % (BNL--E777)

\refjl{Altarev I S \etal 1992}{\PL}{B276}{242--6}

\refjl{Amendolia S R \etal 1986}{\NP}{B277}{168--96}

\refbk{Appelquist T 1980}{}{in {\it Gauge Theories and Experiments at High
  Energies}, ed. K C Brower and D G Sutherland (Scottish University Summer
  School in Physics, St. Andrews)}

\refjl{Appelquist T and Bernard C 1980}{\PR}{D22}{200--13}

\refjl{Bando M, Kugo T and Yamawaki K 1988}{Phys. Rep.}{164}{217--314}

\refjl{Bardeen W A 1969}{\PR}{184}{1848--59}

%\refjl{Barker A \etal (FNAL--E731) 1990}{\PR}{D41}{3546--7}

%\refjl{Barr G D \etal 1990}{\PL}{B242}{523--30} % (CERN--NA31)
\refjl{Barr G D \etal 1992}{\PL}{B284}{440--7;
    {\bf B242} (1990) 523--30}

\refjl{Bell J S and Jackiw R 1969}{\NC}{60A}{47--61}

\refjl{Bernard V \etal 1992}{\NP}{B388}{315--45;
    %{Bernard V, Kaiser N, Kambor J and Mei{\ss}ner U-G 1992}
     \PL\ {\bf B309} (1993) 421--5}
%\refjl{Bernard V \etal 1993a}{\PL}{B309}{421--5}
    %{Bernard V, Kaiser N, and Mei{\ss}ner U-G 1993}

\refjl{Bernard V \etal 1993a}{\PL}{B319}{269--75}
     %{Bernard V, Kaiser N, Mei{\ss}ner U-G and Schmidt A 1993b}

\refjl{Bernard V \etal 1993b}{\ZP}{C60}{111--20}
    %{Bernard V, Kaiser N, and Mei{\ss}ner U-G 1993}

%\refjl{\dash 1994a} %{Bernard V, Kaiser N, and Mei{\ss}ner U-G 1994}
%  {}{}{hep-ph/9411287}
   %{Neutral Pion Photoproduction off Nucleons Revisited}{}{CRN 94-62}

\refjl{\dash 1994}{\ZP}{A348}{317--25; {\it hep-ph}/9411287}
  %{Bernard V, Kaiser N, Mei{\ss}ner U-G and Schmidt A 1994b}

\refjl{Bijnens J 1990}{\NP}{B337}{635--51}

\refjl{\dash 1993a}{Int. J. Mod. Phys.}{A8}{3045--105}

\refjl{\dash 1993b}{\PL}{B306}{343--9}

\refjl{Bijnens J, Bruno C and de Rafael E 1993}{\NP}{B390}{501--41}

%\refjl{Bijnens J, Colangelo G, Ecker G and Gasser J 1994}
%  {Semileptonic Kaon Decays}{}{hep-ph/9411311}

\refjl{Bijnens J, Colangelo G and Gasser J 1994}{\NP}{B427}{427--54}

\refjl{Bijnens J and Cornet F 1988}{\NP}{B296}{557--68}

%\refjl{Bijnens J, Ecker G and Gasser J 1992}{\NP}{B396}{81--118}
%  Radiative Semileptonic K-Decays

%\refbk{Bijnens J, Ecker G and Gasser J 1992}{Introduction to Chiral
%   Symmetry}{in {\it The DA$\Phi$NE Physics Handbook}, Vol. I,
%   eds. L. Maiani, G. Pancheri and N. Paver (Frascati), p~107--14}

\refbk{Bijnens J, Ecker G and Gasser J 1994}{Chiral Perturbation
    Theory}{in {\it The DA$\Phi$NE Physics Handbook} (second edition),
    eds. L. Maiani, G. Pancheri and N. Paver (Frascati)}

\refjl{Bijnens J, Ecker G and A. Pich 1992}{\PL}{B286}{341--7}

\refjl{Bijnens J, Sonoda H and Wise M B 1985}{\NP}{B261}{185--98}

\refjl{Botella F J and Lim C S 1986}{\PRL}{56}{1651--4}

\refjl{Bruno C and Prades J 1993}{\ZP}{C57}{585--93}

%\refbk{Buras A J and Harlander M K 1992}{A Top Quark Story:
%Quark Mixing, CP Violation and Rare Decays
%in the Standard Model}{in {\it Heavy Flavours}, eds. A J Buras
%and M Lindner, Advanced Series on Directions in High Energy Physics
%vol~10 (World Scientific, Singapore) p~58--201}

\refjl{Buras A J \etal 1994}{\NP}{B423}{349--83}

\refjl{Burkhardt H \etal 1987}{\PL}{B199}{139--46} % (CERN--NA31)

\refjl{Callan C, Coleman S, Wess J and Zumino B 1969}{\PR}{177}{2247--50}

\refjl{Cappiello L and D'Ambrosio G 1988}{\NC}{99A}{155--161}

\refjl{Cappiello L, D'Ambrosio G and Miragliuolo M 1993}
{\PL}{B298}{423--31}

\refjl{Chanowitz M S and Gaillard M K 1985}{\NP}{B261}{379--431}

\refjl{Cheng H-Y 1990}{\PR}{D42}{72--84}

\refjl{Chivukula R S, Dugan M J and Golden M 1993}{\PR}{D47}{2930--9}

\refjl{Chivukula R S \etal 1989}{\PL}{B222}{258--62;
    \APNY\ {\bf 192} 93--103}

\refjl{Cho Y \etal 1980}{\PR}{D22}{2688--94}

\refjl{Cohen A, Ecker G and Pich A 1993}{\PL}{B304}{347--52}

\refjl{Coleman S, Wess J and Zumino B 1969}{\PR}{177}{2239--47}

\refjl{Cornwall J M, Levin D N and Tiktopoulos G 1974}{\PR}{D10}
   {1145--67}

\refjl{Crewther \etal 1979}{\PL}{88B}{123--7; Err: {\bf 91B} (1980) 487}
    %{R.J.Crewther, P.Di Vecchia, G.Veneziano and E.Witten}

\refjl{Cronin J A 1967}{\PR}{161}{1483--94}

%\refjl{Dally E B \etal 1980}{\PRL}{45}{232--5}
\refjl{Dally E B \etal 1982}{\PRL}{48}{375--8; {\bf 45} (1980) 232--5}

\refjl{D'Ambrosio G and Espriu D 1986}{\PL}{B175}{237--42}

\refjl{Dashen R 1969}{\PR}{183}{1245--60}

\refjl{\dash 1971}{\PR}{D3}{1879--89}

\refjl{Dashen R and Weinstein M 1969}{\PR}{183}{1261--91}

\refjl{Dawson S and Valencia G 1991}{\NP}{B348}{23--46;
      {\bf B352} 27--44}

\refjl{De R\'ujula \etal 1992}{\NP}{B384}{3--58}

\refjl{Devlin T J and Dickey J O 1979}{\RMP}{51}{237--50}

\refjl{Di Vecchia P 1980}{Acta Phys. Austriaca Suppl.}{22}{341--81}

\refjl{Di Vecchia P \etal 1981}{\NP}{B181}{318--34}
   %{P.Di Vecchia, F.Nicodemi, R.Pettorino and G.Veneziano}

\refjl{Di Vecchia P and Veneziano G 1980}{\NP}{B171}{253--72}

\refjl{Dobado A, Espriu D and Herrero M J 1991}{\PL}{B255}{405--14}

\refjl{Dobado A and Herrero M J 1989}{\PL}{B228}{495--502;
     {\bf B233} 505--11}

\refjl{Dobado A and Pel\'aez J R 1994}{\PL}{B329}{469--78;
  \NP\ {\bf B425} 110--36}

\refjl{Donaldson G \etal 1974}{\PR}{D9}{2960--99}

\refjl{Donoghue J F and Gabbiani F 1994}{hep-ph{\rm /9408390}}{}{}
   %{Amherst preprint}{}{UMHEP--410}

\refbk{Donoghue J F, Golowich E and Holstein B R 1992}
{Dynamics of the Standard Model}{(Cambridge Univ. Press,
    Cambridge)}

\refjl{Donoghue J F, Holstein B R and Valencia G 1987}{\PR}{D35}{2769--75}

\refjl{Donoghue J F, Holstein B R and Wyler D 1992}{\PRL}{69}{3444--7;
    \PR\ {\bf D47} (1993) 2089--97}
%\refjl{\dash 1993}{\PR}{D47}{2089--97}

\refjl{Donoghue J F and Ramirez C 1990}{\PL}{B234}{361--6}

\refjl{Donoghue J F, Ramirez C and Valencia G 1989}{\PR}{D39}{1947--55}

%\refjl{Donoghue J F and Wyler D 1992}{\PR}{D45}{892--909}

\refbk{Ecker G 1990}{Geometrical aspects of the non-leptonic weak
   interactions of mesons}{in Proc. IX Int. Conf.
   on the Problems of Quantum Field Theory (Dubna, 1990),
   ed. M.K. Volkov (JINR, Dubna)}

\refbk{\dash 1993}{Chiral Perturbation Theory}{in
	  {\it Quantitative Particle Physics}, eds. M. Levy \etal
   (Plenum Publ. Co., New York)}

\refjl{\dash 1994a}{Czech. J. Phys.}{44}{405--30}

\refjl{\dash 1994b}{\PL}{B336}{508--17}
  %{Chiral Invariant Renormalization of the
  % Pion-Nucleon Interaction}{}{UWThPh-1994-1}

\refjl{\dash 1995}{Prog. Part. Nucl. Phys.}{35}{(in press)
    [{\it hep-ph}/9501357]}

\refjl{Ecker G, Gasser J, Pich A and de Rafael E 1989a}{\NP}{B321}
{311--42}

\refjl{Ecker G, Gasser J, Leutwyler H, Pich A and de Rafael E 1989b}
{\PL}{B223}{425--32}

\refjl{Ecker G, Kambor J and Wyler D 1993}{\NP}{B394}{101--38}

\refjl{Ecker G, Neufeld H and A. Pich 1992}{\PL}{B278}{337--44;
    \NP\ {\bf B413} (1994) 321--52}
%\refjl{\dash 1994}{\NP}{B413}{321--52}

\refjl{Ecker G and Pich A 1991}{\NP}{B366}{189--205}

\refjl{Ecker G, Pich A and de Rafael E 1987a}{\NP}{B291}{692--719}

\refjl{\dash 1987b}{\PL}{B189}{363--8}

\refjl{\dash 1988}{\NP}{B303}{665--702}

\refjl{\dash 1990}{\PL}{B237}{481--7}

\refjl{Eichten E and Hill B 1990}{\PL}{B234}{511--6}

\refjl{Esposito-Far\`ese G 1991}{\ZP}{C50}{255--74}

\refjl{Espriu D and Herrero M J 1992}{\NP}{B373}{117--68}

\refjl{Espriu D and Matias J 1994}{\NP}{B418}{494--528}
\refjl{\dash 1995}{\PL}{B341}{332--41; {\it hep-ph}/9501279}

\refjl{Espriu D, de Rafael E and Taron J 1990}{\NP}{B345}{22--56;
    Err: {\bf B355} (1991) 278--9}

\refjl{Euler E 1936}{\AP}{26}{398}

\refjl{Euler E and Heisenberg W 1936}{\ZP}{98}{714}

\refjl{Fajfer S and G\'erard J-M 1989}{\ZP}{C42}{425--30}

\refjl{Fearing H W and Scherer S 1994}{hep-ph{\rm /9408346; 9408298}}{}{}
      %{TRI-PP-94-68; TRI-PP-94-64}

\refjl{Flynn J and Randall L 1989}{\PL}{B216}{221--6}

\refjl{Gasiorowicz S and Geffen D A 1969}{\RMP}{41}{531--73}

%\refbk{Gasser J 1991}{Chiral Dynamics}{in Proc. Workshop on Physics
%   and Detectors for DA$\Phi$NE, ed. G. Pancheri (Frascati) p~291}

\refbk{Gasser J 1990}{The QCD vacuum and chiral symmetry}{in
  {\it Hadrons and Hadronic Matter}, eds. D. Vautherin \etal
  (Plenum Publ. Co., New York) p~47}

\refjl{Gasser J and Leutwyler H 1982}{Phys. Rep.}{87}{77--169}

\refjl{\dash 1984}{\APNY}{158}{142--210}

\refjl{\dash 1985}{\NP}{B250}{465--516; 517--38; 539--60}

\refjl{Gasser J, Leutwyler H and Sainio M E 1991}{\PL}{B253}
   {252--9; 260--4}

\refjl{Gasser J, Sainio M E and \v Svarc A 1988}{\NP}{B307}{779--853}

%\refjl{Gell-Mann M 1957}{\PR}{106}{1296--300}
\refjl{Gell-Mann M 1962}{\PR}{125}{1067--84}  % GMO

%\refjl{Gell-Mann M and L\'evy M 1960}{\NC}{16}{705}

\refjl{Gell-Mann M, Oakes R J and Renner B 1968}{\PR}{175}{2195--9}

\refjl{Geng C Q and Ng J N 1990}{\PR}{D42}{1509--20}

\refbk{Georgi H 1984}{Weak Interactions and Modern Particle Theory}
{(Benjamin / Cummings, Menlo Park)}

\refjl{\dash 1990}{\PL}{B240}{447--50}

\refjl{\dash 1991}{\NP}{B363}{301--25}

\refjl{Gilman F J and Wise M B 1979}{\PR}{D20}{2392--407}

\refjl{Goity J L 1987}{\ZP}{C34}{341--5}

\refjl{Goldberger M L and Treiman S B 1958}{\PR}{110}{1178--84}

\refjl{Golden M and Randall L 1991}{\NP}{B361}{3--23}

\refjl{Goldstone J 1961}{\NC}{19}{154}

\refjl{Grinstein B 1990}{\NP}{B339}{253--68}

\refjl{Gross D J and Wilczek F 1973}{\PRL}{30}{1343--6}

\refbk{Gunion J F \etal 1990}{The Higgs hunter's guide}{Frontiers in
  Physics Lecture Note Series (Addison--Wesley, New York)}

\refjl{Harris D A \etal 1993}{\PRL}{71}{3918--21}  % (FNAL-E799)

\refjl{Heiliger P and Sehgal L M 1993}{\PR}{D47}{4920--38}

\refjl{Herczeg P 1983}{\PR}{D27}{1512--7}

\refjl{Herrero M J and Ruiz-Morales E 1994}{\NP}{B418}{431--55;
  {\it hep-ph}/9411207}

\refjl{Holdom B 1991}{\PL}{B258}{156--60; {\bf B259} 329--34}

\refjl{Holdom B and Terning J 1990}{\PL}{B247}{88--92}

\refjl{Isgur N and Wise M B 1989}{\PL}{B232}{113--7; {\bf B237} (1990) 527--30}
  %\refjl{\dash 1990}{\PL}{B237}{527--30}

\refjl{Isidori G and Pugliese A 1992}{\NP}{B385}{437--51}

\refjl{Jamin M and Pich A 1994}{\NP}{B425}{15--38}

\refjl{Jenkins E 1992a}{\NP}{B368}{190--203} %Masses

\refjl{\dash 1992b}{\NP}{B375}{561--81} %Hyperons, Non-leptonic

\refjl{Jenkins E and Manohar A V 1991a}{\PL}{B255}{558--62}

\refjl{\dash 1991b}{\PL}{B259}{353--8}

\refbk{\dash 1992a}{Baryon Chiral Perturbation Theory}{in
  Proc. of the Workshop on Effective Field Theories (Dobog\'ok\"o,
  Hungary, 1991), ed. U.-G. Mei{\ss}ner (World Scientific, Singapore)
  113--37}

\refjl{\dash 1992b}{\PL}{B281}{336--40}

\refjl{Jenkins E \etal 1993}{\NP}{B397}{84--104}  %{CERN-TH.6690/92}

\refjl{Kambor J, Donoghue J F, Holstein B R, Missimer J
   and Wyler D 1992}{\PRL}{68}{1818--21}

\refjl{Kambor J and Holstein B R 1994}{\PR}{D49}{2346--55}

\refjl{Kambor J, Missimer J and Wyler D 1990}{\NP}{B346}{17--64}

\refjl{\dash 1991}{\PL}{B261}{496--503}

\refjl{Kaplan D B and Manohar A V 1986}{\PRL}{56}{2004--7}

\refjl{Krause A 1990}{Helvetica Physica Acta}{63}{3--70}

\refjl{Kroll N M and Ruderman M A 1954}{\PR}{93}{233--8}

\refjl{Langacker P and Pagels H 1973}{\PR}{D8}{4595--619; 4620--7;
          {\bf D10} (1974) 2904--17}
%\refjl{\dash 1974}{\PR}{D10}{2904--17}

\refjl{Lee B W 1964}{\PRL}{12}{83--6}

\refjl{Lee B W, Quigg C and Thacker H B 1977}{\PR}{D16}{1519--31}

\refjl{Leutwyler H 1989}{\NP {\rm B} (Proc. Suppl.)}{7A}{42--58}

\refjl{\dash 1990}{\NP}{B337}{108--18}

\refbk{\dash 1991}{Chiral Effective Lagrangians}{in
   {\it Recent Aspects of Quantum Fields}, eds. H. Mitter and
   M. Gausterer, Lecture Notes in Physics vol~396
   (Springer--Verlag, Berlin)}

\refjl{\dash 1994a}{\APNY}{235}{165--203}

\refjl{\dash 1994b}{hep-ph{\rm /9405330}}{}{}
    %{Masses of the Light Quarks}{}{BUTP-94/8}

\refbk{\dash 1994c}{hep-ph{\rm /9406283}}{}{}
   %{Principles of Chiral Perturbation
   %Theory}{Lectures at the Workshop {\it Hadrons 1994}
   %(Gramado, Brasil), BUTP-94/13}

\refjl{Leutwyler H and Roos M 1984}{\ZP}{C25}{91--101}

\refjl{Leutwyler H and Shifman M A 1990}{\NP}{B343}{369--97}

\refjl{Li L F and Pagels H 1971}{\PRL}{26}{1204--6; {\bf 27} 1089--92}

\refjl{Longhitano A C 1980}{\PR}{D22}{1166--75;
    \NP\ {\bf B188} (1981) 118--54}
%\refjl{\dash 1981}{\NP}{B188}{118--54}

\refjl{Luty M A and White M 1993}{\PL}{B319}{261--8}
  %\refjl{\dash 1994}{\NP}{BXXX}{LBL-33993}

\refjl{Maltman K and Kotchan D 1990}{\it Mod. Phys. Lett.}{A5}{2457--64}

\refjl{Manohar A and Georgi H 1984}{\NP}{B234}{189--212}

\refjl{Mei{\ss}ner U-G 1988}{Phys. Rep.}{161}{213--362}

\refbk{\dash 1992 (editor)}{{\rm Proc.} Workshop on
  Effective Field Theories of the
   Standard Model, {\rm Dobog\'ok\"o, Hungary, 1991}}
   {(World Scientific, Singapore)}

\refjl{\dash U-G 1993}{\RPP}{56}{903--96}

\refjl{\dash 1994}{hep-ph{\rm /9411300}}{}{}
  %{Baryon ChPT A.D. 1994}{}{Bonn preprint TK 94 17}

\refjl{Mohapatra R N 1993}{Prog. Part. Nucl. Phys.}{31}{39--76}

\refjl{Molzon W R \etal 1978}{\PRL}{41}{1213--6}

\refjl{Morozumi T and Iwasaki H 1989}{Progr. Theor. Phys.}{82}{371--9}

\refjl{Nambu Y and Jona-Lasinio G 1961}{\PR}{122}{345--58}

\refjl{Neufeld H 1993}{\NP}{B402}{166--94}

\refjl{Neufeld H and Rupertsberger H 1995}{\ZP}{C}{(in press)
   [Wien preprint UWThPh-1994-15]}

\refjl{Nuyts J 1971}{\PRL}{26}{1604; Err: {\bf 27} 361}

%\refjl{Ohl K E \etal (BNL E845) 1990}{\PRL}{64}{2755--8}

\refjl{Okubo S 1962}{Prog. Theor. Phys.}{27}{949}

\refjl{Pagels H 1975}{Phys. Rep.}{C16}{219}

\refjl{Papadimitriou V \etal 1991}{\PR}{D44}{573--6} % (FNAL E731)

%\refjl{Park T-S, Min D-P and Rho M 1993}{Phys. Rep.}{233}{341--95}

\refjl{Particle Data Group 1994, {\it Review of Particle Properties}}
{\PR}{D50}{1173--825}

\refbk{Pich A 1994}{Introduction to Chiral Perturbation Theory}{
    Lectures at the Fifth Mexican School of Particles and Fields
    (Guanajuato, M\'exico, 1992), AIP Conference Proceedings 317
    (New York) p~95--140}

\refjl{Pich A, Guberina B and de Rafael E 1986}{\NP}{B277}{197--230}

\refjl{Pich A, Prades J and Yepes P 1992}{\NP}{B388}{31--52}

\refjl{Pich A and de Rafael E 1991a}{\NP}{B358}{311--82}

\refjl{\dash 1991b}{\NP}{B367}{313--33}

\refjl{Politzer H D 1973}{\PRL}{30}{1346--9}

\refjl{Prades J and Pich A 1990}{\PL}{B245}{117--21}

\refbk{de Rafael E 1995}{Chiral Lagrangians and Kaon CP-Violation}{ in
  {\it CP Violation and the Limits of the Standard Model}, ed. J.F. Donoghue
  (World Scientific, Singapore) [{\it hep-ph}/9502254]}

\refjl{Riggenbach C, Gasser J, Donoghue J F and Holstein B R 1991}
{\PR}{D43}{127--39}

\refjl{Rosenzweig C, Schechter J and Trahern G 1980}{\PR}{D21}{3388--92}

%\refjl{Schwinger J 1957}{\APNY}{2}{407}

\refjl{Schwinger J 1967}{\PL}{B24}{473--6}

\refjl{Sehgal L M 1988}{\PR}{D38}{808--13;
      {\bf D41} (1990) 161--5}
%\refjl{\dash 1990}{\PR}{D41}{161--5}

\refjl{Shifman M A, Vainshtein A I and Zakharov V I 1978}{\PL}{B78}{443--6}

\refjl{Soldate M and Sundrum R 1990}{\NP}{B340}{1--32}

\refjl{Sugawara H 1964}{Prog. Theor. Phys.}{31}{213--21}

\refjl{Urech R 1995}{\NP}{B433}{234--54}

\refjl{Weinberg S 1966}{\PRL}{17}{616--21}

\refjl{\dash 1967a}{\PRL}{18}{188--91;
    \PR\ {\bf 166} (1968) 1568--77}
%\refjl{\dash 1968}{\PR}{166}{1568--77}

\refjl{\dash 1967b}{\PRL}{18}{507--9}

\dash 1977 in {\it A Festschrift  for I.I. Rabi},
   ed. L Motz (Acad. of Sciences, New York) p~185

\refjl{\dash 1979}{Physica}{96A}{327--40}

\refjl{\dash 1990}{\PL}{B251}{288--92;
      \NP\ {\bf B363} (1991) 3--18}
%\refjl{\dash 1991}{\NP}{B363}{3--18}

\refjl{Wess J and Zumino B 1967}{\PR}{163}{1727--35}

\refjl{\dash 1971}{\PL}{37B}{95--7}

\refjl{Witten E 1980}{\APNY}{128}{363--75}

\refjl{\dash 1983}{\NP}{B223}{422--32}

%%%%%%%%%%%%%%

\Figures

\figure{Feynman diagrams for $K_1^0\to\gamma^*\gamma^*$.
  \label{fig:ksgg}}

\figure{Feynman diagram for the $K_1^0\to \mu^+ \mu^-$ decay.
 The $K_1^0 \gamma^* \gamma^*$ vertex is generated through
 the one-loop diagrams shown in figure~1.  %\ref{fig:ksgg}.
 \label{fig:ksmm}}

\figure{$2\gamma$-invariant-mass distribution for
$K_L\to\pi^0\gamma\gamma$:
$\Or (p^4)$ (\dotted ),
$\Or (p^6)$ with $a_V=0$ (\dashed ),
$\Or (p^6)$ with $a_V=-0.9$ (\full ).
The spectrum is normalized to the 50 unambiguous
events of NA31 (Barr \etal 1992), without acceptance corrections.
 \label{fig:spectrum}}

\figure{Measured (Barr \etal 1992)
$2\gamma$-invariant-mass distribution for
$K_L\to\pi^0\gamma\gamma$ (\full ),
and estimated background (\dashed ).
The experimental acceptance is given by the crosses.
The dotted line simulates the $\Or (p^4)$ ChPT prediction.
 \label{fig:spectrum_NA31}}

%\bye

\input  epsf.tex   %epsfig

\vfill\eject

%%%%%%%%%%%%%% FIGURE %%%%%%%%%%%%%%%%%%%%%
{\midinsert
\centerline{\epsfysize=8.0 true cm\epsfbox{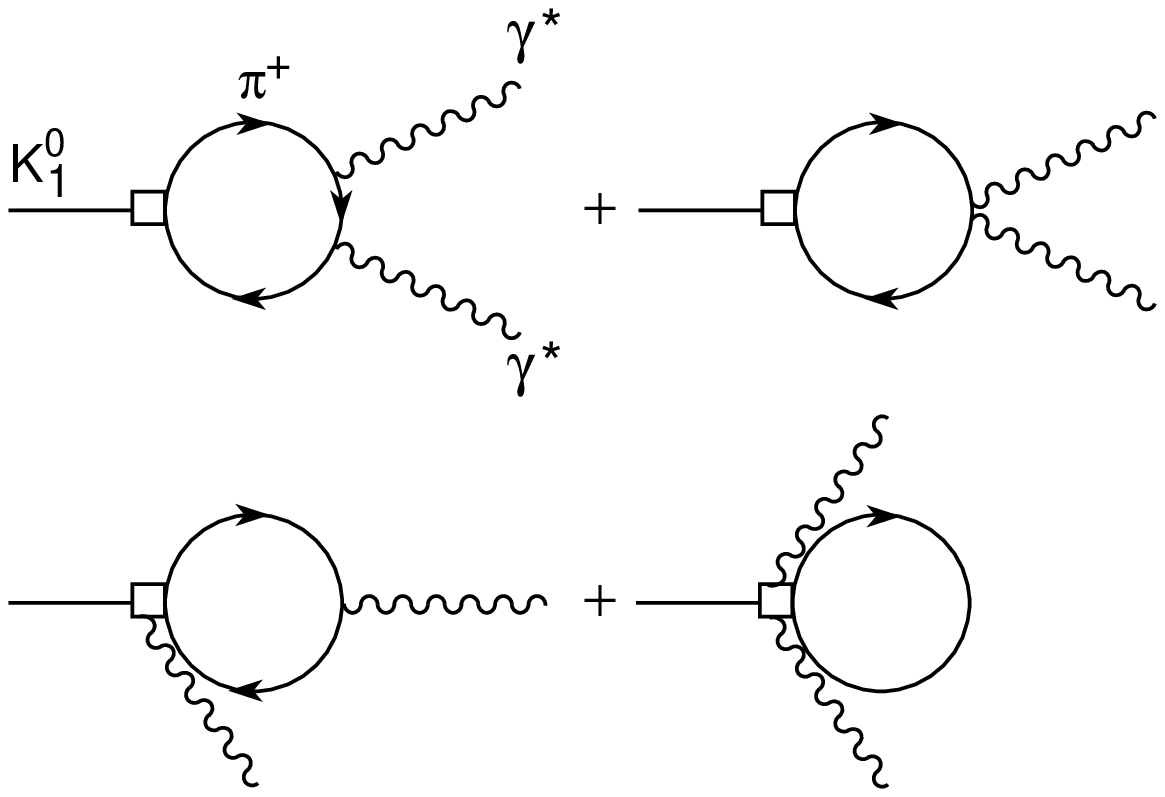}}
     \tenrm\baselineskip=12pt\medskip
\centerline{\noindent{\bf Figure 1.}
 Feynman diagrams for $K_1^0\to\gamma^*\gamma^*$.}
\endinsert}
%%%%%%%%%%%%%%%%%%%% END FIGURE %%%%%%%%%%%%%%%%%%%%%%%%

%\medskip\medskip\medskip\medskip
\vskip 2 true cm
%%%%%%%%%%%%%%%%%%%%%% FIGURE %%%%%%%%%%%%%%%%%%%%%%%%%%%
{\midinsert
\centerline{\epsfysize=5.0 true cm\epsfbox{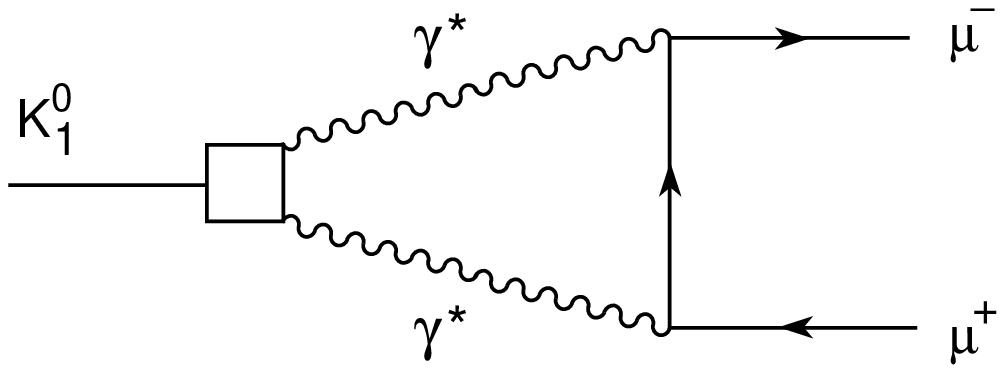}}
     \tenrm\baselineskip=12pt\medskip
\vbox{\noindent{\bf Figure 2.}
Feynman diagram for the $K_1^0\to \mu^+ \mu^-$ decay. The $K_1^0
\gamma^* \gamma^*$ vertex is generated through  the one-loop diagrams
shown in figure~1.}
\endinsert}
%%%%%%%%%%%%%%%%%%%% END FIGURE %%%%%%%%%%%%%%%%%%%%%%%%

\vfill\eject

%%%%%%%%%%%%%%%%%%%%%  Figures %%%%%%%%%%%%%%%%%%%%%%%%%%
{\midinsert
\centerline{\epsfysize=17.0 true cm\epsfbox{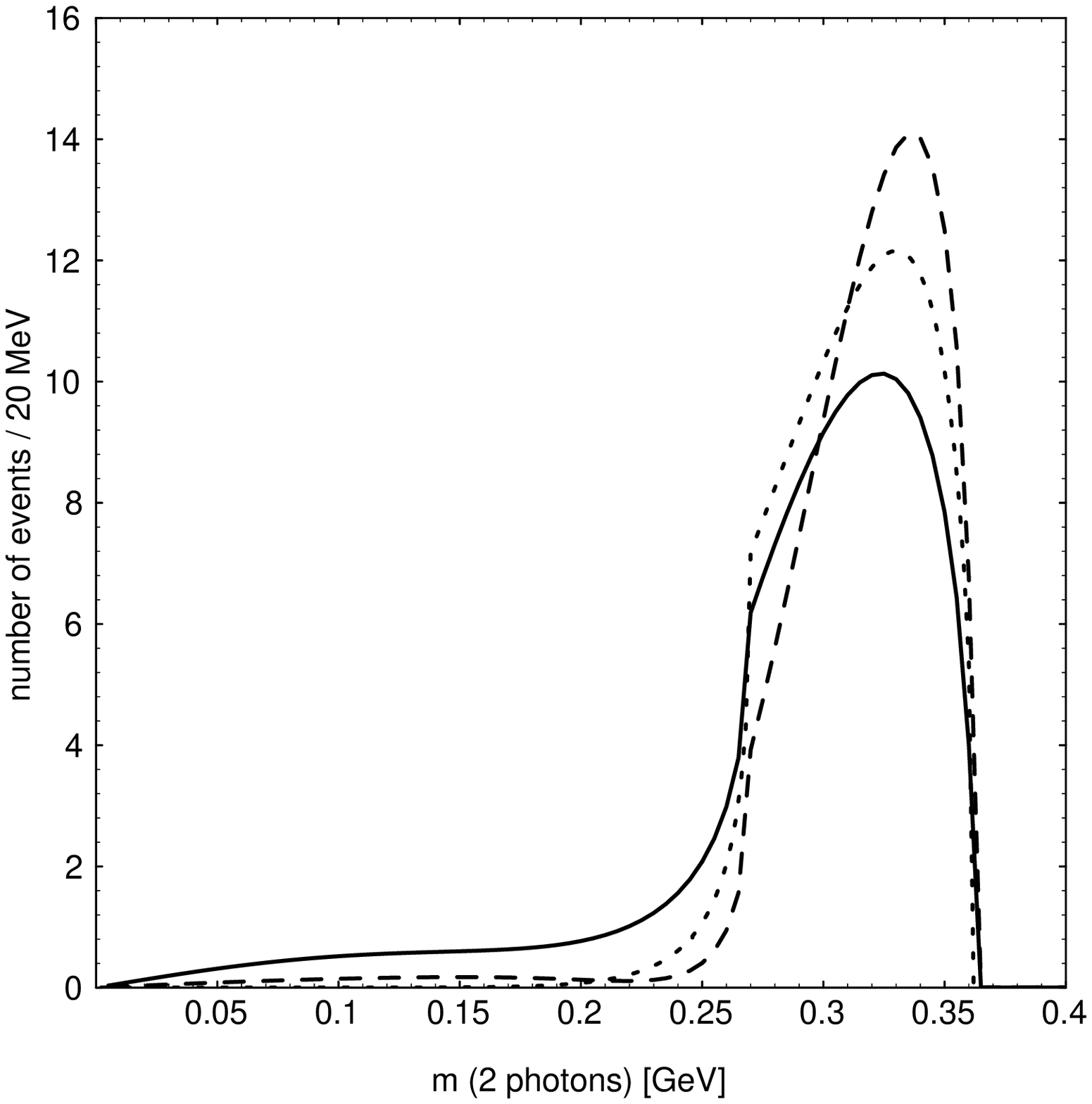}}
   \tenrm\baselineskip=12pt\medskip
\vbox{\noindent{\bf Figure 3.}
$2\gamma$-invariant-mass distribution for
$K_L\to\pi^0\gamma\gamma$:
$\Or (p^4)$ (\dotted ),
$\Or (p^6)$ with $a_V=0$ (\dashed ),
$\Or (p^6)$ with $a_V=-0.9$ (\full ). The spectrum is normalized to
the 50 unambiguous events of NA31 (Barr \etal 1992), without
acceptance corrections.}
\endinsert}
\medskip
{\midinsert
\centerline{\epsfysize=17.0 true cm\epsfbox{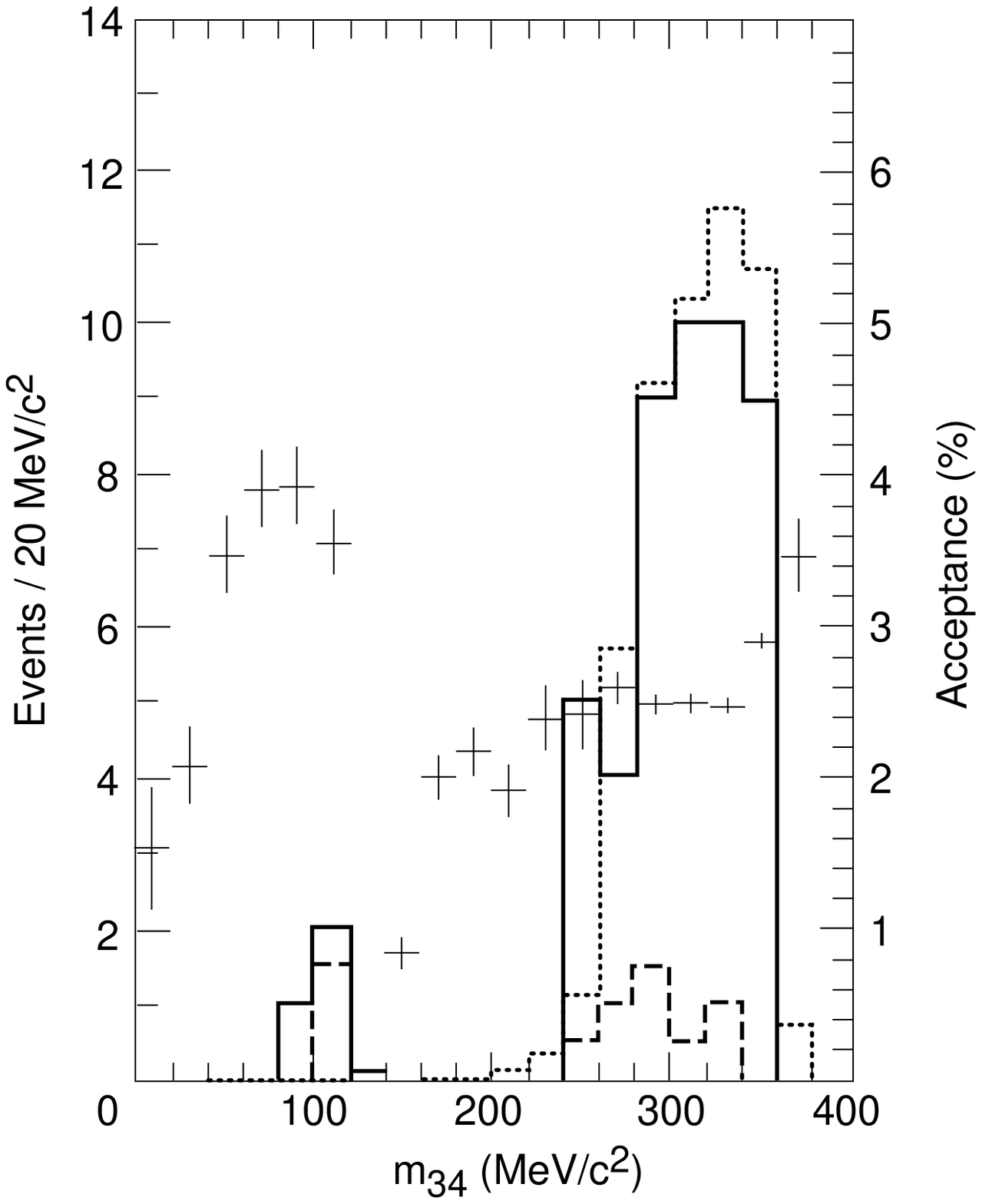}}
    \tenrm\baselineskip=12pt\medskip
\vbox{\noindent{\bf Figure 4.}
Measured (Barr \etal 1992)
$2\gamma$-invariant-mass distribution for
$K_L\to\pi^0\gamma\gamma$ (\full ), and estimated background (\dashed).
The experimental acceptance is given by the crosses. The dotted
line simulates the $\Or (p^4)$ ChPT prediction.}
\endinsert}
%%%%%%%%%%%%%%%%%%%%% End figures %%%%%%%%%%%%%%%%%%%%%%%%%%

\bye